\documentclass[]{./Outils_latex/aa}

\usepackage[utf8]{inputenc} 

\usepackage{natbib}         
\usepackage{graphicx}
\usepackage{longtable}
\usepackage{xcolor}
\usepackage{amssymb}
\usepackage[switch]{lineno} 
\usepackage{hyperref}
\usepackage{rotating}

\bibpunct{(}{)}{;}{a}{}{,} 

\usepackage{soul} 

\hypersetup{
colorlinks = true,
breaklinks = true,
urlcolor = blue,
linkcolor = blue,
citecolor = blue,
}

\newcommand{\ind}[1]{_{\mathrm{#1}}}
\newcommand{\diff}{\mathrm{d}}

\newcommand\Kepler{\emph{Kepler}}
\newcommand\ktwo{K2}
\newcommand\tess{TESS}
\newcommand\corot{CoRoT}
\newcommand\macho{MACHO}
\newcommand\ogle{OGLE}

\newcommand\nunl{\nu\ind{n,\ell}}
\newcommand\numax{\nu\ind{max}}
\newcommand\nmax{n\ind{max}}
\newcommand\dnures{\delta\nu\ind{res}}
\newcommand\Dnu{\Delta\nu}

\newcommand\Teff{T\ind{eff}}
\newcommand\Dnuobs{\Delta\nu\ind{obs}}

\newcommand\eps{\varepsilon}
\newcommand\dol{d\ind{0\ell}}

\newcommand\Vl{V_{\ell}^{2}}
\newcommand\Vlodd{V\ind{\ell,odd}^{2}}
\newcommand\Vleven{V\ind{\ell,even}^{2}}

\newcommand\Gammazero{\langle\Gamma\ind{0}\rangle}
\newcommand\Gammaone{\langle\Gamma\ind{1}\rangle}
\newcommand\Gammatwo{\langle\Gamma\ind{2}\rangle}
\newcommand\Gammal{\langle\Gamma\ind{\ell}\rangle}

\newcommand\Amplzero{\langle A\ind{0,bol} \rangle}
\newcommand\Amplone{\langle A\ind{1,bol} \rangle}
\newcommand\Ampltwo{\langle A\ind{2,bol} \rangle}

\newcommand\tauHeII{\tau_{\mathrm{HeII}}}
\newcommand\tHeII{t_{\mathrm{HeII}}}

\usepackage{amstext}

\begin{document}

\title{Seismic constraints on the internal structure of evolved stars: From high-luminosity RGB to AGB stars\thanks{Full Table~\ref{Table:data_CDS} is only available in electronic form at the CDS via anonymous ftp to cdsarc.u-strasbg.fr(130.79.128.5)}}
\titlerunning{Seismic analysis of evolved giants}

\author{%
 G. Dr\'eau\inst{1},
 B. Mosser\inst{1},
 Y. Lebreton\inst{1,2},
 C. Gehan\inst{3},
 T. Kallinger\inst{4}
}

\institute{
\inst{1} LESIA, Observatoire de Paris, PSL Research University, CNRS, Universit\'e Pierre et Marie Curie,
 Universit\'e Paris Diderot,  F-92195 Meudon, France \\
 email: \texttt{Guillaume.Dreau@obspm.fr}\\
\inst{2} Univ Rennes, CNRS, IPR (Institut de Physique de Rennes) - UMR 6251, F-35000 Rennes, France\\
\inst{3} Instituto de Astrof\'isica e Ci\^encias do Espa\c co, Universidade do Porto, CAUP, Rua das Estrelas, PT4150-762 Porto, Portugal \\
\inst{4} Institut f\"ur Astrophysik, Universit\"at Wien, Türkenschanzstrasse 17, 1180 Vienna, Austria \\
 }


\abstract{The space-borne missions \corot\ and \Kepler\ opened up a new opportunity for better understanding stellar evolution by probing stellar interiors with unrivalled high-precision photometric data. \Kepler\ has observed stellar oscillation for four years, which gave access to excellent frequency resolution that enables deciphering the oscillation spectrum of evolved red giant branch and asymptotic giant branch stars.}
%
{The internal structure of stars in the upper parts of the red and asymptotic giant branches is poorly constrained, which makes
the distinction between red and asymptotic giants difficult. We perform a thorough seismic analysis to address the physical conditions inside these stars and to distinguish them.}
%
{We took advantage of what we have learnt from less evolved stars. We studied the oscillation mode properties of $\sim$ 2.000 evolved giants in a model described by the asymptotic pressure-mode pattern of red giants, which includes the signature of the helium second-ionisation zone. Mode identification was performed with a maximum cross-correlation method. Then, the modes were fitted with Lorentzian functions following a maximum likelihood estimator technique.}
{We derive a large set of seismic parameters of evolved red and asymptotic giants. We extracted the mode properties up to the degree $\ell = 3$ and investigated their dependence on stellar mass, metallicity, and evolutionary status. We identify a clear difference in the signature of the helium second-ionisation zone between red and asymptotic giants. We also detect a clear shortage of the energy of $\ell = 1$ modes after the core-He-burning phase. Furthermore, we note that the mode damping observed on the asymptotic giant branch is similar to that observed on the red giant branch.}
{We highlight that the signature of the helium second-ionisation zone varies with stellar evolution. This provides us with a physical basis for distinguishing red giant branch stars from asymptotic giants. Here, our investigation of stellar oscillations allows us to constrain the physical processes and the key events that occur during the advanced stages of stellar evolution, with emphasis on the ascent along the asymptotic giant branch, including the asymptotic giant branch bump.}

\maketitle

\section{Introduction}



Red giant star seismology has proved to be a good tool for constraining the stellar internal structure with the ultra-high precision photometric data recorded by  \corot\ \citep[Convection, Rotation and planetary Transits,][]{2006ESASP1306...33B}, \Kepler\ \citep{2010Sci...327..977B, 2010PASP..122..131G}, \ktwo\ \citep[Kepler 2,][]{2014PASP..126..398H}, and now \tess\ \citep[Transiting Exoplanet Survey Satellite,][]{2015JATIS...1a4003R}. In the case of evolved giants observed by \Kepler, recent studies have found an equivalence between the solar-like oscillation ridges and the period-luminosity sequences \citep{2013A&A...559A.137M, 2014ApJ...788L..10S, 2020MNRAS.493.1388Y} that have first been identified in the ground-based observations with the microlensing surveys \macho\ \citep[Massive Compact Halo Objects,][]{1999IAUS..191..151W} and \ogle\ \citep[Optical Gravitational Lensing Experiment,][]{2004MNRAS.349.1059W, 2013ApJ...763..103S}. Nevertheless, deciphering the oscillation spectrum of evolved red giant branch (RGB) and asymptotic giant branch (AGB) stars is challenging because it requires long time-series for the modes to be resolved; the lifetime of the modes is longer than one year. Fortunately, with the unrivalled four-year time series of \Kepler , it is now possible to decipher the low-frequency oscillation spectrum of evolved red giants and asymptotic giants in detail. The pressure modes of red giants follow a clear oscillation pattern. The so-called universal pattern (UP) of red giants reads \citep{2011A&A...525L...9M}

\begin{equation}
\label{eq:nunl_asympt}
\nunl^{\mathrm{UP}} = \left(n + \frac{\ell}{2} + \eps - \dol + \frac{\alpha}{2}[n - \nmax]^{2}\right)\Dnu,
\end{equation}
where $n$ is the mode radial order, $\ell$ is the degree, $\eps$ is the acoustic offset that allows locating the radial modes, $\Dnu$ is the mean large frequency separation, which is the mean frequency spacing between consecutive radial modes, $\dol$ is a reduced small separation defined as $\dol = \delta\nu_{0\ell}/\Dnu,$ where $\delta\nu_{0\ell}$ is the small frequency separation between a mode of degree $\ell$ and its neighbouring radial mode, $\alpha = (\diff \log{\Dnu}/\diff n)$ is the curvature term that accounts for the linear dependence of the large frequency separation on the radial order, and $\nmax = \numax/\Dnu$ is the equivalent radial order corresponding to the frequency of the maximum oscillation power $\numax$. The reduced small separations $\dol$ are sensitive to any structure change that impacts the gradient of the sound speed in the deep interior \citep{1986HiA.....7..283G}. These reduced small separations can be used to distinguish different stellar evolutionary stages \citep{1988IAUS..123..295C}.\\


Firstly identified in red giants by \cite{2011Sci...332..205B}, mixed modes that result from the coupling between gravity waves trapped in the stellar core and pressure waves trapped in the stellar envelope carry valuable information on the physical conditions inside the stellar core. The use of mixed modes enables distinguishing core-helium-burning giants and shell-hydrogen-burning giants \citep{2011Sci...332..205B, 2011Natur.471..608B, 2017MNRAS.466.3344E}. However, constraining the internal innermost structure of evolved giants is challenging because their oscillation spectrum only exhibits pure pressure modes. Mixed modes can no longer be identified because the inertia of the g modes in the core becomes too high \citep{2014A&A...572A..11G} and the strength of the coupling between p and g modes decreases \citep{2017A&A...600A...1M}. Despite the absence of mixed modes in evolved RGB and AGB stars, some methods can still be used to distinguish shell-H-burning stars from He-burning stars\footnote{We use the expressions shell-H-burning stars and RGB stars in an equivalent manner. Core-He-burning stars and shell-He-burning stars refer to clump and AGB stars, respectively. He-burning stars indistinctly refer to core-He-burning stars and shell-He-burning stars.}. On the basis of a local analysis, \cite{2012A&A...541A..51K} showed that we can distinguish stars with different evolutionary stages using the central acoustic offset $\eps_{\mathrm{c}}$ \footnote{This central acoustic offset $\eps_{\mathrm{c}}$ is a local measurement of $\eps$ that is computed with the central three radial modes that are closest to $\numax$}.
In addition, \cite{2019A&A...622A..76M} found that He-burning stars have a lower envelope autocorrelation function than their RGB counterparts\footnote{Counterparts refer to stars that have the same $\Dnu$ and $\numax$}, making the separation between these stellar populations possible. \\


The stellar evolution effects reported by \cite{2012A&A...541A..51K} in the acoustic offset $\eps$ can be linked to clear stellar structure differences. The acoustic offset is expected to contain a contribution from the stellar core, hence the signature of structure changes \citep{2000MNRAS.317..141R, 2003A&A...411..215R}. However, it also contains a contribution from the stellar envelope that is dominant \citep{2014MNRAS.445.3685C}. Then, the effects of structure changes in the stellar envelope such as acoustic glitches can be seen in the acoustic offset $\eps$. We recall that a glitch is a sharp structural variation inside the star that causes a modulation in the frequency pattern. The existence of such regions was first predicted \citep{1988IAUS..123..151V, 1990LNP...367..283G} and then confirmed for the Sun \citep{2007MNRAS.375..861H} for main-sequence stars \citep{2012A&A...540A..31M, 2014ApJ...782...18M, 2014ApJ...790..138V, 2016A&A...589A..93D} and for red giants \citep{2010A&A...520L...6M, 2014MNRAS.440.1828B, 2015A&A...579A..84V, 2015A&A...578A..76C}. In stellar interiors, three regions with sharp variations have been studied: the base of the convective envelope, the boundary of the convective core, and the helium second-ionisation zone \citep{1994A&A...283..247M, 2005MNRAS.361.1187M, 2007MNRAS.375..861H, 2016A&A...589A..93D}. In the case of red giants, it has been shown that the dominant glitch has its origin in the helium second-ionisation zone \citep{2010A&A...520L...6M}. The modulation in the mode frequencies has been measured for RGB stars and clump stars \citep{2015A&A...579A..84V}. Vrard et al. (2015) showed that the different modulations between these populations are linked to stellar evolution effects in the local acoustic offset $\eps$.
One of the guidelines of the present work is to perform such an analysis for stars in evolved stages on the RGB and the AGB. \\



Other physical processes can be constrained through the analysis of oscillation spectra, such as mode excitation and damping, especially by measuring the mode amplitudes and the widths. While the physical mechanism causing pressure mode excitation is identified as the Reynolds stresses induced by turbulent convection \citep{1977ApJ...212..243G, 2006A&A...460..183B}, the physical mechanisms behind the mode damping are not fully understood. Nevertheless, recent studies have been conducted to compare modelled and observed mode widths across the Hertzsprung-Russell (HR) diagram \citep{2012A&A...540L...7B, 2017MNRAS.464L.124H, 2018MNRAS.478...69A}. They highlighted that the perturbation of turbulent pressure is the dominant mechanism of mode damping in solar-like pulsators. Several studies have already provided mode widths for main-sequence stars \citep[e.g.][]{2012A&A...543A..54A, 2014A&A...566A..20A, 2017ApJ...835..172L} and red giant stars \citep[e.g.][]{2011A&A...529A..84B, 2012ApJ...757..190C, 2015A&A...579A..83C, 2017MNRAS.472..979H}, but their samples of stars are small. With a larger sample of stars having $\Dnu \in$ [3,15]$\ \mu$Hz, \cite{2018A&A...616A..94V} showed that the pressure mode widths of RGB stars and clump stars are differently distributed and have noticeable mass and temperature dependences. We performed such an analysis for stars in the most evolved stages on the RGB and the AGB. 



In this framework, we analysed the oscillation spectrum of $\sim 2000$ evolved red giants, clump stars, and asymptotic giants observed by the \Kepler\ telescope in detail. We extend the analysis of \cite{2015A&A...579A..84V} and \cite{2018A&A...616A..94V} to the most evolved stages of stars on the RGB and on the AGB. We characterised the pressure modes of evolved stars and the modulation induced by the helium second-ionisation zone in order to obtain seismic constraints for the stellar modelling of evolved red giants and asymptotic giants. \\
This article is organised as follows. In Sect. \ref{sec:dataset} we describe our set of data. In Sect. \ref{sec:Method} we describe the methods we used to extract the seismic parameters from the oscillation spectra, namely the seismic parameters involved in Eq.~\ref{eq:nunl_asympt}, the signature of the helium second-ionisation zone, the visibilities of the modes, the pressure mode widths, and the pressure mode amplitudes. The analysis of these quantities is performed in Sect.~\ref{sec:results}. Finally, Sects.~\ref{sec:discussion} and \ref{sec:conclusion} are devoted to discussion and conclusions, respectively.

\section{Data set}
\label{sec:dataset}



\begin{figure}[ht]
        \begin{minipage}{1.\linewidth}  
                        \rotatebox{0}{\includegraphics[width=1.0\linewidth]{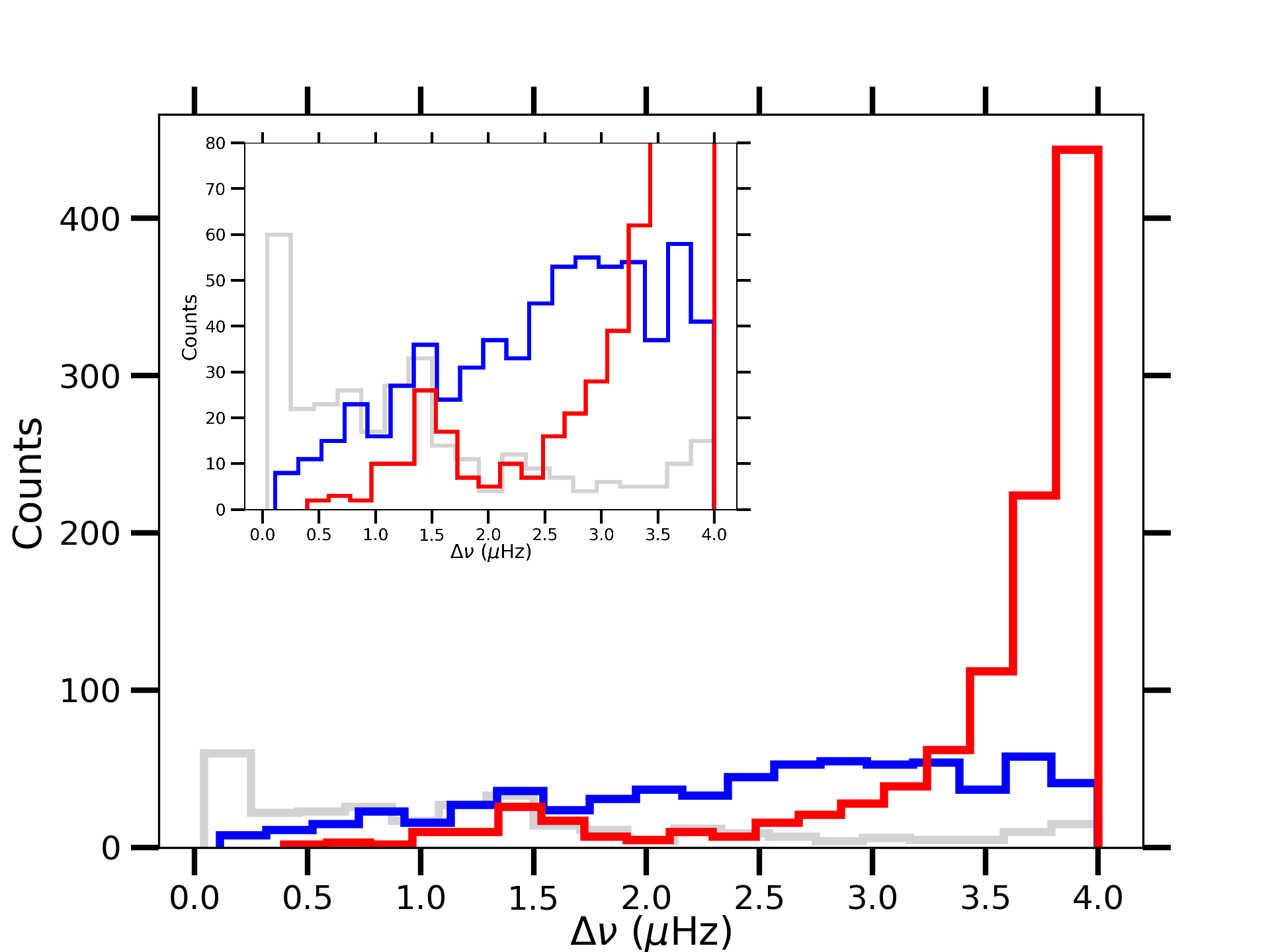}}
                        \rotatebox{0}{\includegraphics[width=1.0\linewidth]{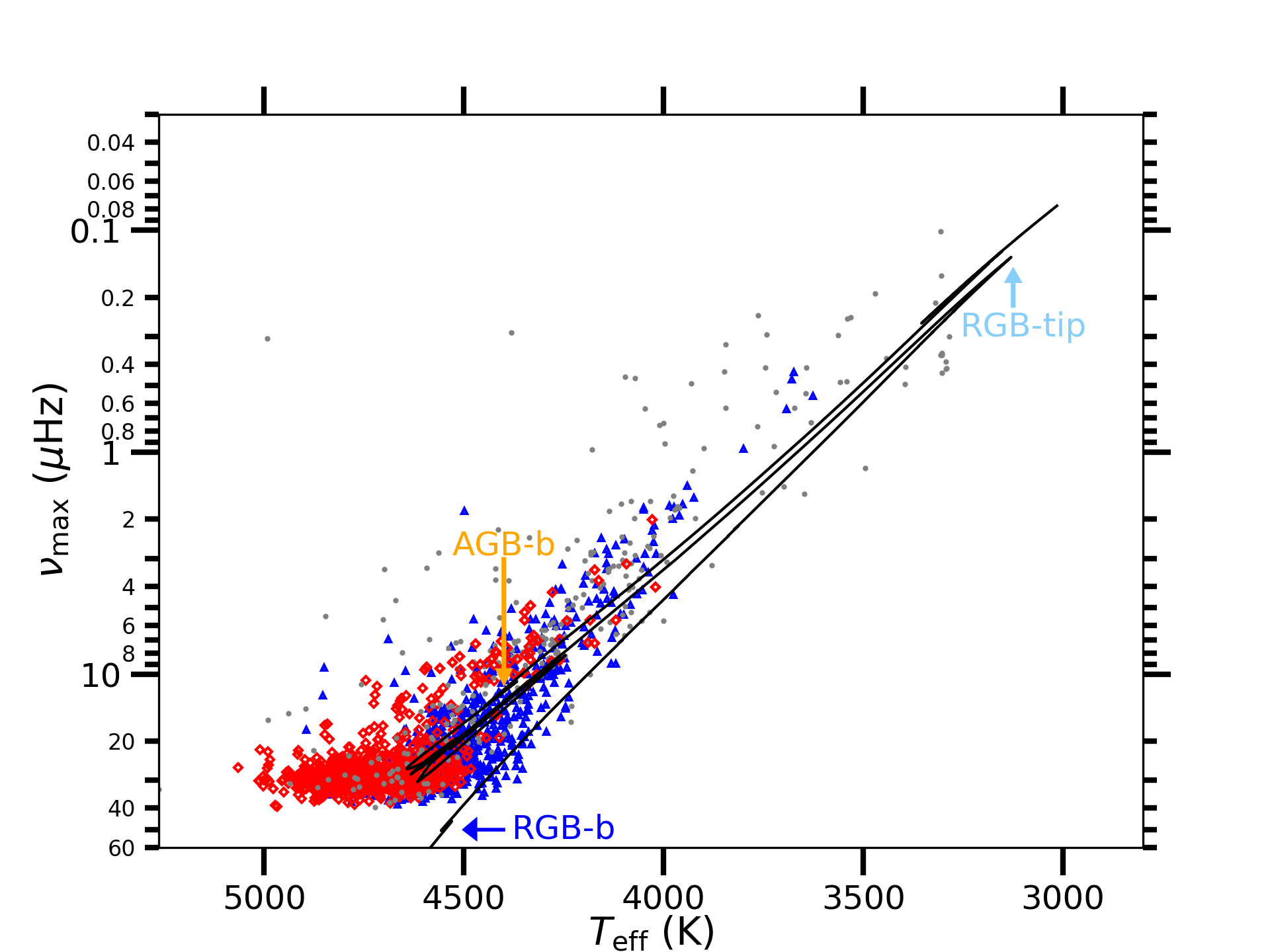}}
        \end{minipage}
        \caption{Upper panel: Distribution of our sample of stars as a function of $\Dnu$, with red giants in blue and He-burning stars in red. Stars with unidentified or uncertain evolutionary stage are plotted in grey. The inset is a zoom-in portion of the large panel. Lower panel: Seismic diagram of our sample of stars with the same colour code as in the upper panel, where 1/$\numax$ is a proxy for the luminosity. The solid black line is the evolutionary track of a 1$M_{\odot}$ model computed with MESA, using the 1M\_pre\_ms\_to\_wd test suite case. Some key events are highlighted: the RGB bump (RGB-b), the luminosity tip of the RGB (RGB-tip), and the AGB bump (AGB-b).
        }
        \label{fig:sample_of_stars}
\end{figure}

We selected the long-cadence data from \Kepler , including the very last data up to quarter Q17. The about 1470-day time-series gives access to a frequency resolution reaching 7.8 nHz. We focus on advanced stages of stellar evolution, including RGB, clump, and AGB giants. 
We selected 2103 stars from \citet{2012A&A...541A..51K} and \citet{2014A&A...572L...5M, 2019A&A...622A..76M} that have $\Dnu \leq 4.0$ $\mu$Hz. We then extracted their $\Dnu$ and $\numax$ from the database of the previous works. The distribution of their $\Dnu$ is shown in Fig.~\ref{fig:sample_of_stars}. The classical properties of these stars, such as their mass and effective temperature, were extracted from the APOKASC catalogue \citep{2014ApJS..215...19P}, which is a survey of \Kepler\ asteroseismic targets complemented by spectroscopic data. More precisely, the stellar masses were computed according to the semi-empirical asteroseismic scaling relation presented in \citet{1995A&A...293...87K} as corrected by \cite{2018ApJS..239...32P}. The correcting factor was computed star by star and is a function of the stellar parameters. For some stars, the classical properties are not listed in the APOKASC catalogue either because no asteroseismic data were returned for them or because the power spectra were too noisy. This concerns roughly 5\% of our sample of stars, with half of this fraction being associated with very low $\Dnu$-values (i.e. $\Dnu \leq 0.5\ \mu$Hz). In this case, we nevertheless obtained rough estimates of the stellar mass and effective temperature using semi-empirical and empirical scaling relations implying both the frequency at the maximum oscillation power $\numax$ and large frequency separation $\Dnu$ \citep{1995A&A...293...87K, 2010A&A...509A..77K, 2010A&A...517A..22M}. \\
In order to identify the evolutionary status, we used two classification methods. The first method is based on the estimate of differences between RGB stars and He-burning stars in the pressure-mode pattern, mainly through the acoustic offset $\eps$ \citep{2012A&A...541A..51K}. The second method is based on the estimate of differences in the envelope autocorrelation function \citep{2019A&A...622A..76M}. However, the disagreement between these two classification methods rapidly grows at low $\Dnu$. For example, we reach 35\% disagreement for 112 stars having $\Dnu \leq 1.0$ $\mu$Hz. Accordingly, we decided to only retain the evolutionary status so obtained if both classification methods agree.


\section{Method}
\label{sec:Method}

Acoustic modes dominate in the oscillation spectrum of evolved RGB and AGB stars. Gravity-dominated mixed modes start to disappear in the oscillation spectrum when $\Dnu \leq 3\ \mu$Hz because of their high radiative damping and inertia \citep{2009A&A...506...57D}. The oscillation pattern of evolved stars can then be described by the asymptotic expression of the frequency of acoustic modes (Eq.~\ref{eq:nunl_asympt}).

\subsection{Adjusting the mode frequencies $\nunl$}
\label{sec:Dnu_adjust}


\subsubsection{Best-matching template spectrum}
\label{sec:best_matching_template_spectrum}

The first step to be performed is the identification of the modes in the oscillation spectrum, which is ensured by using Eq.~\ref{eq:nunl_asympt}. First, we refined the analysis of the observed spectrum as follows. The background component that is dominated by the stellar granulation \citep{2008Sci...322..558M} was parametrised in the vicinity of $\numax$ by a power law of the form

\begin{equation}
\label{eq:background}
    B(\nu) = B_{\mathrm{max}}\left(\frac{\nu}{\numax}\right)^{\alpha_{B}},
\end{equation}
where $B_{\mathrm{max}}$ and $\alpha_{B}$ are free parameters \citep{2012A&A...537A..30M}. Then, we divided the observed spectrum by the background contribution. For the sake of visibility, we reduced the stochastic appearance of the oscillation pattern by smoothing the spectrum with a Gaussian function, for which the full width at half maximum (FWHM) is $\mathrm{FWHM} = d_{02}\Dnu/4$. We estimated $d_{02}$ following an iterative process, starting with a rough estimate extracted from the scaling relation $d_{02} = 0.162 - 0.013\log \Dnu$ \citep{2013A&A...550A.126M}.\\
Second, we built a template spectrum composed of radial, dipole, and quadrupole pressure modes located at the pressure mode frequencies derived from Eq.~\ref{eq:nunl_asympt}. The seismic parameters $\eps$ and $\dol$ were set following a scaling relation of the form $A + B\ \log{\Dnu}$, where the guess values of $A$ and $B$ were taken from \cite{2013A&A...559A.137M}. Then, the modes were modelled by Lorentzian functions. The heights of the Lorentzian functions were fixed by the underlying power excess distribution, which we modelled by a Gaussian function centred on the frequency of maximum oscillation power $\numax$ \citep{2012A&A...537A..30M}. Furthermore, the curvature term was set following the curvature of the red giant radial oscillation pattern as follows \citep{2013A&A...550A.126M}:

\begin{equation}
\label{eq:curvature}
\alpha = 0.015 \ \Dnu^{-0.32}.
\end{equation}



We did not adjust the parameters in the expression of $\alpha$ since precise measurement of $\alpha$ is not crucial \citep{2015A&A...579A..84V}.
Finally, we found the best-matching template spectrum by computing the maximum cross-correlation with the smoothed spectrum. Then, the observed mode frequencies were identified at the local maxima close to the optimised frequency pattern. When mixed dipole modes were present, the most intense of the closest modes of the expected pure-pressure mode was adjusted. We report that the best-matching template spectrum is less reliable when $\Dnuobs \leq 0.4\ \mu$Hz. In this case, most spectra do not exhibit a clear and intense pattern of $\ell = {0, 1, 2}$ modes, making the mode identification difficult. Nevertheless, a mode identification could be performed for these stars by vertically stacking their power spectrum with increasing $\numax$ \citep[see e.g.][]{2020MNRAS.493.1388Y}.

\subsubsection{Detection thresholds}

Because of the stochastic nature of the modes, some modes are not sufficiently intense to be detected. 
Once the best-matching template spectrum was found with the method described in Sect.~\ref{sec:best_matching_template_spectrum}, we obtained a set of candidate modes. Then, in order to reduce biases, we applied the robust detection method of \citet{2006ESASP1306..377A} to this set.
To this end, the most intense modes were selected by evaluating the $\mathrm{S_{N}}$ function, which corresponds to the most restrictive detection threshold in terms of height-to-background ratio in the power spectrum \citep{2006ESASP1306..377A}. $\mathrm{S_{N}}$ reads
\begin{equation}
\label{eq:first_threshold}
\mathrm{S_{N}} = - \left( \frac{s_{\mathrm{det}}}{\ln\left( P_{\mathrm{H_{1}}}\right)} + 1\right),
\end{equation}
where $P_{\mathrm{H_{1}}}$ is the probability of accepting that the observed peak is a mode and $s_{\mathrm{det}}$ is the rejection level relative to noise. We chose $P_{\mathrm{H_{1}}}$ such that the height-to-background ratio $S_{\rm{N}}$ reached 20 when $s_{\rm{det}} = 8,$ and the rejection level $s_{\mathrm{det}}$ was defined by

\begin{equation}
\label{eq:second_threshold}
s_{\mathrm{det}} \approx \ln (T) + \ln(\Dnu) - \ln(p_{\mathrm{det}}),
\end{equation}
where $T$ is the observation time in units of $10^6$ s, $\Dnu$ is given in $\mu$Hz, and $p_{\mathrm{det}}$ is the rejection probability that we kept equal to 5\%.
Second, we retained the candidate mode frequencies that were close to the expected pressure-mode frequencies with a less restrictive height-to-background ratio, which is given by Eq.~\ref{eq:second_threshold}.
Owing to the small amplitudes of the $\ell = 3$ modes due to geometric cancellation, we used a less restrictive  detection threshold for the $\ell = 3$ modes. The threshold for selecting $\ell = 3$ modes is $25\%$ lower than for the other degrees.

\subsection{Glitch inference}

When the best-matching template spectrum is found, we can search for the signature of glitches. To extract the signature of the helium second-ionisation zone in evolved giants, we followed the same technique as \citet{2015A&A...579A..84V} for less evolved giants.
As the oscillation spectrum of evolved red giants shows a limited number
of radial orders, we calculated the frequency difference considering all degrees as follows:

\begin{equation}
\label{eq:Dnu_local}
\Dnu_{n,\ell} = \nu_{n+1,\ell} - \nu_{n,\ell},
\end{equation}
which is different from Eq. 4 of \citet{2015A&A...579A..84V} because it is only based on radial modes. The frequency reference for these local large frequency separations was taken as the mid-point between consecutive mode frequencies. 
We isolated the glitch signature $\delta_{n,\ell}^{\mathrm{g,obs}}$ by computing the difference between the measured and the expected local large frequency separations according to the universal pattern (Eq.~\ref{eq:nunl_asympt})

\begin{equation}
\label{eq:glitch_signature_obs}
\delta_{n,\ell}^{\mathrm{g,obs}} = \Dnu_{n,\ell} - \Dnu^{\mathrm{UP}}_{n,\ell},
\end{equation}
with $\Dnu^{\mathrm{UP}}_{n,\ell} = \left(1+\alpha\left(n - n_{\mathrm{max}} + \frac{1}{2}\right)\right) \Dnu$ \citep{2013A&A...550A.126M}. We then fitted a damped oscillatory component of $\delta^{\mathrm{g,obs}}_{n,\ell}$ according to

\begin{equation}
\delta^{\mathrm{g,obs}}_{n,\ell} = \mathcal{A}\left(\frac{\numax}{\nu}\right)^{2}\Dnu \cos\left(\frac{2\pi\left(\nu-\numax\right)}{\mathcal{G}\Dnu} + \Phi\right),
\end{equation}
where $\mathcal{A}$ and $\mathcal{G}$ are the amplitude and the period of the modulation expressed in units of $\Dnu$, respectively, and $\Phi$ is the phase centred on $\numax$ \citep{2015A&A...579A..84V}. Many studies have used a more complicated function for the amplitude of the modulation. As we are restricted by the low number of observed modes, we preferred to use a simple frequency-dependent amplitude as was used before in the study of the base of the solar convective zone \citep{1994A&A...283..247M}.  \\

In evolved giants, quadrupole modes essentially behave as pure pressure modes. The case of dipole modes is complicated: They are most often reduced to a pressure-dominated mixed mode or to a cluster of modes very close to the pressure-dominated mixed mode. Because in most cases we have no way to identify the mixed-mode pattern, in practice we also consider dipole modes as pure pressure modes. This hypothesis is discussed below. We can add them in the fit of the glitch modulation without deteriorating the fit of the modulation. Then, the Nyquist criterion, which states that the frequency of the modulation must be strictly less than half the sample rate, writes $\mathcal{G} \geq 1$ instead of $\mathcal{G} \geq 2$ when only radial modes are used.
When $\Dnu \gtrapprox 3\ \mu$Hz, dipole modes are no longer pure-pressure modes. It has been shown that adding the dipole modes of lowest inertia in each $\Dnu$ range could bias the fit of the modulation, especially for the least evolved red giants \citep{2014MNRAS.440.1828B, 2020MNRAS.497.1008D}. Nevertheless, for the range of $\Dnu$ we consider here, \citet{2014MNRAS.440.1828B} reported that the use of dipole modes of lowest inertia remarkably improves the robustness of the fit. When mixed modes are present, we then took into account the most intense dipole mode of the modes closest to the expected location of the pure-pressure mode.


\subsection{Computation of mode visibilities}
\label{sec:computation_mode_visibilities}

\begin{figure}[t]
        \includegraphics[width=1.0\linewidth]{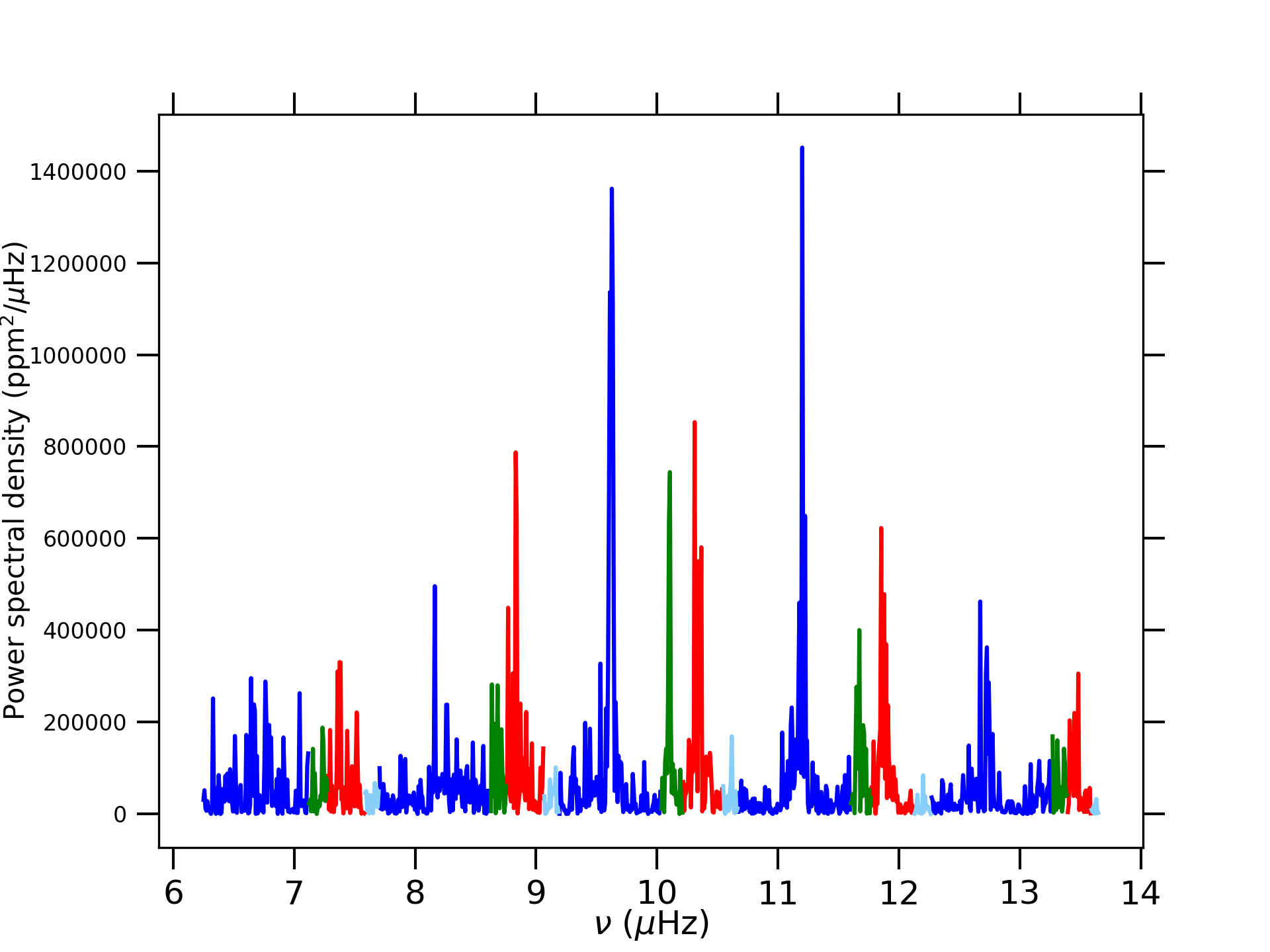}
        \caption{Oscillation spectrum of the star KIC 2695975 ($\Dnu = 1.538~\mu$Hz, and $\numax = 10.11~\mu$Hz), with an emphasis on the spectral range where the power spectral density is integrated for each mode. Red, blue, green, and light blue are associated with radial, dipole, quadrupole, and octupole modes, respectively. This star has been classified as an RGB star with the two identification methods adopted in this work.
        }
        \label{fig:colour_spectrum_KIC_2695975}
\end{figure}

We investigated the energy distribution among modes of different degree $\ell$ in the case of evolved stars. The technique we used to compute the mode visibilities is described in \cite{2012A&A...537A..30M}. First, we computed the total mode energy, noted $A_{\ell}^{2}(n)$, for which the radial order $n$ lies between the lowest and highest observed radial orders. This was done by subtracting the background component and integrating the power spectral density over the whole spectral range where the mode is expected, that is, around the p-mode frequency inferred from Eq.~\ref{eq:nunl_asympt} (see Table~\ref{tab:visibility_boundaries} and Fig.~\ref{fig:colour_spectrum_KIC_2695975}). Then we computed the visibility $\Vl$ of a mode of degree $\ell$ as


\begin{table}
        \caption{Boundaries for the integration of the power spectral density.}
                        \begin{tabular}{ccc}
                            \hline
                                \hline
                                  &  \bf $\nu_{\mathrm{inf}}(n,\ell)$ &  $\nu_{\mathrm{sup}}(n,\ell)$\\  
                                \hline
                                $\ell = 0$ & $(\nu_{n,0} + \nu_{n-1,2})/2$& $(3\ \nu_{n-1,3} + \nu_{n,0})/4$ \\
                                
                                $\ell = 3$ &$(3\ \nu_{n-1,3} + \nu_{n,0})/4$& $(7\ \nu_{n-1,3} - \nu_{n,0})/6$ \\
                                
                                $\ell = 1$ &$(7\ \nu_{n-1,3} - \nu_{n,0})/6$& $(4\  \nu_{n-1,2} - \nu_{n,0})/3$ \\
                                
                                $\ell = 2$ &$(4\ \nu_{n-1,2} - \nu_{n,0})/3$& $(\nu_{n,0} + \nu_{n-1,2})/2$ \\
                                \hline
                        \end{tabular}
                        \\
                        \textbf{Notes:} The boundaries are equivalent to the mid-point between consecutive modes, except when $\ell = 1$ and $\ell = 3$ modes are involved. This is illustrated in Fig.~\ref{fig:colour_spectrum_KIC_2695975}. The boundary between the $\ell = 1$ and the $\ell = 3$ modes is chosen close to the $\ell = 3$ mode frequency. We made this choice to avoid any confusion between a $\ell = 3$ mode and the neighbouring dipole mixed-modes because the dipole mixed-modes extend up to the $\ell = 3$ modes. 
        \label{tab:visibility_boundaries}
\end{table}

\begin{equation}
\label{eq:visibility}
\Vl = \frac{\langle A_{\ell}^{2}\rangle}{\langle A_{0}^{2}\rangle},
\end{equation}
where $\langle A_{\ell}^{2}\rangle$ is the squared amplitude of the mode of degree $\ell$.

When mixed modes are present, the procedure was the same: The energy $A_{\ell}^{2}(n)$ corresponds to the total energy of the mixed modes associated with the radial order $n$. The errors on the visibilities were computed from the errors on the boundary frequencies listed in Table~\ref{tab:visibility_boundaries}: The energy contained in the 1$\sigma$ error region of the boundary frequencies is interpreted as the error on the parameter $A_{\ell}^{2}(n)$.

\subsection{Mode fitting}
\label{sec:mode_fitting_method}

The mode amplitudes and widths derived from the fit of the modes provide unique constraints on the mode excitation and damping. In this context, we adopted a frequentist approach. The modes were fitted with Lorentzian profiles following the maximum likelihood estimator technique described in \cite{1994A&A...289..649T}. The fit was performed radial order by radial order, so that we have three modes at most to fit per iteration. Owing to their very low amplitudes, $\ell = 3$ modes cannot be fitted. Radial, dipole, and quadrupole modes were fitted on top of the background using

\begin{equation}
\mathcal{L}(n) = \sum_{\ell = {0,1,2}}\frac{H_{n,\ell}}{1 + \left(2\frac{\nu - \nunl}{\Gamma_{n,\ell}}\right)^{2}} + B(\nu),
\end{equation}
where $H_{n,\ell}$, $\nunl$, and $\Gamma_{n,\ell}$ are the height, frequency, and width of the mode of radial order $n$ and degree $\ell$, respectively. We point out that the background was extracted separately, and we kept it fixed when fitting the modes.
The mode amplitude can be deduced from the mode height and the width by

\begin{equation}
\label{eq:amplitude_height_link}
A_{n,\ell} = \sqrt{H_{n,\ell}\pi\Gamma_{n,\ell}}.
\end{equation}

Because of the low signal-to-noise ratios, the presence of mixed modes, and the stochastic excitation, some modes were not correctly fitted. The measurements were rejected when the width was too close to the frequency resolution (i.e. when $\Gamma_{n,l} \leq 1.1\delta\nu_{\rm{res}}$) or when the width was overestimated (i.e. when $\Gamma_{n,l} \geq \Dnu/7$).
When mixed modes were present, we fitted the closest mixed modes to the expected pure-pressure mode. Then, following \citet{2014ApJ...781L..29B}, \citet{2015A&A...579A..31B} and \citet{2018A&A...618A.109M}, we inferred the mode width and the mode amplitude that the mode would have if it were purely acoustic through

\begin{equation}
\label{eq:mixed_correction}
\Gamma_{n,\ell}^{\mathrm{p}} = \frac{\Gamma_{n,\ell}}{1-\zeta} \ \mathrm{and} \ A_{n,\ell}^{\mathrm{p}} = \frac{A_{n,\ell}}{\sqrt{1-\zeta}},
\end{equation}
where $\zeta$ depends on the inertia of the fitted mixed mode. Characterising the mixed-mode pattern is beyond the scope of this work. However, we estimated the mode inertia that is defined in \citet{2018A&A...618A.109M}, for example, using scaling relations (Eq.~17 and 18 from \citet{2017A&A...600A...1M} for the coupling factor $q$ and the database from \citet{2016A&A...588A..87V} for the period spacings $\Delta\Pi_{1}$).

We finally computed the mean mode amplitude $\langle A_{\ell} \rangle$ using the three p modes of degree $\ell$ closest to $\numax$. We corrected the wavelength dependence of the photometric variation integrated over the \Kepler\ bandpass according to

\begin{equation}
\label{eq:amplitude_bolometric_correction}
\langle A_{\ell,\mathrm{bol}} \rangle = \langle A_{\ell} \rangle \left( \frac{\Teff}{T_{K}}\right)^{0.80},
\end{equation}
where $T_{K} = 5934$ K \citep{2011A&A...531A.124B}. The average mode width $\langle\Gamma_{\ell}\rangle$ was computed as the weighted mean of the three p modes of degree $\ell$ closest to $\numax$, where the mode amplitude was used as weight \citep[see e.g.][]{2018A&A...616A..94V}.


\section{Results}
\label{sec:results}

In this section, we characterise the oscillation spectrum of evolved giants as precisely as possible. We compare our measurements with previous studies that focused on less evolved stages and with theoretical predictions.

\subsection{Acoustic offset $\eps$ and reduced small separations $\dol$}

\begin{figure*}[ht]
        \begin{minipage}{1.0\linewidth}  
                \rotatebox{0}{\includegraphics[width=0.5\linewidth]{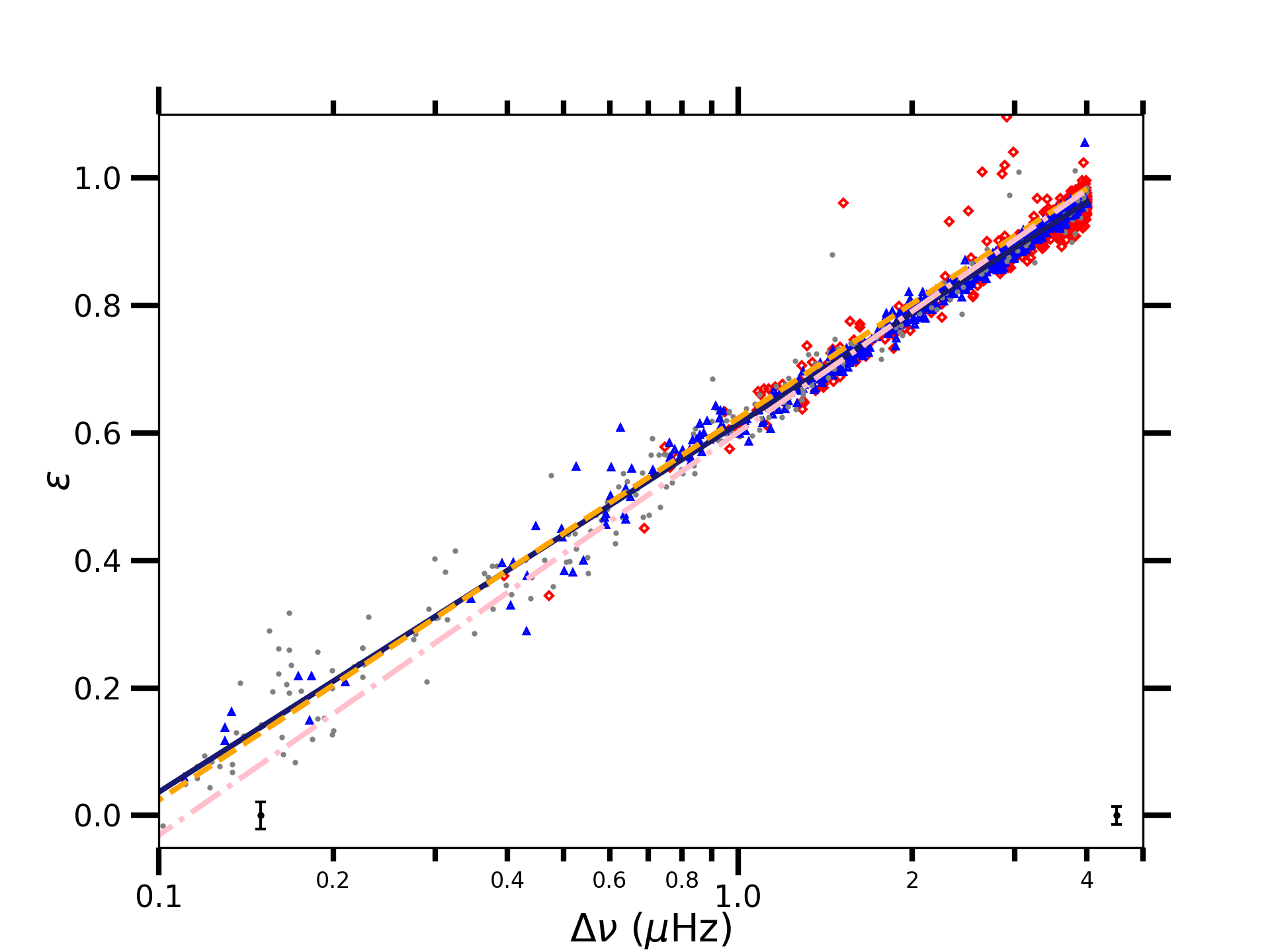}}
                \rotatebox{0}{\includegraphics[width=0.5\linewidth]{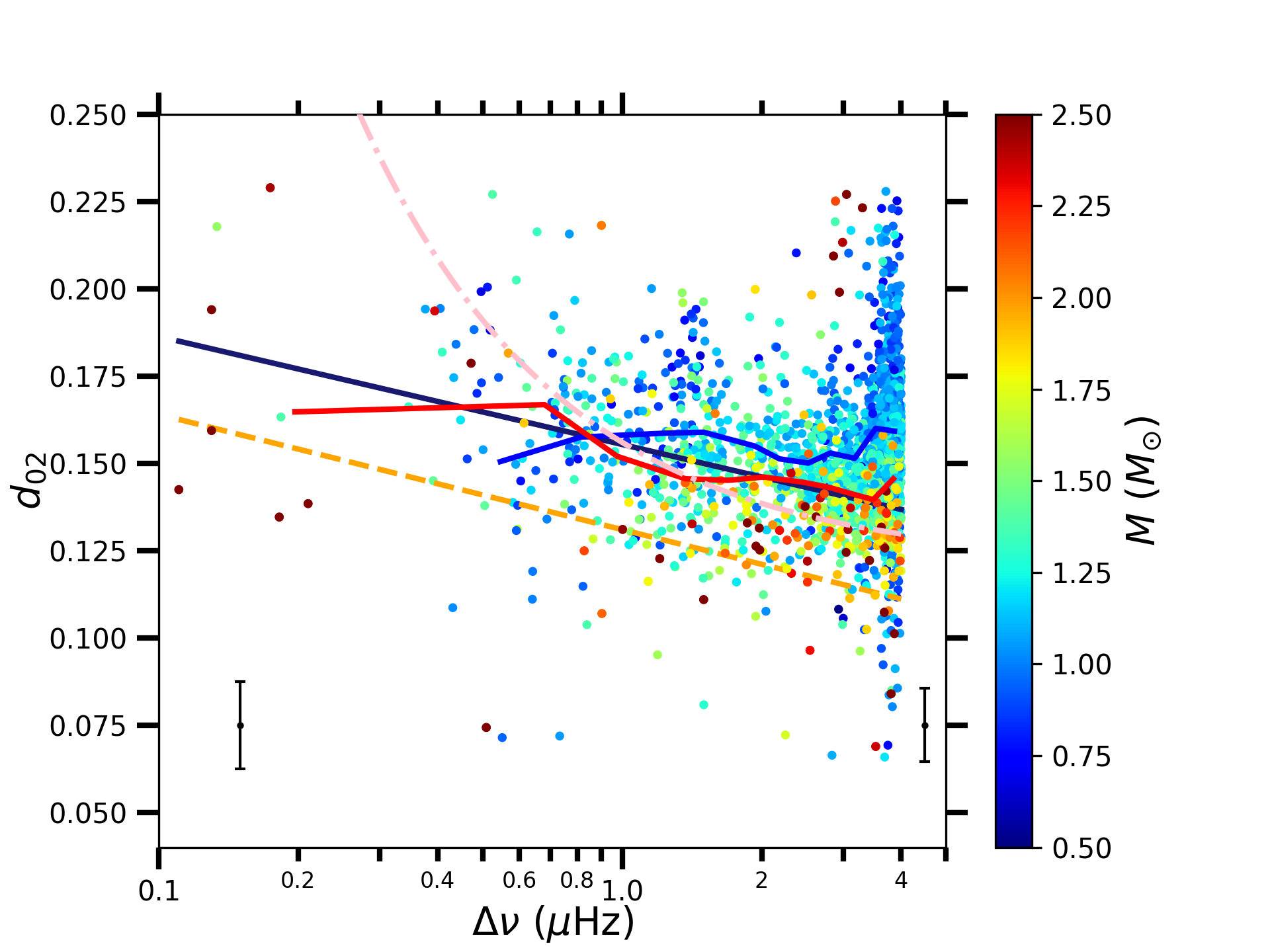}}

\end{minipage}
        \begin{minipage}{1.0\linewidth}  
                \rotatebox{0}{\includegraphics[width=0.5\linewidth]{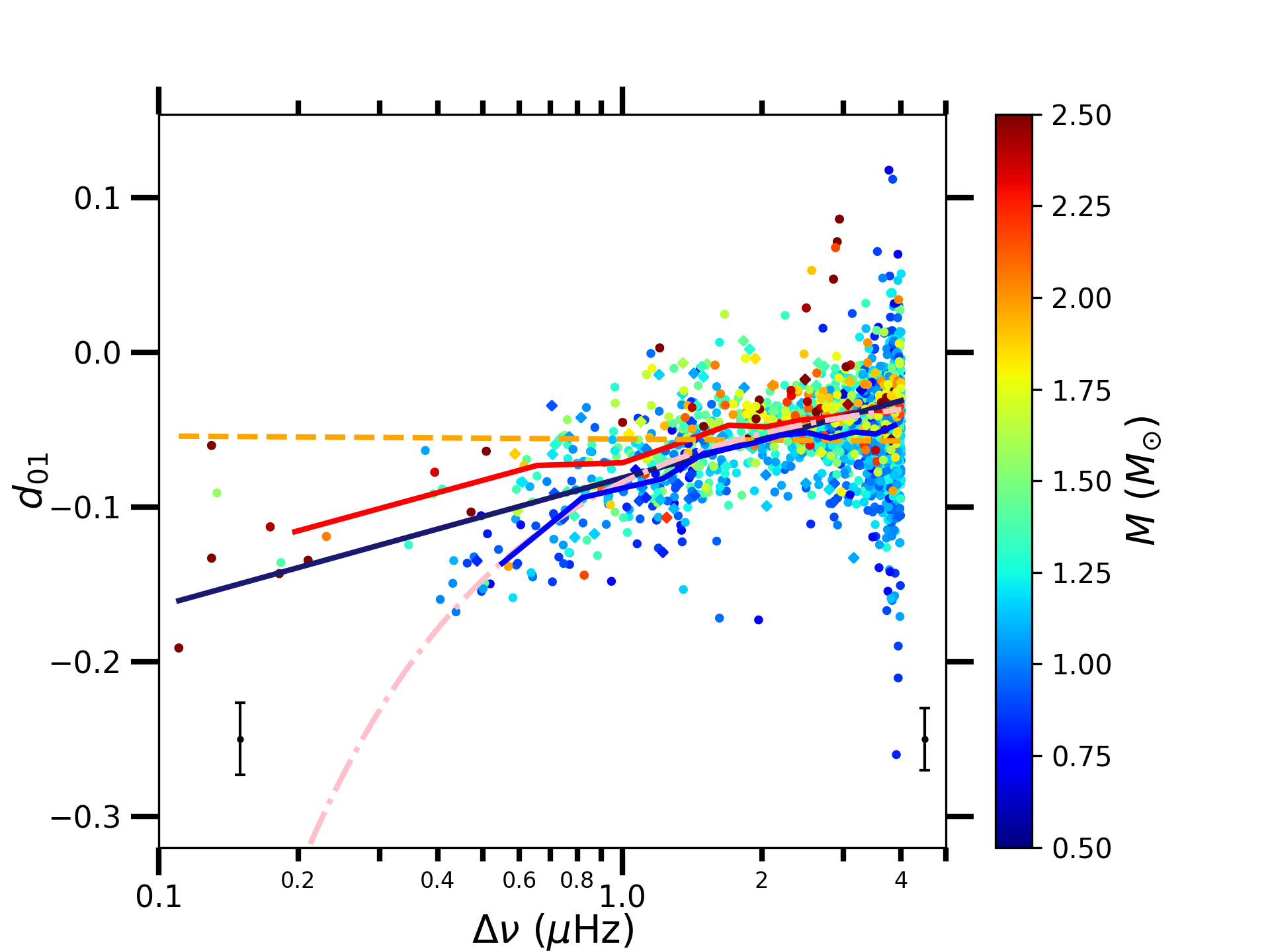}}
                \rotatebox{0}{\includegraphics[width=0.5\linewidth]{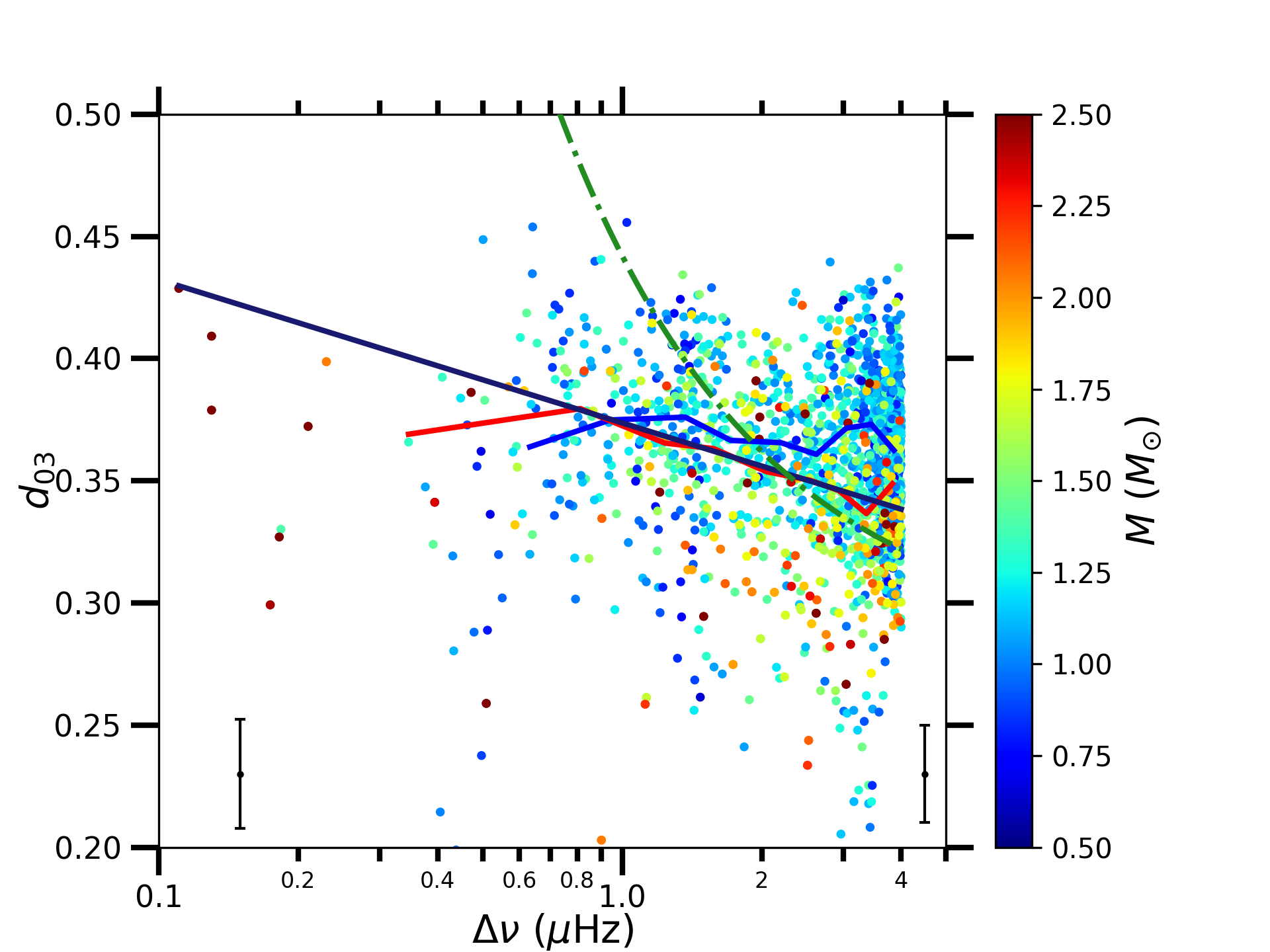}}

\end{minipage}
        \caption{Seismic parameters of the asymptotic pattern of red giants (Eq.~\ref{eq:nunl_asympt}) after adjusting $\Dnu$ with the best-matching template following the procedure described in Sect.~\ref{sec:Method}. Upper left panel: Acoustic offset $\eps$ as a function of $\Dnu$, where blue triangles indicate RGB stars and red diamonds He-burning stars. Stars with either an unidentified evolutionary stage or disagreement between the two classification methods described in Sect.~\ref{sec:dataset} are represented in grey. Upper right panel: Reduced small separation $d_{02}$ as a function of $\Dnu$; the stellar mass is colour-coded. Bottom panels: Same labels as for the upper right panel, but for $d_{01}$ and $d_{03}$ as a function of $\Dnu$. The solid blue and red lines are the median values in 0.4 $\mu$Hz $\Dnu$ bins for low-mass stars ($M \leq 1.2~M_{\odot}$) and for high-mass stars ($M \geq 1.2~M_{\odot}$), respectively. The dashed orange lines represent the scaling relations from \cite{2013A&A...559A.137M}, and the solid dark blue lines are the scaling relations derived in this study (listed in Table~\ref{Table:scaling_relation_parameter_mode_frequencies}). The dot-dashed pink and dark green lines correspond to the scaling relations for less evolved stars from \citet{2012ApJ...757..190C} and \citet{2010ApJ...723.1607H}, respectively. Mean error bars estimated at low $\Dnu$ ($\Dnu \leq 1.0~\mu$Hz) and at high $\Dnu$ ($\Dnu \geq 1.0~\mu$Hz) are represented at the bottom of each panel.
        }
        \label{fig:seismic_parameters}
\end{figure*}

From the fit of the spectrum described by Eq.~\ref{eq:nunl_asympt} we derived the global acoustic offset $\eps$ and the reduced small separations $\dol$ associated with the detected modes (see Fig.~\ref{fig:seismic_parameters}). Scaling relations were adjusted to our sets of seismic parameters in the form $A_{\ell} + B_{\ell}\ \log\left( \Dnu \right)$, where $A_{\ell}$ and $B_{\ell}$ are free parameters that are summarised in Table~\ref{Table:scaling_relation_parameter_mode_frequencies}, and $\Dnu$ is given in $\mu$Hz.

\begin{table}

\caption{Fit of the seismic parameters $\eps$ and $\dol$, and the dimensionless glitch parameters}
\begin{tabular}{rrrr}
\hline
\hline
$\ell$ & & $A_{\ell}\qquad$ & $B_{\ell}\qquad$\\
\hline
0 & $\eps$ & $0.614 \pm 0.002$ & $0.578 \pm 0.003$\\
1 &$d_{01}$ & $-0.081 \pm 0.002$ & $0.083 \pm 0.005$\\
2 &$d_{02}$ & $0.156 \pm 0.001$ & $-0.031 \pm 0.003$\\
3 &$d_{03}$ & $0.374 \pm 0.002$ & $-0.059 \pm 0.005$\\

\hline
 & & $C\qquad$ & $D\qquad$ \\
\hline
 & $\mathcal{A}$ & $0.072 \pm 0.003$ & $-0.411 \pm 0.006$\\
 & $\mathcal{G}$ & $1.879 \pm 0.001$ & $0.045 \pm 0.002$\\

  \hline
  \label{Table:scaling_relation_parameter_mode_frequencies}
\end{tabular}
\\
\textbf{Notes:} The fits were performed for RGB stars alone. The acoustic offset and the reduced small separations were fitted by a linear fit $A_{\ell} + B_{\ell}\ \log\left( \Dnu \right),$ while the glitch parameters were fitted by a power law $C\Dnu^{D}$.
\end{table}

\subsubsection{Acoustic offset $\eps$}
The oscillation spectrum of radial modes is depicted by
the global acoustic offset $\eps$ as shown in Fig.~\ref{fig:seismic_parameters}. The trend that we observe for RGB stars is similar to what has been obtained in previous studies \citep{2013A&A...559A.137M, 2020MNRAS.493.1388Y}.
With our method, which uses a global fit of the oscillation pattern, we derive similar values of $\eps$ for RGB and He-burning stars, in contrast to \citet{2012A&A...541A..51K}, who used a local approach.
As they showed, the glitches have limited effect on global measurements of the seismic parameters but they affect local measurements considerably. In Sect. ~\ref{sec:results:HeII} we investigate the local effects on the mode frequencies by studying the modulation left by the helium second-ionisation zone in p-mode frequencies.



\subsubsection{Reduced small separations $\dol$}

Theoretical models predict that the effects of stellar evolution are reflected in the reduced small separations $\dol$. They are sensitive to any internal structure change that affects the gradient of the sound speed \citep{1980ApJS...43..469T, 2003A&A...411..215R} in the deep interiors. While a star ascends the RGB, the stellar core contracts but does not undergo important structure changes. Therefore the reduced small separations vary only slowly along the RGB. \\

 Fig.~\ref{fig:seismic_parameters} shows that $d_{01}$ decreases when $\Dnu$ decreases, as observed by previous observational studies on less evolved stars \citep{2010ApJ...723.1607H, 2012ApJ...757..190C, 2013A&A...559A.137M}. This points out the fact that during stellar evolution, dipole p modes approach the doublet formed by $\ell = 0$ and $\ell = 2$ modes, in agreement with theoretical models \citep{2010ApJ...721L.182M, 2014ApJ...788L..10S}. The variation of $d_{01}$ during late stellar evolution can be linked to the location of the turning points of $\ell = 1$ modes. By examining the structure of low-mass red giant models, \cite{2010ApJ...721L.182M} found that $d_{01}$ takes negative values when the turning points of $\ell = 1$ modes are deep in the convective envelope. This is exactly what we observe and allows us to extend the interpretation made for RGB stars to AGB stars, which have negative $d_{01}$. Stellar models of \cite{2010ApJ...721L.182M} also predict that core-He-burning stars have both positive and negative $d_{01}$, and that the turning points of $\ell = 1$ modes are located inside the radiative region. The determination of $d_{01}$ in clump stars is more difficult because the observed large spread in $d_{01}$ mainly reflects the presence of mixed modes that perturb the adjustment of the acoustic dipole modes.





The reduced small separation $d_{02}$ is sensitive to the structure differences between core He-burning stars and RGB stars: We report that $d_{02}$ is larger on average for core He-burning stars than for RGB stars, as has been reported by \cite{2012A&A...541A..51K}. We note a clear mass effect: the lower the mass, the larger $d_{02}$. The first evidence of this mass dependence in red giants has been discussed in \cite{2010ApJ...723.1607H}, in agreement with the theoretical models \citep{2012ASSP...26...23M}. We find that this mass dependence is also visible for $d_{01}$ and $d_{03}$, as predicted by the theoretical models of \citet{2010ApJ...721L.182M}, despite the presence of mixed modes that cause the values of these parameters to become more scattered. However, Montalb\'an and collaborators did not discuss the origin of this mass dependence. Further work is therefore needed to physically understand this behaviour. \\

As for $\ell = 3$ modes (see the bottom panel of Fig.~\ref{fig:seismic_parameters}), we note that the reduced small separation $d_{03}$ increases when $\Dnu$ decreases, as shown by the observations of \Kepler\ \citep{2010ApJ...723.1607H} and stellar models \citep{2010ApJ...721L.182M}. This expresses the fact that the $\ell = 3$ modes approach the left-hand side of $\ell = {0, 2}$ modes during stellar evolution. Further theoretical work is needed to investigate and understand this behaviour. \\

\subsection{Signature of the helium second-ionisation zone}
\label{sec:results:HeII}

The results obtained after fitting the modulation left by the helium second-ionisation zone in $\Dnu_{n,\ell}$ are shown in Fig.~\ref{fig:Ampl_G_phi_Dnu}. In Table~\ref{Table:scaling_relation_parameter_mode_frequencies} we present the scaling relations found for the dimensionless amplitude $\mathcal{A}$ and period $\mathcal{G}$, computed for RGB stars alone, in the form $C\Dnu^{D}$, where $C$ and $D$ are free parameters and $\Dnu$ is given in $\mu$Hz.


\subsubsection{Modulation amplitude $\mathcal{A}$}

\begin{figure*}[ht]
        \begin{minipage}{1.0\linewidth}  
                \rotatebox{0}{\includegraphics[width=0.5\linewidth]{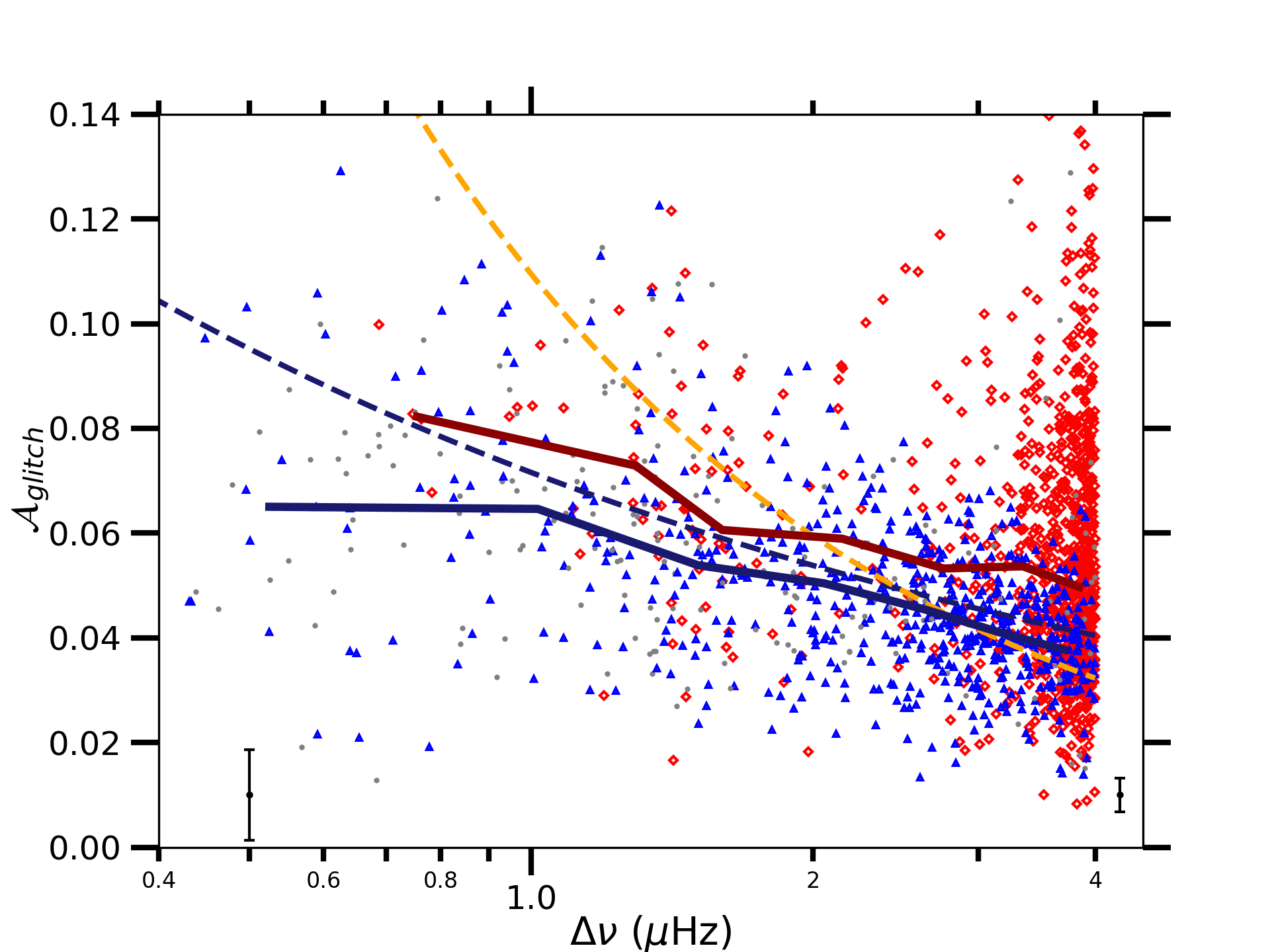}}
                \rotatebox{0}{\includegraphics[width=0.5\linewidth]{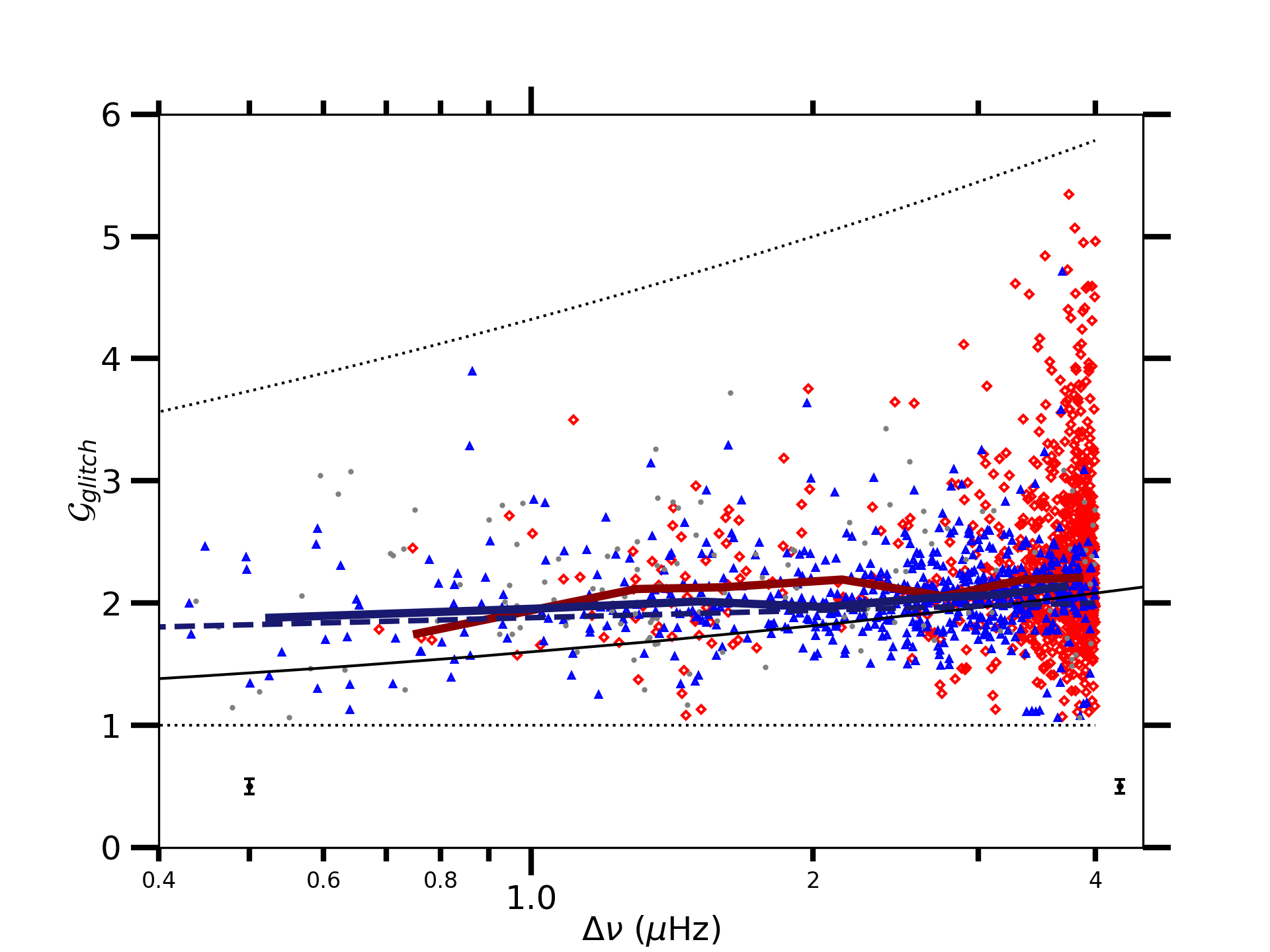}}

\end{minipage}
        \begin{minipage}{1.0\linewidth}  
                \rotatebox{0}{\includegraphics[width=0.5\linewidth]{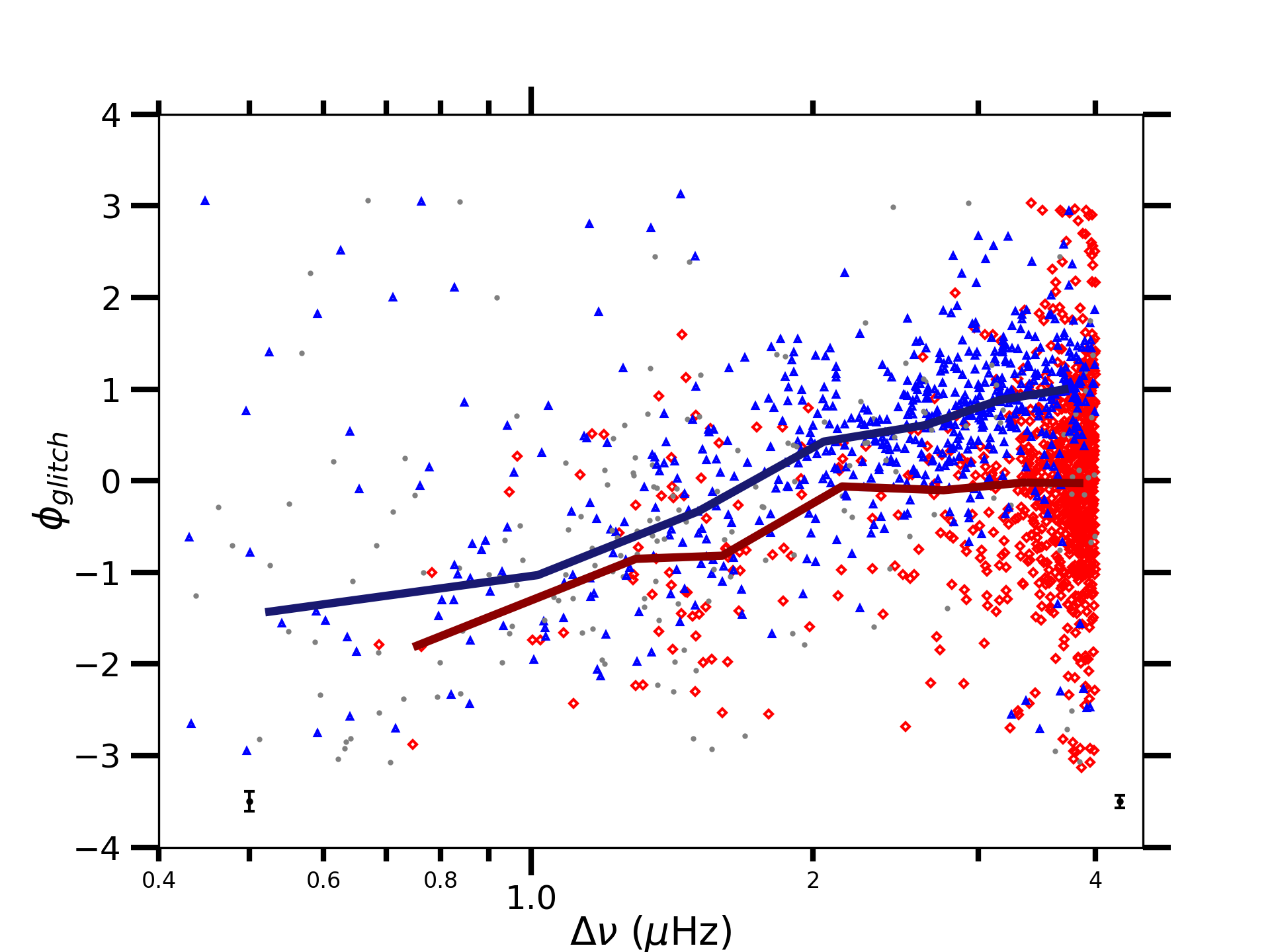}}
                \rotatebox{0}{\includegraphics[width=0.5\linewidth]{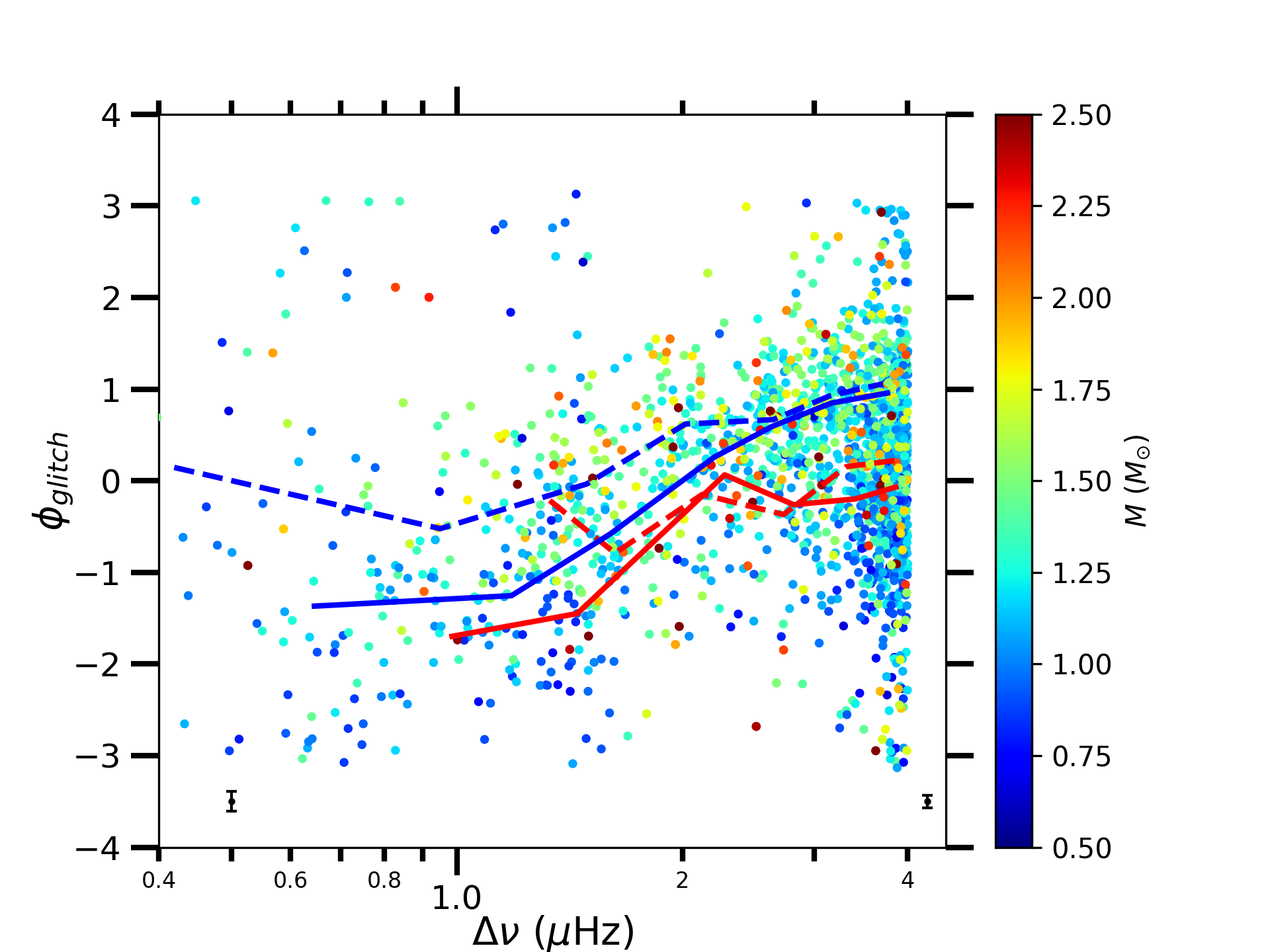}}

\end{minipage}
        \caption{Modulation amplitude $\mathcal{A}$, modulation period $\mathcal{G}$, and modulation phase $\Phi$ as a function of $\Dnu$, with the same labels as in Fig.~\ref{fig:seismic_parameters}. The thick solid lines are the median values in 0.5 $\mu$Hz $\Dnu$ bins, shown in blue for RGB stars and in red for He-burning stars. In the upper panels, dashed blue lines are the fits presented in Table~\ref{Table:scaling_relation_parameter_mode_frequencies}. In the upper left panel, the dashed orange line is the fit obtained for less evolved stars \citep{2015A&A...579A..84V}, corrected with a factor that accounts for the differences between the methods we used to fit the modulation as described in Sect.~\ref{sec:results:HeII}. In the upper right panel, dotted black lines delimit the domain of reliable measurements of $\mathcal{G}$, which are described in Sect.~\ref{sec:Method}, and the thin solid line is the modulation period inferred from MESA models for a 1 $M_{\odot}$ star and starting from the RGB up to the AGB. The error bars are computed in the same way as in Fig.~\ref{fig:seismic_parameters}. In the lower right panel, the stellar mass is colour-coded, and medians are represented by solid lines for low-mass stars ($M \leq 1.2M_{\odot}$) and by dashed lines for high-mass stars ($M \geq 1.2M_{\odot}$). Because $\Phi$ varies with stellar evolution, we calculated the medians for RGB and He-burning stars separately. We show them in blue for RGB and in red for He-burning stars.
        }
        \label{fig:Ampl_G_phi_Dnu}
\end{figure*}

When a star ascends the RGB or AGB ($\Dnu \lesssim 3~\mu$Hz for the early AGB), the dimensionless amplitude of the modulation $\mathcal{A}$ notably increases. During the clump phase, it is more difficult to conclude because of the large spread of the amplitudes. We verified that most of the stars with $\mathcal{A} \geq 0.09$ also have a dim \Kepler\ magnitude, hence their oscillation spectrum is
recorded with a low signal-to-noise ratio, so that the measurement of $\mathcal{A}$
is quite noisy. Globally, the amplitude of the modulation is larger for He-burning stars than for RGB stars. This is consistent with the results presented in \cite{2015A&A...579A..84V} between clump and RGB stars having $\Dnu \geq 3.0\ \mu$Hz. We note that the values of the modulation amplitude are larger in our work than in \cite{2015A&A...579A..84V}, as expected from the different methods with which $\Dnu_{n,\ell}$ and $\delta_{\mathrm{g,obs}}(n,\ell)$ were computed. To compare our results with those of Vrard et al. (2015), we then estimated how different the modulation amplitudes are between the two methods. We note that our modulation amplitudes are 1.8 times larger on average than those extracted in \cite{2015A&A...579A..84V}. Then, we multiplied the fit reported in the latter study by 1.8 and compared it with ours, as plotted in the upper left panel of Fig.~\ref{fig:Ampl_G_phi_Dnu}.
Our results are also consistent with stellar evolution models that indicate that the difference observed in the modulation amplitude $\mathcal{A}$ between RGB and He-burning phases is correlated with a difference of temperature and density at the level of the helium second-ionisation zone \citep{2014MNRAS.445.3685C}.


\subsubsection{Modulation period $\mathcal{G}$}

We note that the modulation period $\mathcal{G}$ slightly decreases throughout the stellar evolution (Fig.~\ref{fig:Ampl_G_phi_Dnu}). This means that the helium ionisation zone slowly sinks into the stellar interior during evolution, as predicted by stellar models \citep[Fig.~6 of][]{2014MNRAS.440.1828B}. The typical period does not globally differ between He-burning stars and their RGB counterparts. Our measurements were compared to the results derived with the stellar evolution code Modules for Experiments in Stellar Astrophysics (MESA) using the 1M\_pre\_ms\_to\_wd test suite case \citep{2011ApJS..192....3P, 2013ApJS..208....4P, 2015ApJS..220...15P, 2018ApJS..234...34P, 2019ApJS..243...10P}. In Appendix \ref{appendix:link_period_G_location_heII} we describe how we extracted the modulation period $\mathcal{G}$ from stellar models.
Stellar models indicate that RGB stars and He-burning stars of the same mass and same large separation should have the same modulation period $\mathcal{G}$, in agreement with observations. However, He-burning stars have more scattered $\mathcal{G}$ values than their RGB counterparts. The large spread does not appear to stem from the presence of mixed modes because \cite{2015A&A...579A..84V} also reported a spread like this for clump and RGB stars, although they only used radial modes in the modulation fits. As reported for the modulation amplitude $\mathcal{A}$, the spread is rather well explained by the dim magnitudes, hence by the low signal-to-noise ratios in the oscillation spectra.

\subsubsection{Modulation phase $\Phi$}

The modulation phase $\Phi$ differs depending on the evolutionary stage. By letting the phase vary in the interval $[-\pi, +\pi]$, we observe that He-burning stars globally show a negative phase difference compared to their H-burning counterparts. This difference has been reported by \cite{2015A&A...579A..84V} for clump and RGB stars. The authors showed that the phase difference is related to the difference in $\varepsilon$ reported in the study of \cite{2012A&A...541A..51K} between clump and RGB stars. The link between $\Phi$ and $\eps,$ which depends on the evolutionary stage, is discussed in Sect.~\ref{sec:discussion}. Similarly to the modulation amplitude $\mathcal{A}$ and the modulation period $\mathcal{G}$, the spread of the modulation phase $\Phi$ is larger for He-burning stars than for H-burning stars. We verified that the spread of $\Phi$ becomes important when the \Kepler\ magnitude exceeds 11. The large spread of $\Phi$ could then be explained by low signal-to-noise ratios in the oscillation spectra.

\subsubsection{Mass dependence of the glitch parameters}




%

We also investigated the stellar mass dependence of the glitch modulation parameters. 
We find evidence of a mass dependence for the modulation amplitude, which varies as $\mathcal{A}_{\mathrm{RGB}} \propto \Dnu^{-0.41 \pm 0.01} M^{-0.34 \pm 0.02}$ on the RGB with a similar dependence during He-burning phases. 
Conversely, the modulation period is weakly correlated with the stellar mass on the RGB and follows $\mathcal{G}_{\mathrm{RGB}} \propto \Dnu^{-0.05 \pm 0.01} M^{-0.04 \pm 0.01}$ , while it is practically independent of the stellar mass during the He-burning phase. 
The mass dependence of the modulation phase $\Phi$ is illustrated in Fig.~\ref{fig:Ampl_G_phi_Dnu}. We note a negative phase difference between low-mass and high-mass RGB stars for $\Dnu \leq 2.0~\mu$Hz. The lack of data for He-burning stars at low $\Dnu$ prevents us from drawing any conclusion. In case of less evolved stars, \citet{2015A&A...579A..84V} did not find any correlation between the stellar mass and the glitch parameters, except for the modulation phase for clump stars. These mass dependences remain empirical, and further theoretical work is needed to determine their physical basis.


\subsection{Mode widths}

\begin{figure*}[htbp]
        \begin{minipage}{1.0\linewidth}  
                \rotatebox{0}{\includegraphics[width=0.50\linewidth]{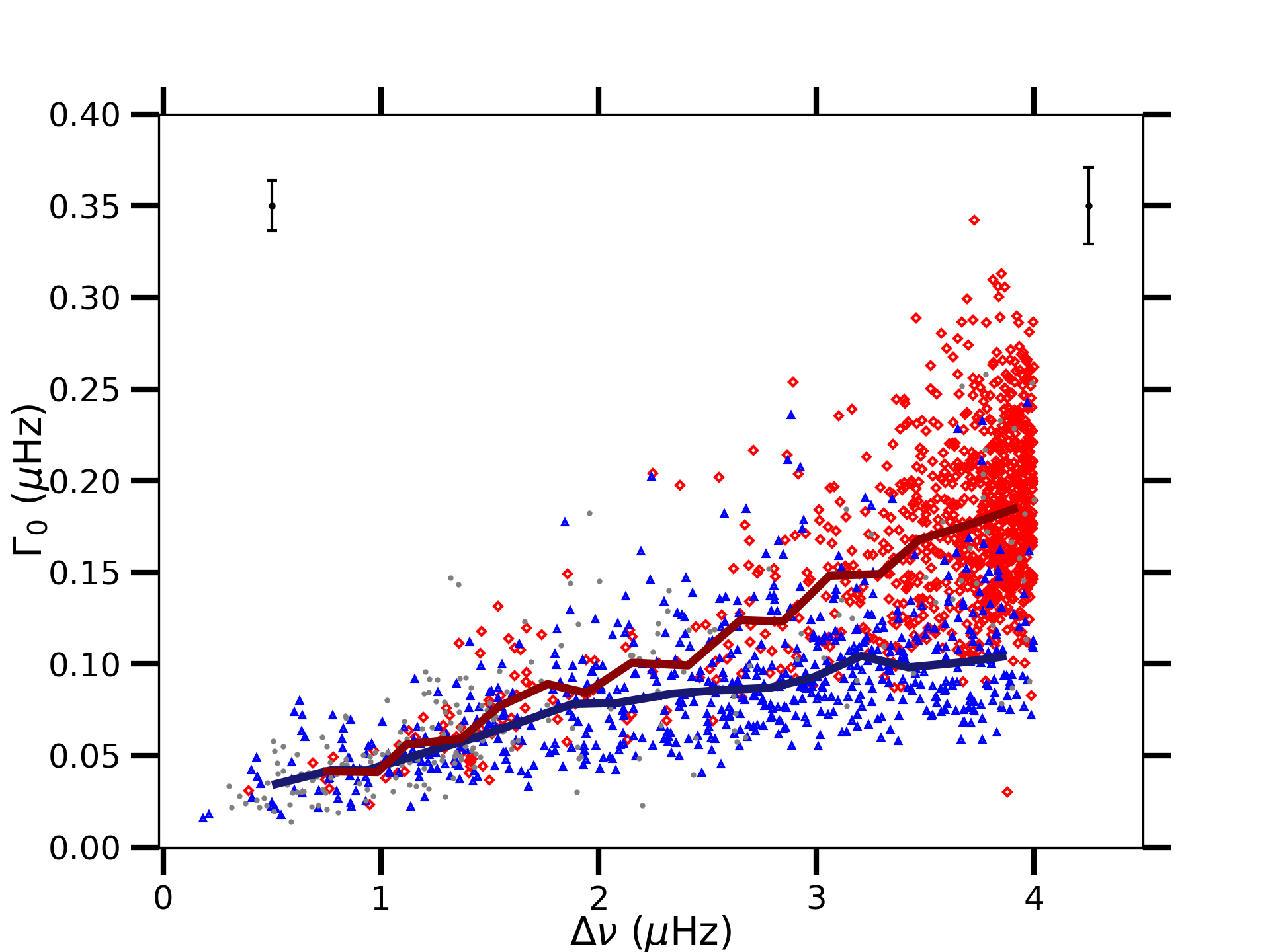}}
        \rotatebox{0}{\includegraphics[width=0.50\linewidth]{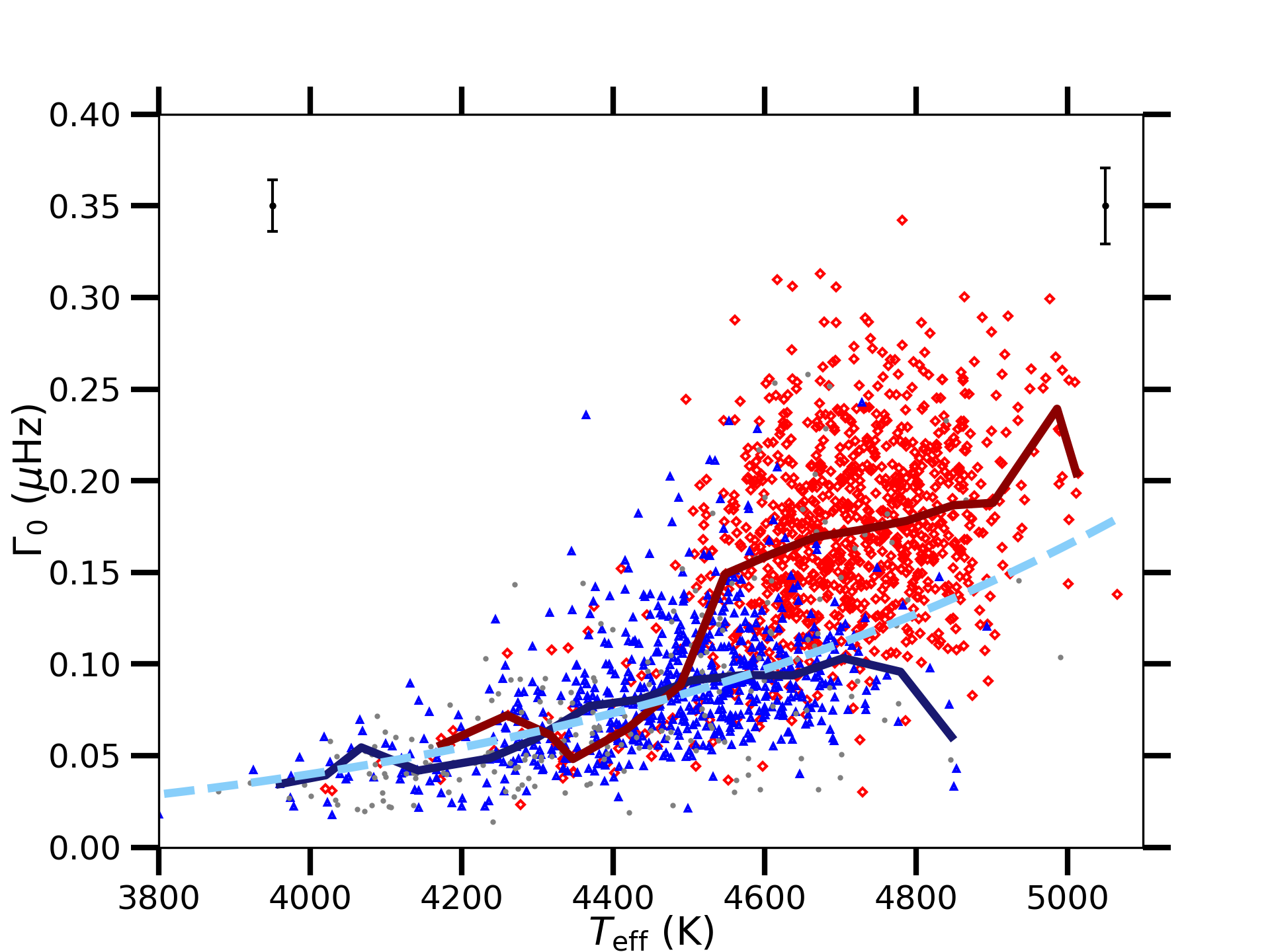}}
\end{minipage}
    \begin{minipage}{1.0\linewidth}  
                \rotatebox{0}{\includegraphics[width=0.50\linewidth]{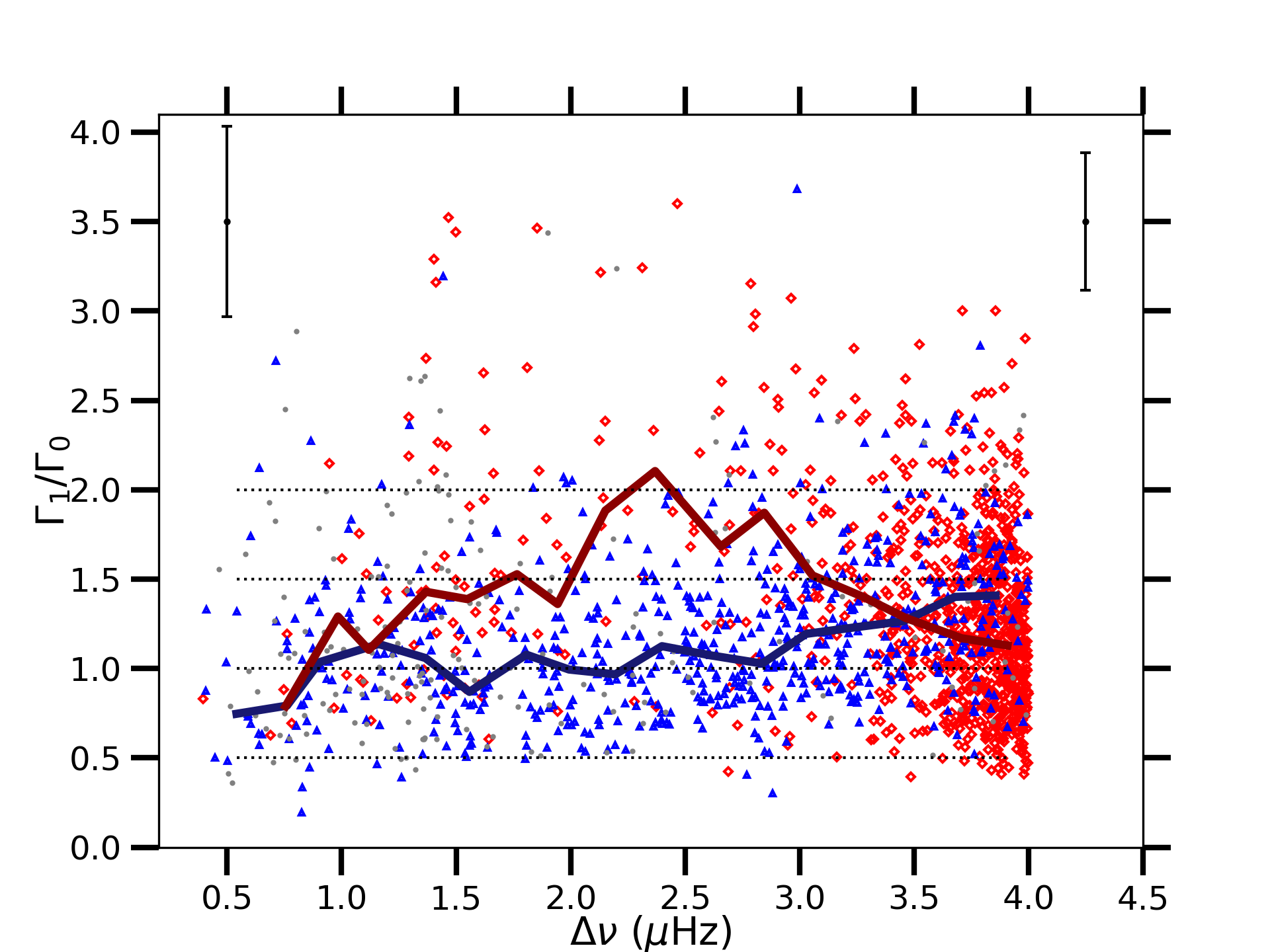}}
        \rotatebox{0}{\includegraphics[width=0.50\linewidth]{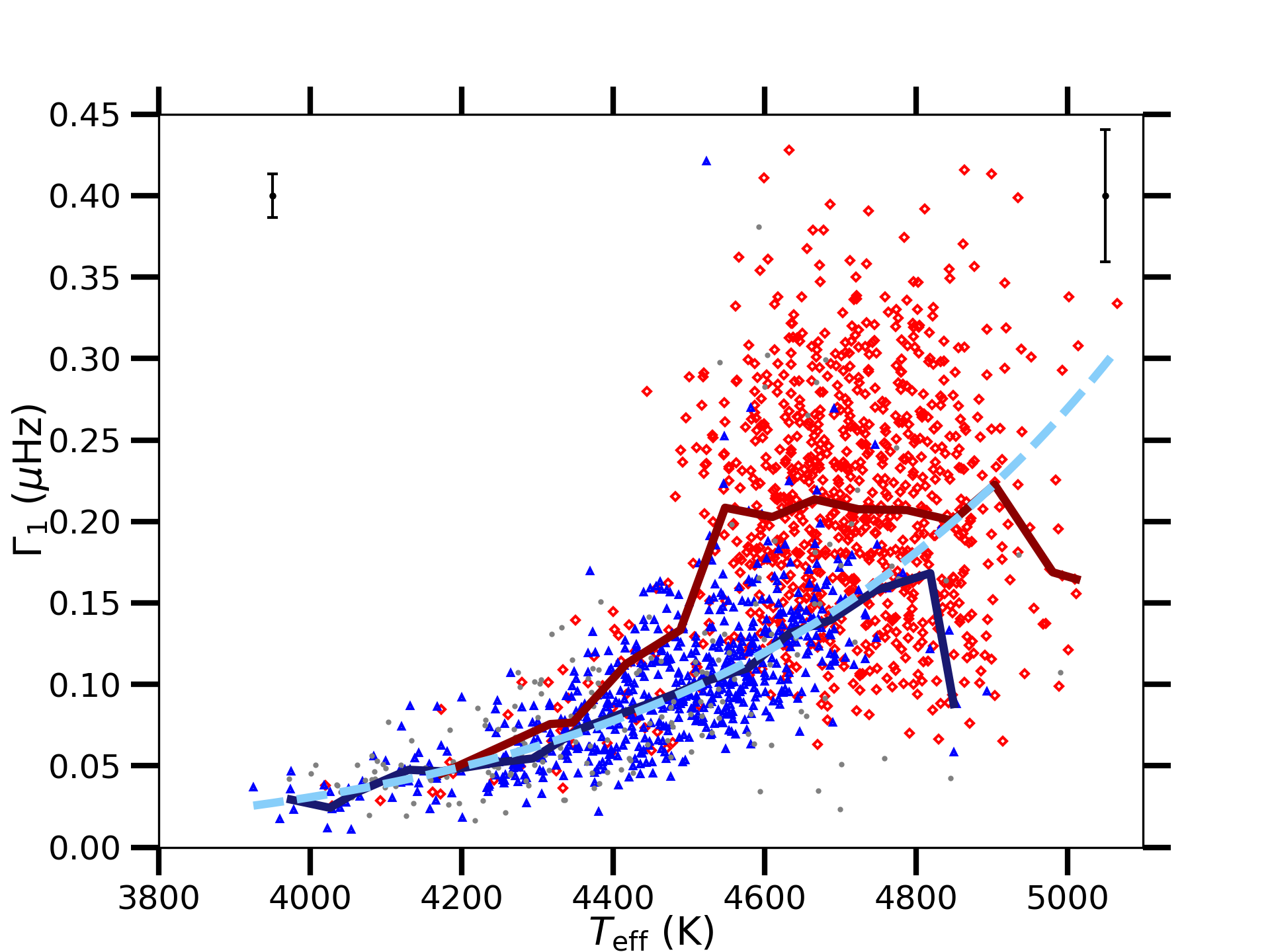}}
\end{minipage}
        \begin{minipage}{1.0\linewidth}  
            \rotatebox{0}{\includegraphics[width=0.50\linewidth]{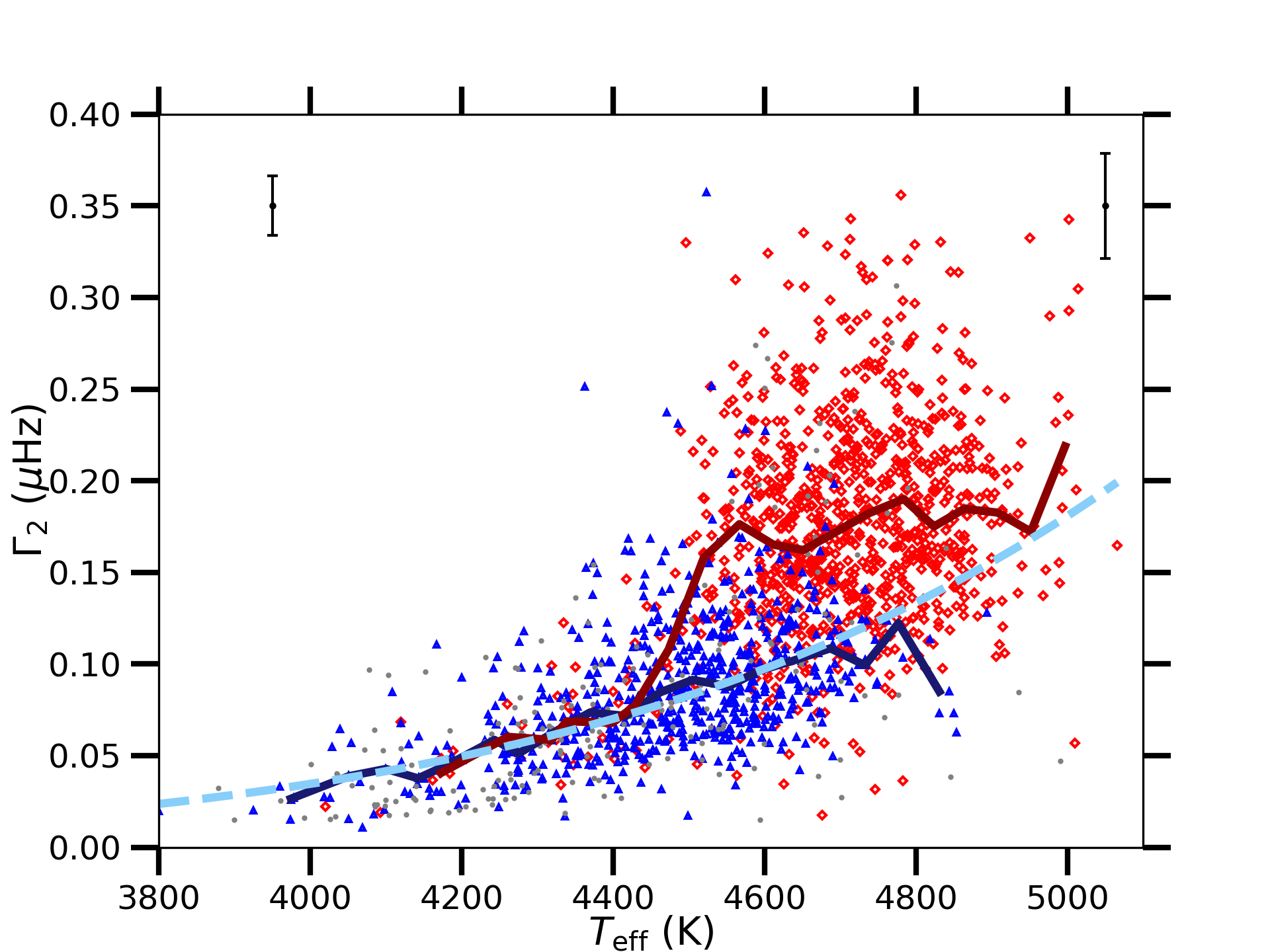}}
            \rotatebox{0}{\includegraphics[width=0.50\linewidth]{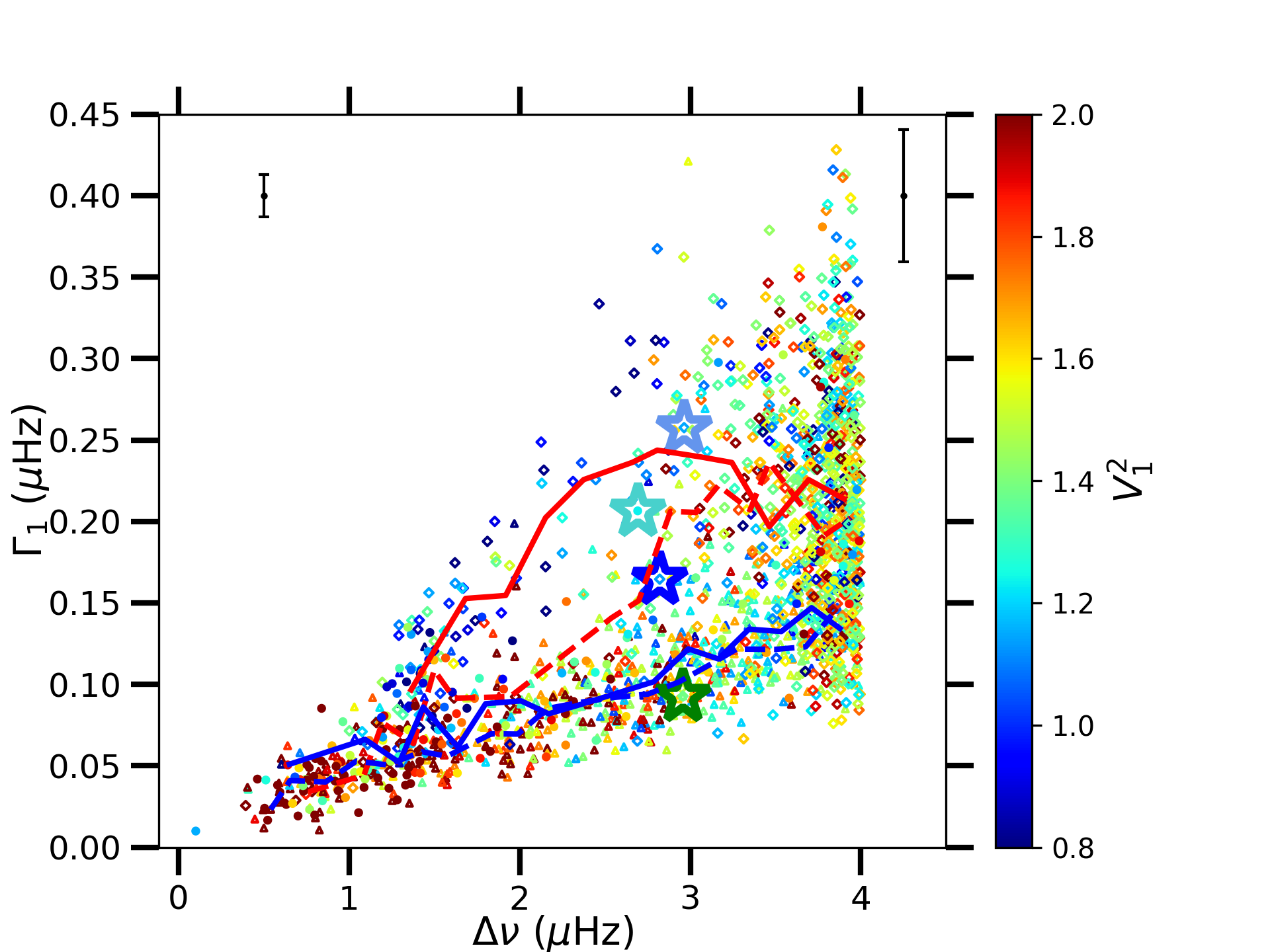}}
\end{minipage}
        \caption{Upper panels: $\Gammazero$ as a function of $\Dnu$ and $\Teff$. Middle panels: Ratio of $\Gammaone$ and $\Gammazero$ as a function of $\Dnu$ and $\Gammaone$ as a function of $\Teff$. For convenience, horizontal dotted black lines are plotted at specific values of 0.5, 1.0, 1.5, and 2.0. Bottom left panel: $\Gammatwo$ as a function of $\Teff$. The colours and symbols are the same as in Fig.~\ref{fig:Ampl_G_phi_Dnu}. Mean error bars on the widths have been computed both at low $\Teff$ ($\Teff \leq 4200$K) and at high $\Teff$ ($\Teff \geq 4200$K). These limits are equivalent to the limits in $\Dnu$ chosen in Fig.~\ref{fig:seismic_parameters}. The fits presented in Table~\ref{Table:scaling_relation_mode_width_mode_amplitude} are plotted with dashed light blue lines for RGB stars. Bottom right: $\Gammaone$ as a function of $\Dnu$ with the dipole mode visibilities colour-coded. The solid and dashed lines correspond to the median values for low-visibility dipole modes ($V_{1}^{2} \leq 1.5$) and for high-visibility dipole modes ($V_{1}^{2} \geq 1.5$), respectively, in blue for RGB stars and in red for He-burning stars. The turquoise, dark blue, light blue and green stars are the individual stars KIC 6847371, KIC 11032660, KIC 5461447, and KIC 6768042, respectively. They are studied in Sect.~\ref{sec:energy_equipartition} to test the reliability of the measurements of the dipole mode width. The median values are computed in $0.2$ $\mu$Hz wide $\Dnu$ bins and in $50$ K wide $\Teff$ bins.
        }
        \label{fig:Gamma_l_Dnu_T_eff}
\end{figure*}

The mode widths $\langle\Gamma_{\ell}\rangle$ were fitted by the function 

\begin{equation}
\label{eq:scaling_rel_modewidth}
\langle\Gamma\ind{\ell}\rangle =  a_{\ell}\left(\frac{\Teff}{4800\mathrm{K}}\right)^{b_{\ell}},
\end{equation}
where $a_{\ell}$ and $b_{\ell}$ are free parameters. The fits are presented in Fig.~\ref{fig:Gamma_l_Dnu_T_eff} and summarised in Table~\ref{Table:scaling_relation_mode_width_mode_amplitude}.\\

We note that clump stars globally have larger radial mode widths with a larger spread than those observed for RGB stars, as mentioned in previous studies \citep{2012ApJ...757..190C, 2018A&A...616A..94V}. However, when core-He-burning ends and the star ascends the AGB ($\Dnu \lesssim 3~\mu$Hz), the radial mode widths decrease and become comparable to measurements made on the RGB. \\

In Fig.~\ref{fig:Gamma_l_Dnu_T_eff} we compare the dipole mode widths $\Gammaone$ to the radial mode widths $\Gammazero$. On the RGB, we note that $\Gammaone$ values are globally 20\% higher than $\Gammazero$ above $\Dnu \geq 3.5 \ \mu$Hz, while they are globally similar below. For He-burning stars, the $\ell = 1$ modes have larger widths than the $\ell = 0$ modes above $\Dnu \geq 1.5 \ \mu$Hz. We identified three reasons that might explain this behaviour. First, as mentioned in Sect.~\ref{sec:Method}, we applied the correction expressed by Eq.~\ref{eq:mixed_correction} to $\Gammaone$ when the fitted modes are mixed modes. However, the term $\zeta$ is close to $1$, therefore the correction to $\Gammaone$ introduces large uncertainties on the inferred dipole p-mode widths. Second, most of the unexpectedly high $\Gammaone$ values are in fact highly perturbed by mixed modes. Gravity-dominated mixed modes can only be observed if the condition 
\begin{equation}
\label{eq:condition_g_dominated}
    \mathcal{N} \leq \frac{1}{4q}\left( \frac{\pi}{2} \frac{\Gamma_{0}}{\dnures} - 5\right)
\end{equation}
is met \citep{2018A&A...618A.109M}, where $\mathcal{N} = \Dnu/(\nu^{2}\Delta\Pi_{1})$ is the number of gravity modes per radial order $n$, $\Delta\Pi_{1}$ is the period spacing, $q$ is the coupling factor, $\Gamma_{0}$ is the radial mode width, and $\dnures$ is the frequency resolution. Using typical values of $q$ \citep[see e.g.][]{2017A&A...600A...1M} and $\Gammazero$, we can infer that the right-hand side term of Eq.~\ref{eq:condition_g_dominated} is close to $20$ at $\Dnu \sim 3~\mu$Hz for He-burning stars. Then, Eq.~\ref{eq:condition_g_dominated} is hardly verified and only p-dominated modes are mainly visible. In these cases, the mixed modes are so close that the fits rather reproduce several confused mixed modes than a unique pure pressure mode. Third, we note that all the highest values of $\Gammaone$ are systematically associated with low $\ell = 1$ mode visibilities in the interval $\Dnu \in [1.5, 2.5] \ \mu$Hz. These dipole modes with a low amplitude are unexpectedly large and are further discussed in Sect.~\ref{sec:energy_equipartition}. The comparison between $\Gammatwo$ and $\Gammazero$ is not discussed here because $\Gammazero \sim \Gammatwo,$ as expected.\\



We also investigated the temperature dependence of $\Gammal$ (Fig.~\ref{fig:Gamma_l_Dnu_T_eff}). The fits performed on each stellar population (cf. Table~\ref{Table:scaling_relation_mode_width_mode_amplitude}) indicate that $\Gammal$ and $\Teff$ are strongly correlated, regardless of the degree $\ell$. 
\citet{2018A&A...616A..94V} also reported that $\Gammazero$ is correlated with $\Teff$ for less evolved giants, but this correlation is not as pronounced as in the present study. A strong correlation like this is expected across the HR diagram according to theoretical work \citep{2012A&A...540L...7B}.

\begin{table}
\caption{Scaling relations for the mode widths and for the mode amplitudes}
\begin{tabular}{rrrr}
\hline
\hline
 & Population & $a_{\ell}\qquad$ & $b_{\ell}\qquad$\\
\hline
$\Gammazero$ ($\mu$Hz) & RGB & & $10.8^{*}$\\
 & RGB & $0.13 \pm 0.02$ & $6.36 \pm 0.37$\\
$\Gammaone$ ($\mu$Hz) & RGB & $0.18 \pm 0.01$ & $9.73 \pm 0.34$\\
$\Gammatwo$ ($\mu$Hz) & RGB & $0.14 \pm 0.02$ & $7.41 \pm 0.40$\\
 \hline
  &  & $c_{\ell}\qquad$ & $d_{\ell}\qquad$\\
 \hline
$\Amplzero$ (ppm) & $M \leq 1.2M_{\odot}$ & $1013 \pm 20$ & $-0.64 \pm 0.02$\\
 & $M \geq 1.2M_{\odot}$ & $902 \pm 36$ & $-0.68 \pm 0.03$\\

$\Amplone$ (ppm) & $M \leq 1.2M_{\odot}$ & $928 \pm 19$ & $-0.59 \pm 0.02$\\
 & $M \geq 1.2M_{\odot}$ & $853 \pm 35$ & $-0.64 \pm 0.03$\\

$\Ampltwo$ (ppm) & $M \leq 1.2M_{\odot}$ & $1090 \pm 31$ & $-0.74 \pm 0.02$\\
 & $M \geq 1.2M_{\odot}$ & $1031 \pm 30$ & $-0.75 \pm 0.02$\\
 \hline
\end{tabular}
\label{Table:scaling_relation_mode_width_mode_amplitude}
\\
\textbf{Notes:} The mode widths $\langle\Gamma\ind{\ell}\rangle$ and the mode amplitudes $\langle A \ind{\mathrm{bol},\ell}\rangle$ are fitted by Eq.~\ref{eq:scaling_rel_modewidth} and Eq.~\ref{eq:scaling_rel_mode_amplitude}, respectively. (*) The exponent $b_{\ell = 0}$ indicated in the first row for $\Gammazero$ is the value expected on the RGB \citep{2012A&A...540L...7B}.
\end{table}

\subsection{Mode amplitudes}

\begin{figure}[htbp]
        \begin{minipage}{1.\linewidth}  
                \rotatebox{0}{\includegraphics[width=1.0\linewidth]{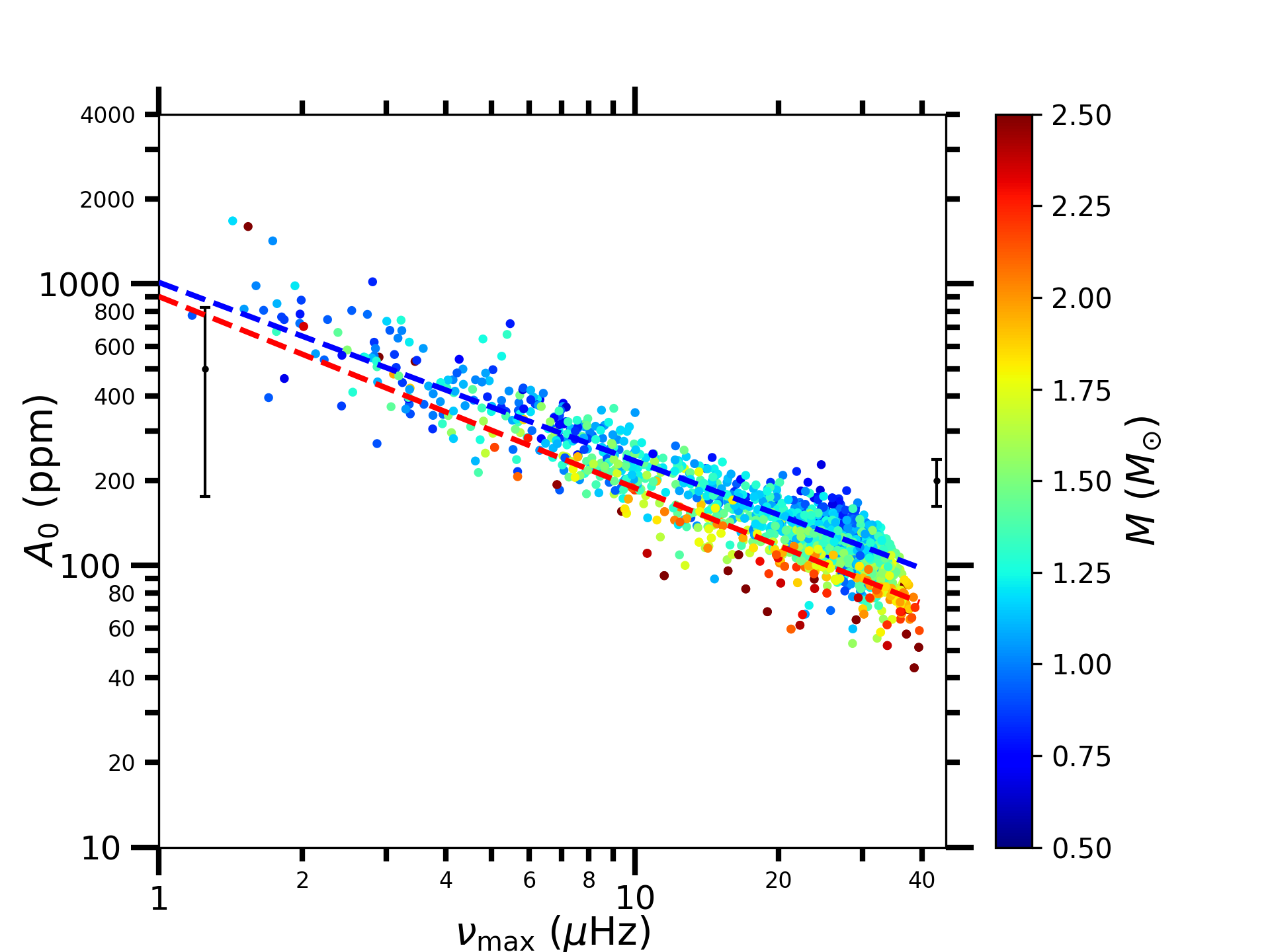}}
        \end{minipage}
        \caption{Radial mode amplitude $\Amplzero$ computed from Eq.~\ref{eq:amplitude_bolometric_correction} as a function of $\numax$ with the stellar mass colour-coded. The dashed lines are the fits presented in Table~\ref{Table:scaling_relation_mode_width_mode_amplitude}, shown in blue for low-mass stars ($M \leq 1.2~M_{\odot}$) and in red for high-mass stars ($M \geq 1.2~M_{\odot}$). The error bars are computed in the same way as in Fig.~\ref{fig:Gamma_l_Dnu_T_eff}.
        }
        \label{fig:Ampl_mode_0_nu_max}
\end{figure}

The radial mode amplitude $\Amplzero$ defined in Eq.~\ref{eq:amplitude_bolometric_correction} is plotted as a function of $\numax$ in Fig.~\ref{fig:Ampl_mode_0_nu_max} and was adjusted by the scaling relation

\begin{equation}
\label{eq:scaling_rel_mode_amplitude}
\langle A \ind{\ell, \mathrm{bol}}\rangle = c_{\ell}\ \numax^{d_{\ell}},
\end{equation}
where $c_{\ell}$ and $d_{\ell}$ are free parameters and $\numax$ is given in $\mu$Hz. The radial mode amplitudes follow the same trend as highlighted in recent studies \citep[e.g.][]{2011ApJ...743..143H, 2011ApJ...737L..10S, 2012A&A...537A..30M, 2018A&A...616A..94V}. The radial mode amplitude does not differ between RGB stars and He-burning stars. For both stellar populations, the radial mode amplitude follows a power law with an exponent roughly equal to $-0.70$. Furthermore, the previous studies reported a clear mass dependence regardless the evolutionary stage: the higher the mass, the lower the radial mode amplitude (see Fig.~\ref{fig:Ampl_mode_0_nu_max}).

\subsection{Mode visibilities}
\label{sec:mode_visibilities_results}

\begin{figure*}[!ht]
        \begin{minipage}{1.0\linewidth}  
                \rotatebox{0}{\includegraphics[width=0.5\linewidth]{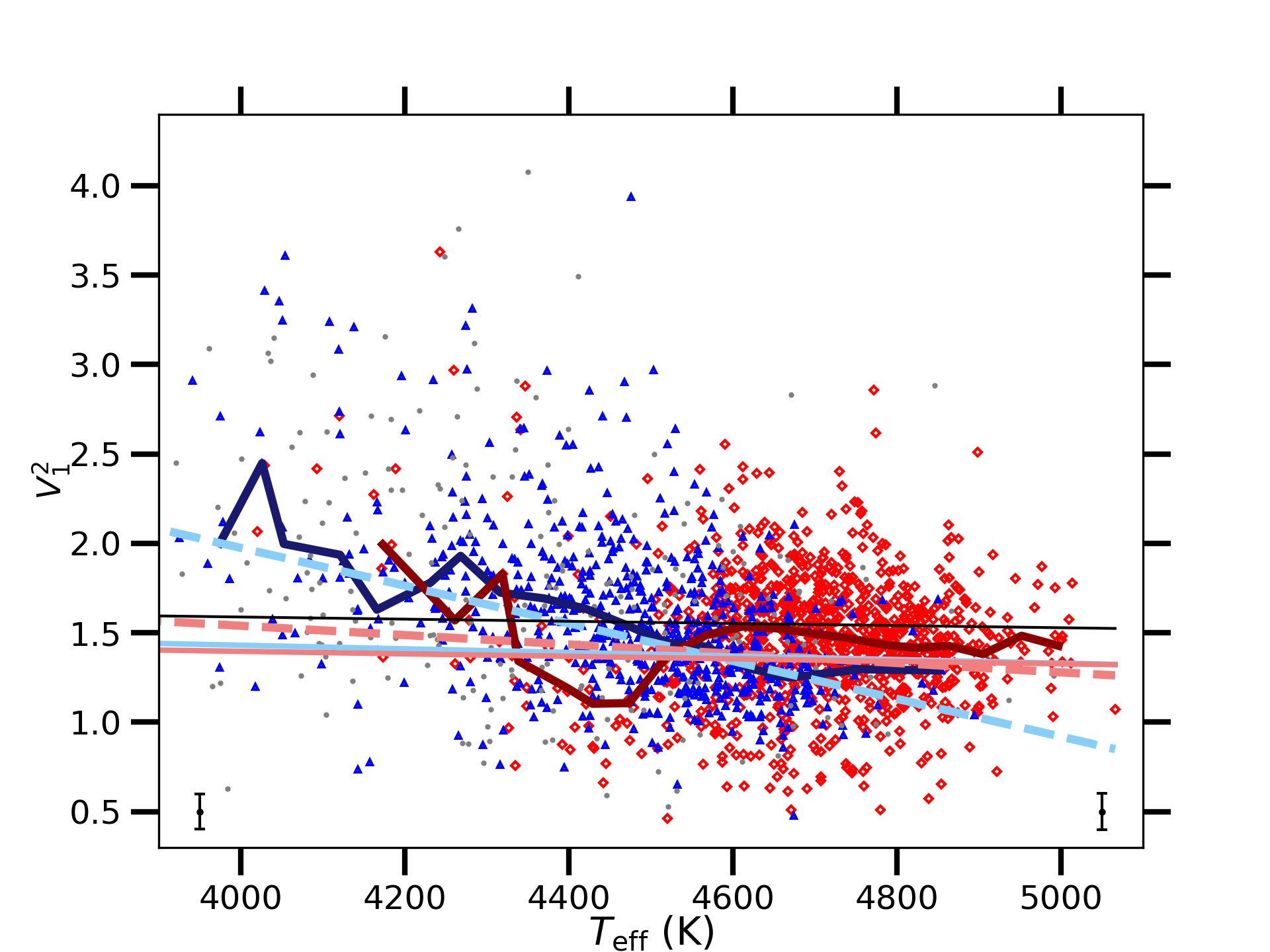}}
        \rotatebox{0}{\includegraphics[width=0.5\linewidth]{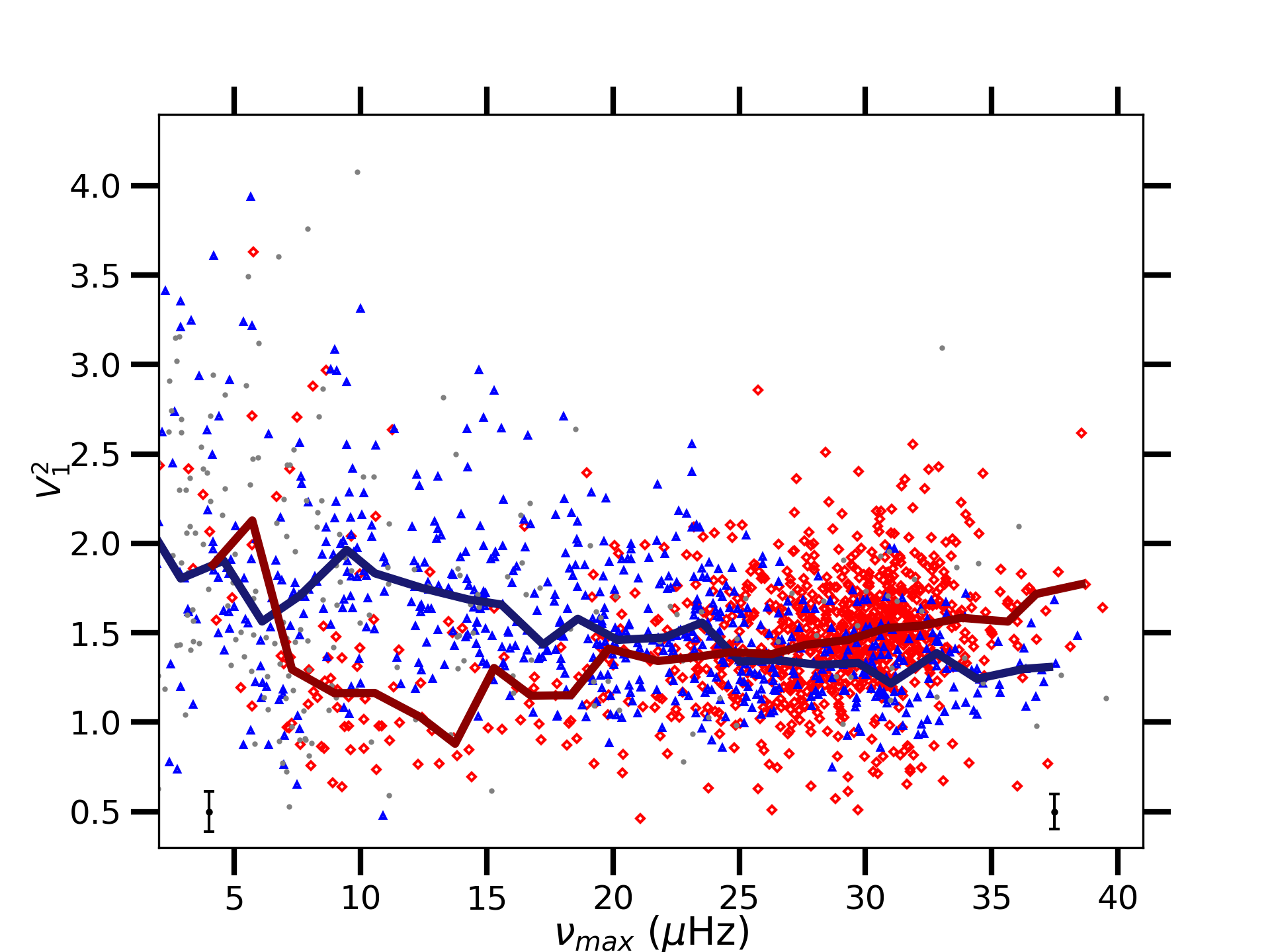}}
\end{minipage}
    \begin{minipage}{1.\linewidth}  
                \rotatebox{0}{\includegraphics[width=0.5\linewidth]{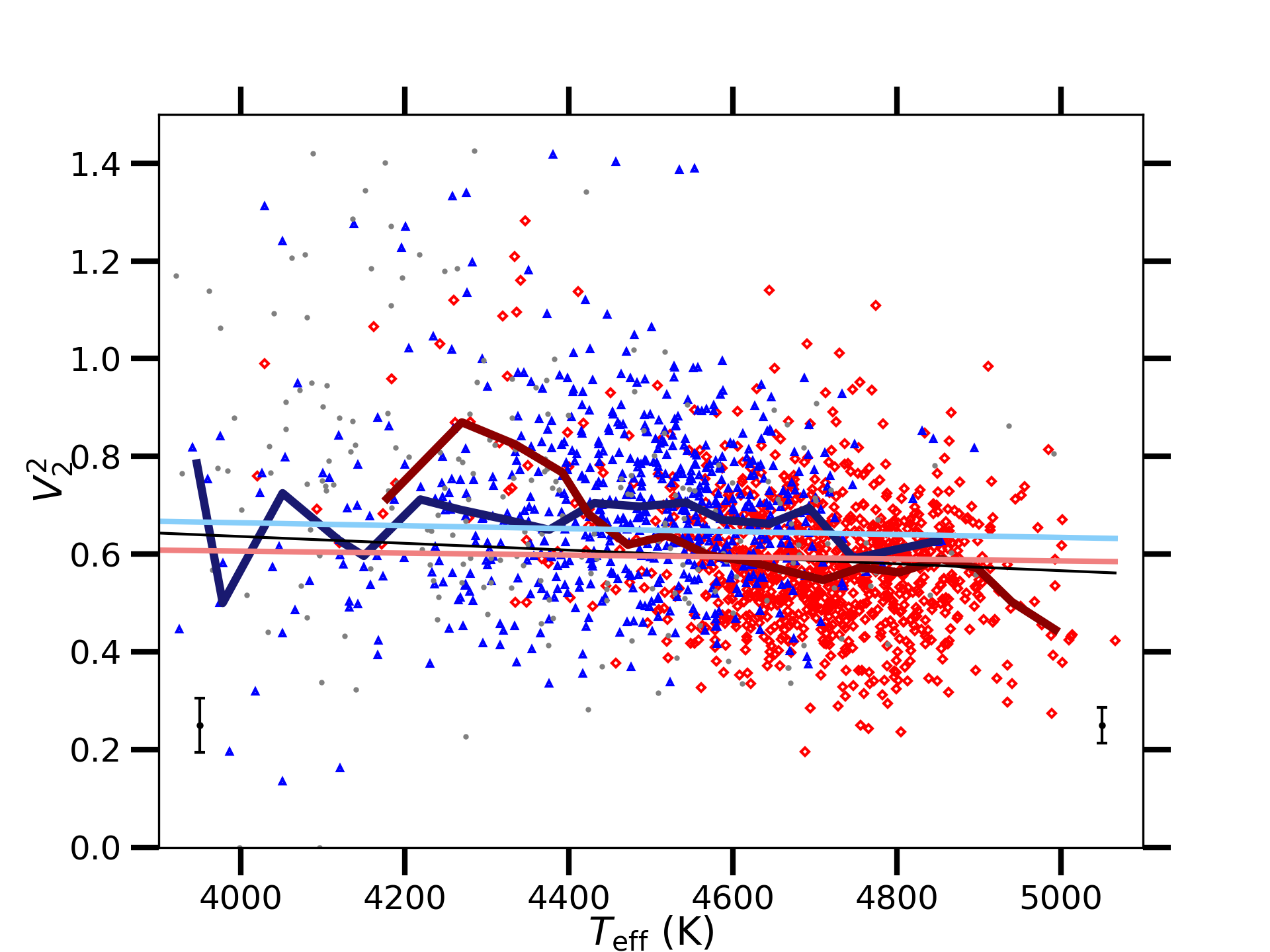}}
                \rotatebox{0}{\includegraphics[width=0.5\linewidth]{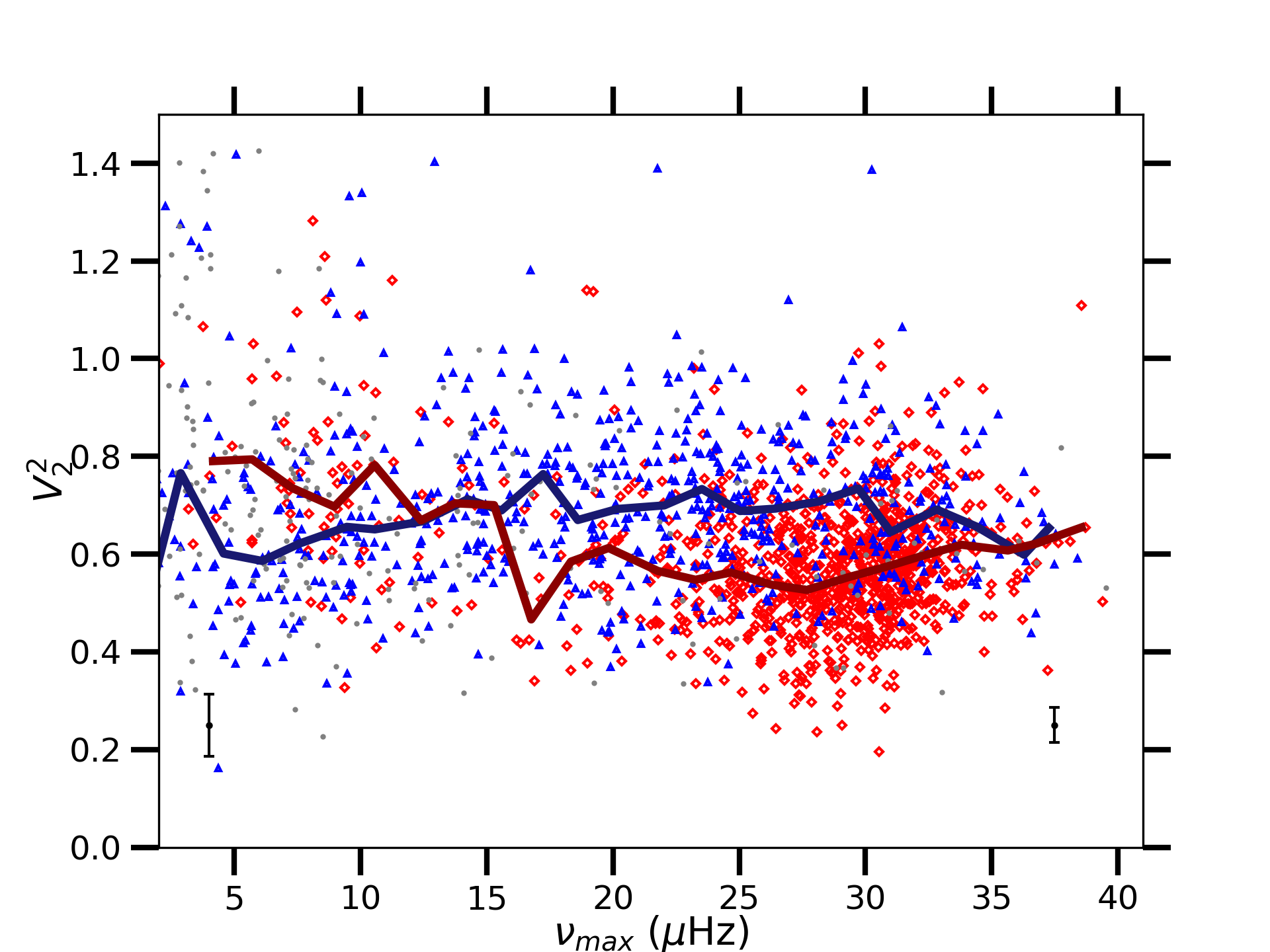}}
\end{minipage}
        \begin{minipage}{1.\linewidth}
                \rotatebox{0}{\includegraphics[width=0.5\linewidth]{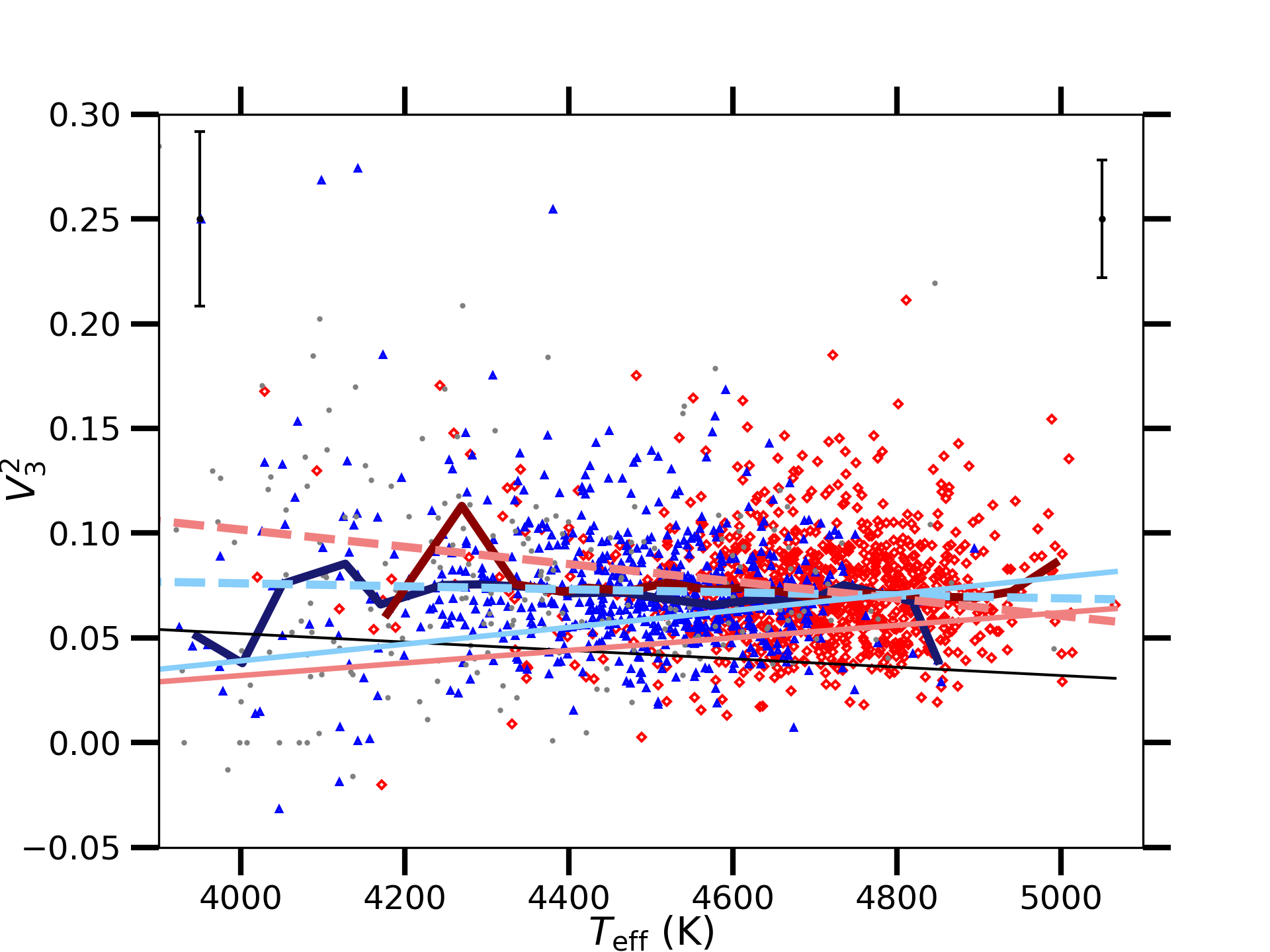}}
\end{minipage}
        \caption{Upper panels: Visibility of the $\ell = 1$ modes as a function of $\Teff$ in the left panel and of $\numax$ in the right panel. The colours and symbols are the same as in Fig.~\ref{fig:Gamma_l_Dnu_T_eff}. The error bars are computed in the same way as in Fig.~\ref{fig:Gamma_l_Dnu_T_eff}. Similarly, error bars on the visibilities are given for both low $\numax$ ($\numax \leq 4.5~\mu$Hz) and high $\numax$ ($\numax \geq 4.5~\mu$Hz). The dashed lines are the fits presented in Table~\ref{Table:scaling_relation_mode_visibilities}, in light blue for RGB stars and in light red for He-burning stars. The thin solid light blue and light red lines are the fits obtained for less evolved stars \citep{2012A&A...537A..30M} for RGB stars and for He-burning stars, respectively. The thin solid black line is the theoretical prediction \citep{2011A&A...530A..97B}. Middle panels: Same labels as in the upper panels, but for the visibility of $\ell = 2$. Lower left panel: Same labels as in the upper left panel, but for the visibility of $\ell = 3$ modes. The median values are computed in $50$ K wide $\Teff$ bins and in $1.5$ $\mu$Hz wide $\numax$ bins.
        }
        \label{fig:Vl_Teff_numax}
\end{figure*}

The energy distribution between modes of different degree $\ell$ can be studied through the mode visibilities (Eq.~\ref{eq:visibility}). They are presented in Fig.~\ref{fig:Vl_Teff_numax} and were fitted by the linear function

\begin{equation}
\label{eq:scaling_rel_visibilities}
\Vl = \alpha + \beta (\Teff - 4800K),
\end{equation}
where $\alpha$ and $\beta$ are free parameters and $\Teff$ is given in K (see Table~\ref{Table:scaling_relation_mode_visibilities}).

\begin{table}
\caption{Fits of the mode visibilities presented in Fig.~\ref{fig:Vl_Teff_numax}.}
\begin{tabular}{rrrr}
\hline
\hline
 & Population & $\alpha$ & $\beta\ (10^{3}\ \mathrm{K}^{-1})$\\
\hline

$V_{1}^{2}$ & Solar-like* & 1.54 & -0.06 \\
 & RGB & $1.13 \pm 0.02$ & $-1.006 \pm 0.012$\\
 & He-burning & $1.33 \pm 0.05$ & $-0.261 \pm 0.013$\\
\hline
$V_{2}^{2}$ & Solar-like* & 0.58 & -0.07 \\
 & RGB & $0.66 \pm 0.12$ & $-0.148 \pm 0.058$\\
 & He-burning & $0.55 \pm 0.01$ & $-0.390 \pm 0.006$\\
\hline
$V_{3}^{2}$ & Solar-like* & 0.036 & -0.02 \\
 & RGB & $0.07 \pm 0.02$ & $-0.007 \pm 0.005$\\
 & He-burning & $0.07 \pm 0.01$ & $-0.041 \pm 0.006$\\
  \hline
\end{tabular}
\label{Table:scaling_relation_mode_visibilities}
\\
\textbf{Notes:} The mode visibilities are fitted by Eq.~\ref{eq:scaling_rel_visibilities}. (*) The expected coefficients are derived for solar-like oscillators, including RGB stars \citep{2011A&A...531A.124B}.
\end{table}

\label{sec:mode_visibilities}

We verified that the high values of $V_{1}^{2}$ and $V_{2}^{2}$ can be explained by very weak radial mode amplitudes. For some He-burning stars, the dipole mixed modes extend up to the frequency range where $\ell = 3$ modes are located. When mixed modes are too close to the $\ell = 3$ modes, a fraction of the energy associated with mixed modes can be accidentally accounted for as part of the energy of $\ell = 3$ modes. Consequently, some $V_{1}^{2}$ values may be underestimated, and $V_{3}^{2}$ is inevitably overestimated. In the case of less evolved stars, \cite{2012A&A...537A..30M} suggested that the scatter in $V_{1}^{2}$ could be related to the conditions that govern the coupling between g modes and p modes, giving rise to mixed modes. In the case of He-burning stars, this could explain the spread we obtain because these stars clearly exhibit mixed modes when $\Dnu \gtrsim 3.0\ \mu$Hz.

Although we note a large spread for the mode visibilities, it is clear that the non-radial mode visibilities increase when $\Teff$ decreases both for RGB stars and He-burning stars, as expected from theoretical predictions \citep{2011A&A...531A.124B}. The only exception is the visibility of $\ell = 1$ modes in He-burning stars. This is due to the presence of several dipole modes with very low visibilities in the interval $\Teff \in [4200, 4500]$K, as reflected by the gap between the medians computed for RGB and He-burning stars. The mode visibilities in evolved stars similarly behave as in less evolved stars, except in the case of $\ell = 3$ modes, since we note that $V_{3}^{2}$ increases towards low $\Teff$ , whereas \cite{2012A&A...537A..30M} observed the opposite trend. \\





The visibility of dipole modes $V_{1}^{2}$ is represented as a function of $\numax$ in the upper right panel of Fig.~\ref{fig:Vl_Teff_numax}. We observe a clear difference between RGB stars and He-burning stars in the interval $\numax \in$ [7, 20] $\mu$Hz, with He-burning stars having weaker $V_{1}^{2}$ than their RGB counterparts. In parallel, we previously reported that $\Gamma_{1}$ is greater for He-burning stars in this interval (Fig.~\ref{fig:Gamma_l_Dnu_T_eff}). It is certain that mixed modes perturb the extraction of the pure pressure dipole mode widths when $\Dnu \geq 1.5$ $\mu$Hz, but the presence of low-visibility dipole modes reflects a shortage of dipole mode energy, which could be linked to a higher dipole mode damping, hence to a higher dipole mode width. We study this question in Sect.~\ref{sec:discussion}. Furthermore, we find that quadrupole modes have larger amplitudes in the H-burning phases than on the He-burning phases in the interval $\numax \in [15, 35]\mu$Hz, as represented in the middle right panel of Fig.~\ref{fig:Vl_Teff_numax}. This difference may be linked to the mixed character of the quadrupole modes, which is more pronounced during the clump phase in the interval $\numax \in [15, 35]\mu$Hz.

Nevertheless, we note a large spread of the dipole mode visibilities.
The dipole mode visibilities of He-burning stars become comparable with those measured on the RGB at low $\Dnu$ ($\Dnu \leq 1.5$ $\mu$Hz), when mixed modes disappear in the oscillation spectrum. The physical mechanisms that govern the coupling between the p-mode and the g-mode cavities might therefore be linked to the observation of low dipole mode visibilities. Even if the presence of depressed modes in advanced stages of stellar evolution is not clear, the simultaneous presence of low dipole mode visibilities and dipole mixed modes could help to identify the physical processes that cause the depressed modes in less evolved stages, which are still under debate \citep{2015Sci...350..423F, 2016Natur.529..364S, 2016ApJ...824...14C, 2017A&A...598A..62M}.

\begin{figure}[htbp]
        \begin{minipage}{1.\linewidth}  
                \rotatebox{0}{\includegraphics[width=1.0\linewidth]{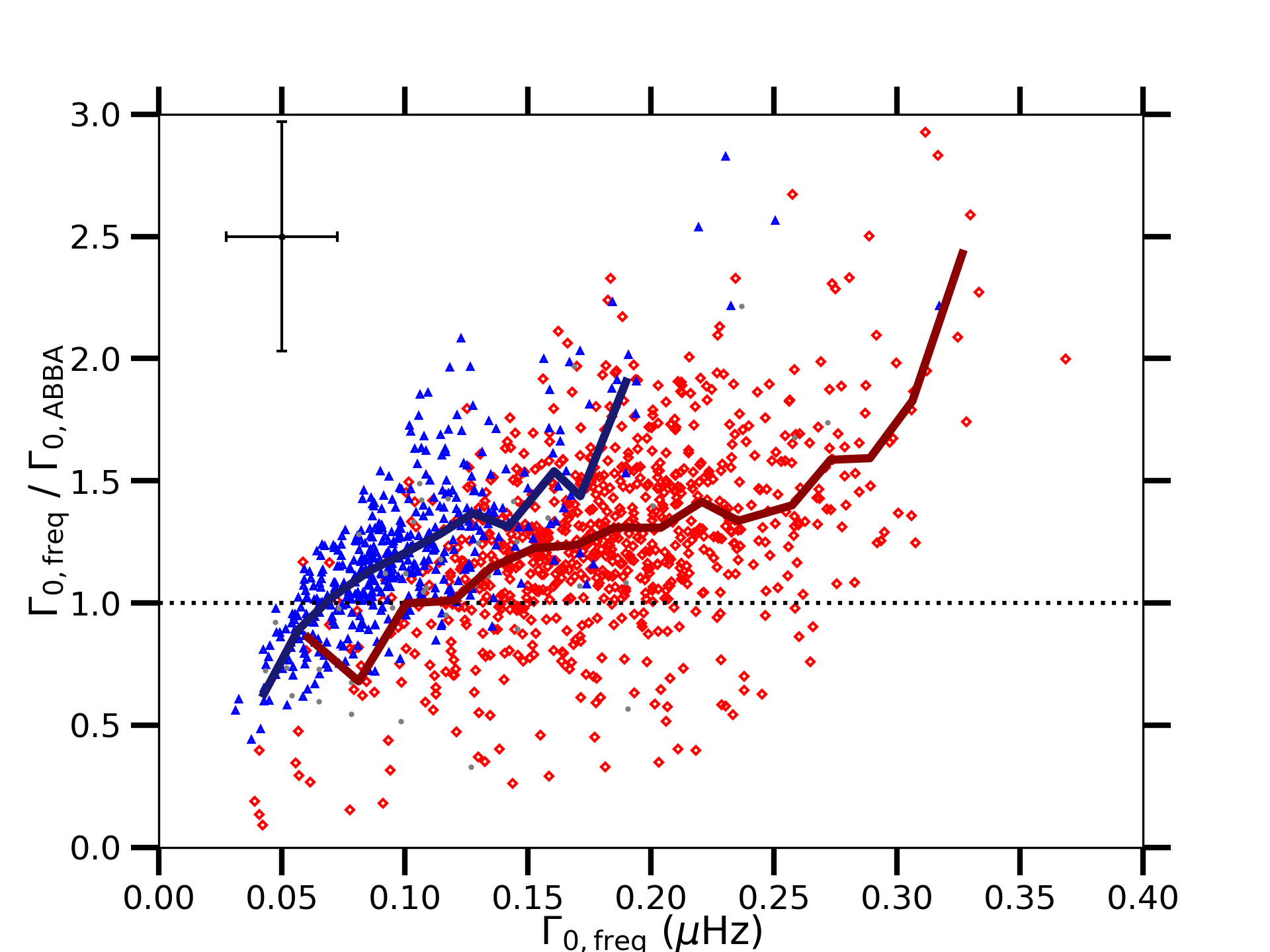}}
                \rotatebox{0}{\includegraphics[width=1.0\linewidth]{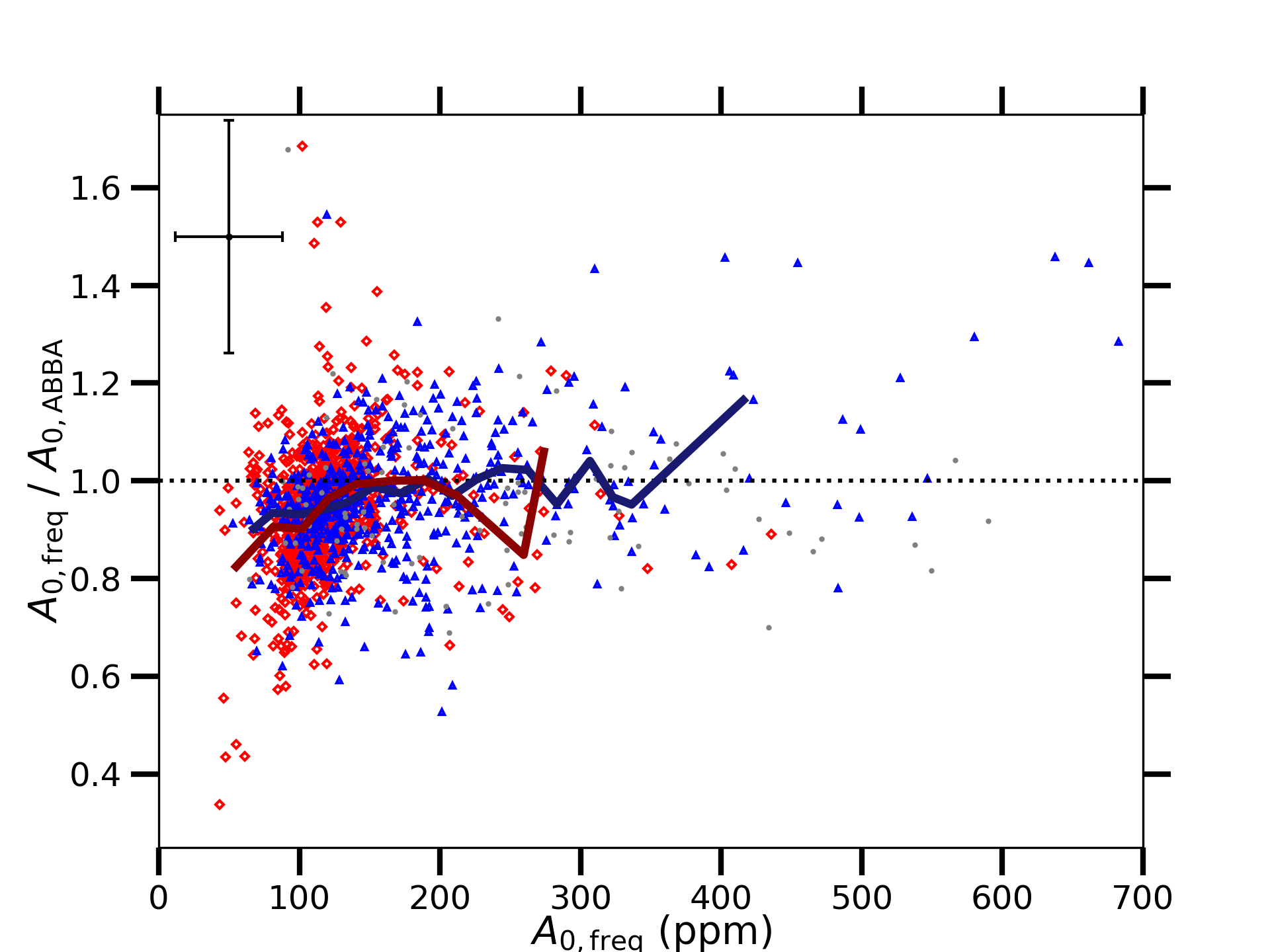}}
        \end{minipage}
        \caption{Top panel: Ratio of the average radial mode widths $\Gammazero$ obtained in this study and those obtained with a Bayesian method \citep{2019arXiv190609428K}. The median values are computed in $0.015$$\mu$Hz wide $\Gamma_{0,\mathrm{freq}}$ bins. Bottom panel: Same as in the upper panel, but for the average radial mode bolometric amplitude $\Amplzero$. The colours and symbols are the same as in Fig.~\ref{fig:Ampl_G_phi_Dnu}. The median values are computed in $20$ ppm $A_{0,\mathrm{freq}}$ bins. The dotted line represents the 1:1 agreement. Mean error bars are represented at the top of each panel.
        }
        \label{fig:comparison_peak_bagging_methods}
\end{figure}

\subsection{Comparison with other peak-bagging methods}
\label{sec:comparison_peak_bagging_methods}

The measurements inferred from our frequentist peak-bagging are compared with those\footnote{https://github.com/tkallinger/KeplerRGpeakbagging} of the automated Bayesian peak-bagging algorithm A\rotatebox[origin=c]{180}{$\mathrm{B}$}BA \citep{2019arXiv190609428K}, which uses the Bayesian nested sampling algorithm MULTINEST \citep{2009MNRAS.398.1601F}. The average radial mode widths and bolometric amplitudes derived in the Bayesian approach were computed in the same way as in Sect.~\ref{sec:mode_fitting_method}. The comparison is shown in Fig.~\ref{fig:comparison_peak_bagging_methods}. \\
The radial mode widths $\Gammazero$ derived with our frequentist peak-bagging are globally larger than those obtained with A\rotatebox[origin=c]{180}{$\mathrm{B}$}BA by about 25\%. This overestimate is frequency dependent because it increases for higher values of $\Gammazero$. Conversely, our radial bolometric mode amplitudes are weakly underestimated by about 5\% with respect to the A\rotatebox[origin=c]{180}{$\mathrm{B}$}BA values. \cite{2018A&A...616A..94V} also reported that the radial mode width was overestimated by about 10\% in the frequentist approach with respect to a Bayesian approach. Moreover, we approximated the background component by Eq.~\ref{eq:background} around $\numax$ , while \citet{2019arXiv190609428K} modelled it with two super-Lorentzian functions \citep{2014A&A...570A..41K}. The background parametrisation has a non-negligible impact on the mode fitting, and stellar background bias is one of the main sources of frequency-dependent systematic errors in the measurements of mode widths and heights \citep{2014A&A...566A..20A}. The way that the stellar background was modelled may therefore partly explain the differences we find between measurements.

\section{Discussion}
\label{sec:discussion}

\subsection{Stellar classification at advanced stages}

Using a global measurement of the large separation $\Dnu$ and the oscillation pattern of red giants (Eq.~\ref{eq:nunl_asympt}), we did not find any difference in the acoustic offset $\eps$ between RGB stars and He-burning stars. In parallel, we highlighted a difference in the signature of the helium second-ionisation zone between these stellar populations, especially in the modulation phase $\Phi$. This phase difference locally affects the measurement of $\Dnu$ according to Eq.~\ref{eq:glitch_signature_obs}. A local change in $\Dnu$ can be linked to a local change in $\eps$ by differentiating Eq.~\ref{eq:nunl_asympt}, leading to 

\begin{equation}
\label{eq:link_deps_dDnu}
\delta\eps = - (n + \eps)\frac{\delta \Dnu}{\Dnu}.
\end{equation}
In the case of RGB and clump stars, \cite{2015A&A...579A..84V} showed that the values of $\delta\eps$ inferred from $\delta (\log\Dnu)$ that is related with the helium second-ionisation zone match the typical difference in $\eps$ between RGB and clump stars. They identified the glitch signatures as the physical basis of the stellar population identification method based on the acoustic offset $\eps$. We extended the conclusions raised by \cite{2015A&A...579A..84V} to more advanced evolutionary stages, that is, between RGB and He-burning stars, including clump and AGB stars. The difference in the local measurements of $\eps$ between RGB and AGB stars reported by \cite{2012A&A...541A..51K} is caused by the different glitch signature of the helium second-ionisation zone, especially for the modulation phase $\Phi$.

\subsection{AGB bump}

\begin{figure*}[htbp]
        \begin{minipage}{1.\linewidth}  
                \rotatebox{0}{\includegraphics[width=0.5\linewidth]{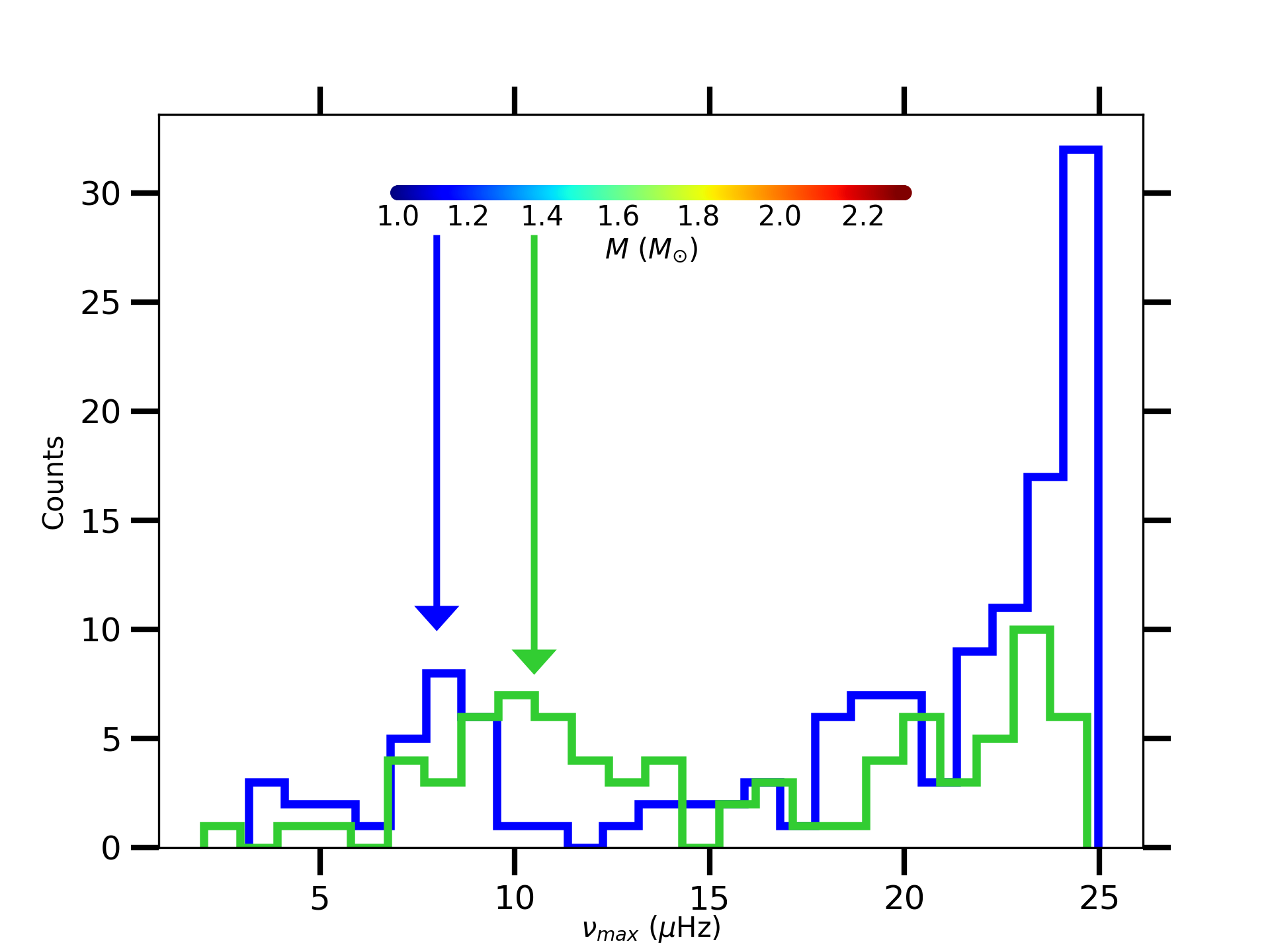}}
                \rotatebox{0}{\includegraphics[width=0.5\linewidth]{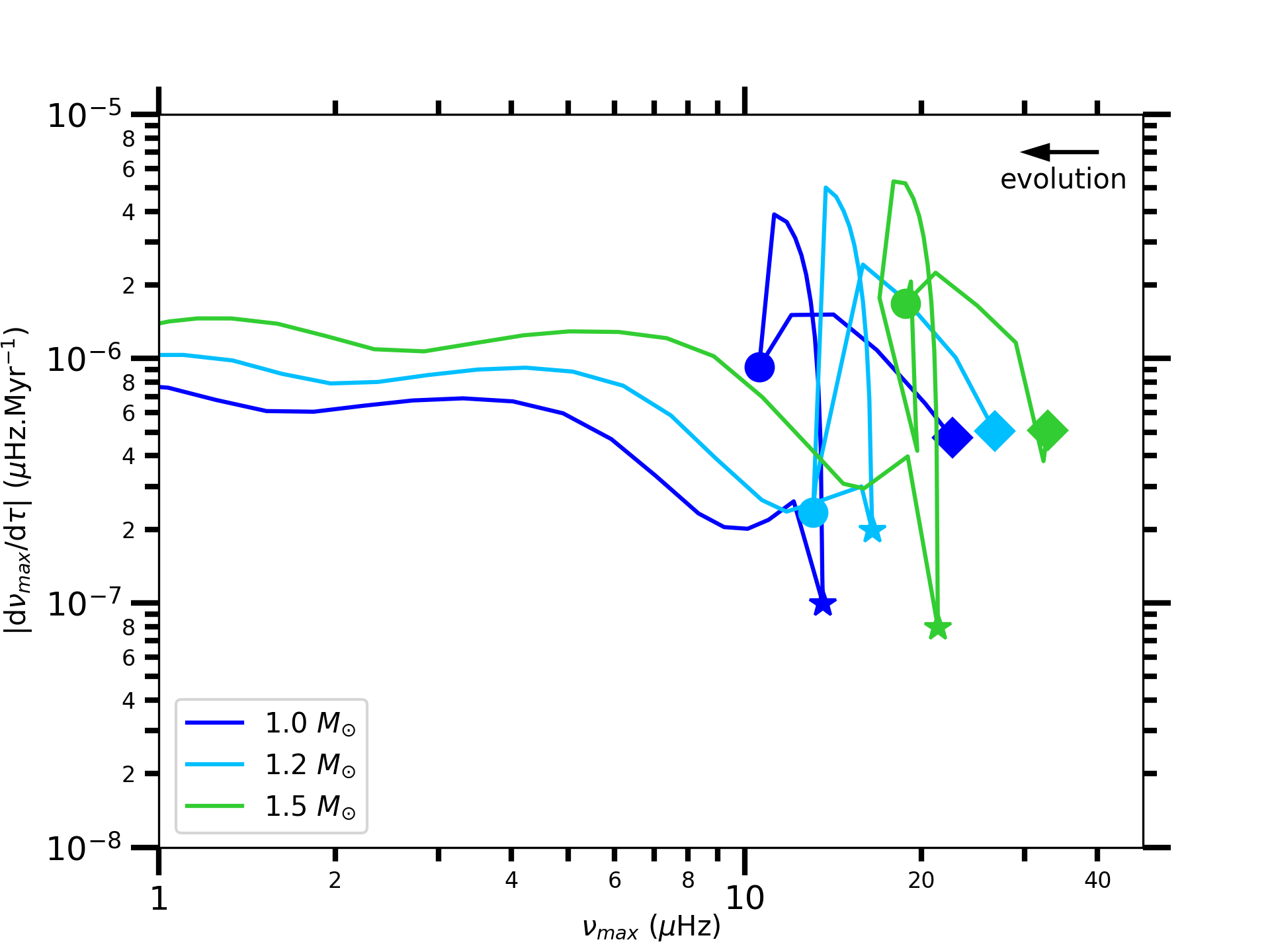}}
        \end{minipage}
        \caption{Left panel: Distribution of He-burning stars in terms of $\numax$. In blue we show the number of low-mass He-burning stars ($M \leq 1.2\ M_{\odot}$), and in green we show the number of high-mass He-burning stars ($M \geq 1.2\ M_{\odot}$). The colour bar indicates the location in $\numax$ where we expect the AGB bump for a given mass following Eq.~\ref{eq:scaling_relation_nu_max_L} and adopting $\Teff = 4800\left(\numax/40.0\right)^{0.06}$ \citep{2010A&A...517A..22M}. The blue and green arrows roughly indicate the location of the AGB bump for low-mass and high-mass stars, respectively, which is characterised by a local excess of stars. Right panel: Evolution speed $\mathrm{d \numax}/\mathrm{d}\tau$, where $\tau$ is the stellar age, as a function of $\numax$ for different stellar masses. The models computed with MESA start from the end of the clump phase, which is marked by a diamond. The start and the end of the AGB bump are marked by a circle and a star, respectively.
        }
        \label{fig:hist_low_mass_high_mass}
\end{figure*}

After leaving the clump phase where helium-burning takes place in the core, the star enters the AGB phase. During the early asymptotic giant branch (eAGB) and in the case of low-mass and intermediate-mass stars, two turning-backs of the evolutionary track can be seen in a narrow interval of luminosity, similarly to what can be seen during the RGB bump: This is the so-called AGB bump (AGBb). It is caused by the onset of the shell-He burning and was first identified in the Large Magellanic Cloud colour-magnitude diagram \citep{1998ApJ...495L..43G}. The AGBb is observationally characterised by a local excess of stars in the luminosity distribution of stellar populations. Such an increment has been identified at $\log\left(L/L_{\odot}\right) \sim 2.2$ \citep{2015MNRAS.453.2290B}. For a star of $M = 1.0\ M_{\odot}$ and $\Teff = 4500K$, this is equivalent to $\numax \sim 8\ \mu$Hz according to the scaling relation \citep{1995A&A...293...87K}

\begin{equation}
\label{eq:scaling_relation_nu_max_L}
\frac{\numax}{\nu\ind{max,\odot}} = \frac{M}{M\ind{\odot}}\left( \frac{L}{L\ind{\odot}} \right)^{-1} \left( \frac{\Teff}{T\ind{eff,\odot}} \right)^{7/2}.
\end{equation}
Accordingly, we selected He-burning stars that left the clump phase (i.e. $\numax \lesssim 25~\mu$Hz). Their distribution as a function of $\numax$ is shown in Fig.~\ref{fig:hist_low_mass_high_mass}. By tracking stellar evolution towards low $\numax$, we note a depleted region followed by a peak for low mass-stars (at $\numax \sim 8~\mu$Hz) and for high-mass stars (at $\numax \sim 11~\mu$Hz). The depleted region could be explained by a difference in the evolution speed.
We have computed models with the MESA code, using the 1M\_pre\_ms\_to\_wd test suite case to investigate the evolution speed between the end of the clump phase and the ascent on the AGB. The results are presented in the right panel of Fig.~\ref{fig:hist_low_mass_high_mass}. For a given mass, we note that the evolution is faster between the end of the clump phase and the start of the AGBb than right after the AGBb since the variation of $\numax$ with time is more important before the AGBb. The fast evolution speed before the AGBb results in a small statistical probability to meet low-mass stars in the interval $\numax \in [8,15]\mu$Hz and high-mass stars in the interval $\numax \in [14,18]\mu$Hz. Investigating the AGBb in depth is part of our future work.

\subsection{A strong damping during the eAGB phase?}
\label{sec:energy_equipartition}

Very low degree modes have similar eigenfunctions in the stellar outer layers, so that they are excited in similar conditions and show similar power spectral densities. However, as mentioned in Sect.~\ref{sec:mode_visibilities_results}, many He-burning stars have very low dipole mode visibilities below $\numax = 20~\mu$Hz. In parallel, we found that most of the He-burning stars with low dipole mode visibilities have larger dipole mode widths. These low dipole mode visibilities reflect a lack of energy that could be linked to a strong dipole mode damping. Accordingly, we analysed the correlation between low visibility and large damping of dipole modes in detail by fitting the mixed-mode pattern during the early-AGB phase.

To this end, we considered single stars that have been identified as eAGB stars according to the classification method of \cite{2014A&A...572L...5M}. We selected five eAGB stars that have both low visibility dipole modes and a mixed-mode pattern clear enough to fit individual mixed modes and measure their widths (KIC 6847371, 11032660, 5461447, 10857623, and 6768042). We compared these widths to the pure-pressure dipole-mode width with Eq.~\ref{eq:mixed_correction}. The results shown in Appendix \ref{appendix:individual_fits_mixed_modes} are unfortunately not unequivocal. Three of these eAGB stars present a strong dipole-mode damping, which is within the 1$\sigma$ uncertainty for KIC 6847371 and KIC 5461447 and within the $2\sigma$ uncertainty for KIC 11032660. Nevertheless, this is not what we observe for KIC 10857623 and KIC 6768042. We face several constraints to reduce the uncertainties (or to fit other spectra), such as a low signal-to-noise ratio,  a high ratio $\zeta$ of the mode inertia in the core and the total mode inertia (which leads to high uncertainties through Eq.~\ref{eq:mixed_correction}), limited frequency resolution, rotational splittings, and buoyancy glitch signature. We therefore tested another method to process these constraints together.
\\

\begin{figure}[htbp]
        \includegraphics[width=1.0\linewidth]{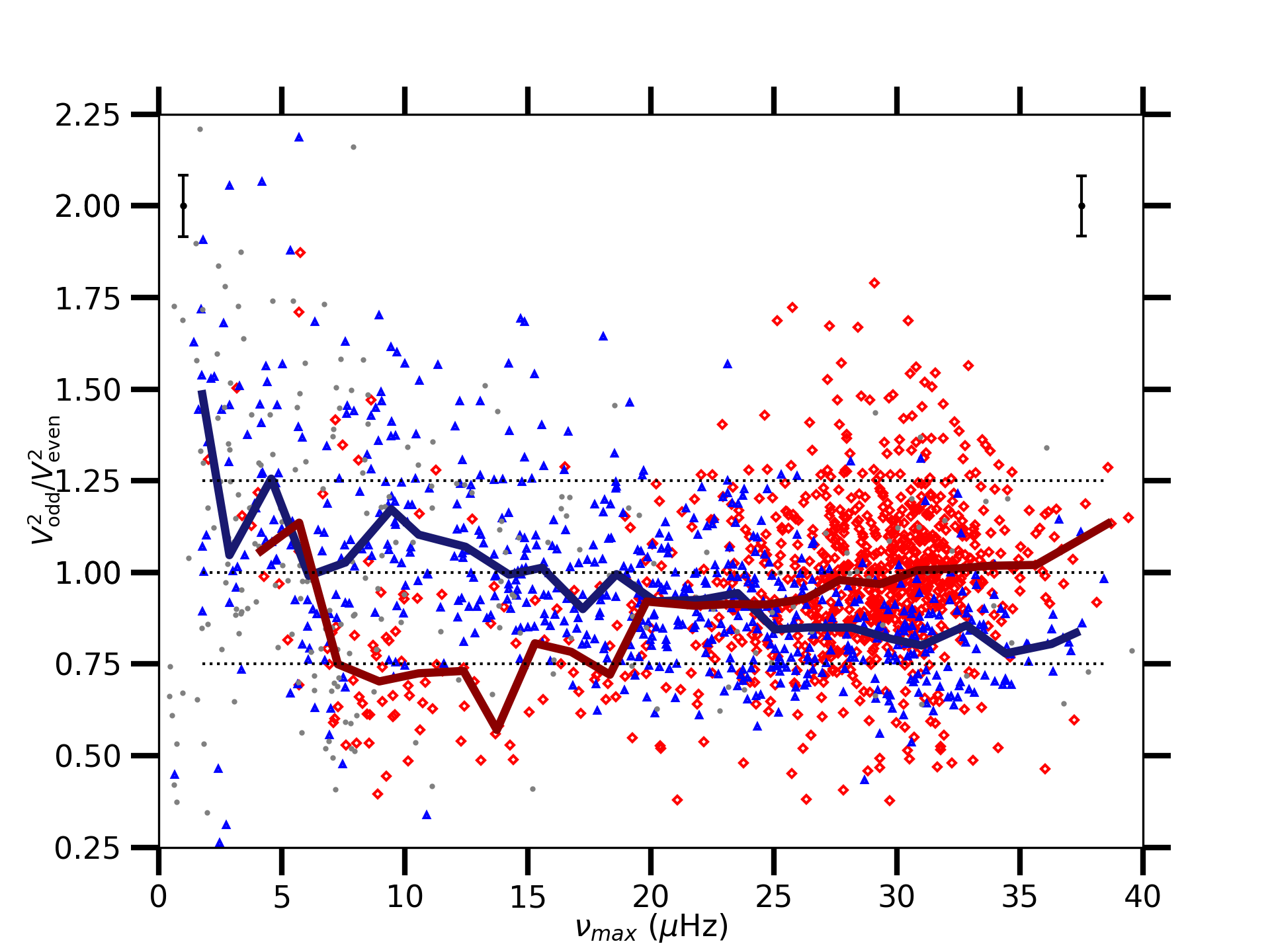}
        \caption{Ratio of the mode visibilities of odd and even degrees as a function of $\numax$. Same labels as in Fig.~\ref{fig:Ampl_G_phi_Dnu}. For convenience, horizontal dashed black lines are plotted at specific values of 0.75, 1.00, and 1.25. The error bars are computed in the same way as in Fig.~\ref{fig:seismic_parameters}. The median values are computed in $1.5~\mu$Hz $\numax$ bins.
        }
        \label{fig:Vodd_Veven_nu_max}
\end{figure}

In a star in spherical equilibrium (thus non-rotating and without magnetic field), we expect the energy equipartition between modes of even and odd degrees to be satisfied. The lack of energy observed for dipole modes can be studied by comparing the energy of even-degree modes to that of odd-degree modes. To this end, we computed the ratio between the visibilities of odd and even degrees, 

\begin{equation}
\label{eq:ratio_V_odd_even}
\frac{\Vlodd}{\Vleven} = \frac{V_{1}^2 + V_{3}^{2}}{1 + V_{2}^2}.
\end{equation}
Modes that have a degree of same parity are close one to each other, as illustrated in Fig.~\ref{fig:colour_spectrum_KIC_2695975}. 
As a result, studying the global contribution of the energy then limits the impact of the energy leakage between individual degrees. We use the measurement of the odd/even visibility ratio as an indicator of the variation of the visibilities of $\ell = 1$ modes (Fig.~\ref{fig:Vodd_Veven_nu_max}). The variation is in fact  dominated by the dipole modes for two reasons. On the one hand, the visibility of the octupole modes is very low for geometrical reasons. On the other hand, the quadrupole modes essentially behave as pure pressure modes for evolved giants, and always have a more pronounced pressure character than dipole modes. Below $\numax = 20\ \mu$Hz, the energy equipartition seems to be invalid for He-burning stars. The dipole mode visibility is weaker than predicted by theory for He-burning stars. This lack of dipole mode energy is linked to a large dipole mode damping according to our study, and invalidates the energy equipartition between the different low-degree modes.\\

For RGB stars we can note that the visibility of $\ell = 1$ modes is globally lower at high $\numax$ (see Fig.~\ref{fig:Vl_Teff_numax}) and is especially lower than the expected value (1.54), which makes $\Vleven$ greater than $\Vlodd$ above $\numax = 25\ \mu$Hz. This difference was first observed by \citet{2012A&A...537A..30M} for less evolved red giants and can theoretically be explained by a difference of dipole mode damping \citep{2012A&A...539A..83D}. The radiative damping causes an energy loss at the envelope base that decreases when the star ascends the RGB \citep[see Figs. 6 and 7]{2012A&A...539A..83D}. For a star with an initial mass $M_{0} = 2M_{\odot}$, the energy loss by gravity wave emission at the envelope base is expected to cancel out when $\numax \simeq 28~\mu$Hz on the RGB. This is consistent with our observations because we note that $V_{1}
^{2}$ increases with evolution below $\numax \leq 28~\mu$Hz (see the upper right panel of Fig.~\ref{fig:Vl_Teff_numax}). On the RGB, radiative damping may explain the damping of energy of dipole modes observed at high $\numax$. On the AGB, the dipole mode visibility evolves in the opposite direction. Given that the evanescent region between the g-mode and the p-mode cavities grows while the star ascends the AGB, mixed modes become less visible, and consequently, the only visible dipole modes are trapped in the envelope, as observed on the RGB. It might therefore be relevant to investigate the radiative damping in order to determine whether it can explain the damping of dipole modes on the AGB. \\

\section{Conclusion}
\label{sec:conclusion}

So far, we have performed the first exhaustive study of the seismic analysis of evolved giants, including $\sim$ 2.000 stars ascending the RGB towards the luminosity tip and He-burning stars both in the clump phase and ascending the AGB. We successfully characterised the oscillation spectrum of stars with $\Dnu \geq 0.5$ $\mu$Hz and extracted the radial, dipole, and quadrupole mode parameters. By investigating the signature of the helium second-ionisation zone, we identified the physical origin on which the classification method based on $\eps$ and presented in \citet{2012A&A...541A..51K} relies at low $\Dnu$, that is, between RGB and AGB stars. We found that the amplitude and phase of the modulation introduced in the mode frequencies differ in RGB and in He-burning stars, that is, in core-He-burning and AGB stars. Work is in progress to investigate these differences with modelling. These differences affect local measurements of $\eps$ and enable classifying RGB and He-burning stars. Thus, we extended the work of \citet{2015A&A...579A..84V}, who drew the same conclusions, but considering RGB stars versus clump stars. As a consequence, we now have two methods relying on the same physical basis to decipher stellar evolution effects in evolved giant stars. On the one hand, we can adopt a local analysis where the signature of the helium second-ionisation zone is included in the acoustic offset $\eps$. In this case, the possible values of $\eps$ reflect stellar evolution effects. On the other hand, we can adopt a global analysis where the values taken by $\eps$ are squeezed together and the stellar evolution effects in $\eps$ fade. However, in this case, we can still emphasise the stellar evolution effects by considering an additional term in Eq.~\ref{eq:nunl_asympt} representing the signature of the helium second-ionisation zone on mode frequencies. \\

Having access to seismic diagnoses of evolved giants is promising for the understanding of stellar evolution, especially during the AGBb. The AGBb is expected to occur at $\log \left( L/L_{\odot}\right) \sim 2.2$ after the core He-burning phase. The investigation of the AGBb will be approached in a forthcoming paper. Furthermore, we highlighted that after the core He-burning phase, (i) the evolution is faster for low-mass stars, (ii) the dipole mode energy decreases, and (iii) the pressure-mode damping slowly becomes comparable to that measured on the RGB. This suggests that other physical processes need to be investigated in order to understand the mode damping and the observed visibilities as soon as core He-burning stops, that is, when the core becomes radiative again.


\section*{Acknowledgements}

The authors are grateful to the anonymous referee who helped them in improving this paper with constructive suggestions. G.D thanks P. Houdayer, R. Samadi and M. Vrard for fruitful discussions that helped improving this work. C.G. is supported by FCT - Fundaç\~ao para a Ci\^encia e a Tecnologia through national funds (PTDC/FIS-AST/30389/2017), by FEDER – Fundo Europeu de Desenvolvimento Regional through COMPETE2020 - Programa Operacional Competitividade e Internacionalizaç\~ao (POCI-01-0145-FEDER-030389), and by FCT/MCTES through national funds (PIDDAC) by these grants UIDB/04434/2020 and UIDP/04434/2020.

\begin{appendix}

\section{Location of the helium second-ionisation zone}
\label{appendix:link_period_G_location_heII}

The modulation period $\mathcal{G}$ defined in Eq.~\ref{eq:glitch_signature_obs} can be linked to the location of the helium second-ionisation zone. The structural variations caused by the helium glitch can be seen in the first adiabatic exponent profile, defined by

\begin{equation}
\label{eq:gamma_1_def}
\gamma_{1} = \left(\frac{\diff \log P}{\diff \log \rho }\right)_{\mathrm{s}},
\end{equation}
where $P$ and $\rho$ are the pressure and the density, respectively, and the subscript $s$ indicates that the derivative is taken at constant entropy. It is commonly assumed that the signature of the helium glitch arises from the dip in the $\gamma_{1}$ profile caused by the helium second-ionisation zone \citep{2005MNRAS.361.1187M, 2007MNRAS.375..861H}. In stellar models, we therefore took the acoustic radius at this local minimum as the location of the helium second-ionisation zone, for instance, $\tHeII$, defined by

\begin{equation}
\label{eq:acoustic_radius_def}
\tHeII = \int_{0}^{r_{\mathrm{HeII}}}{\frac{\diff r}{c_{\mathrm{s}}(r)}}.
\end{equation}
In this expression, $r_{\mathrm{HeII}}$ is the distance of the local minimum from the centre of the star, and $c_{\mathrm{s}}$ is the adiabatic sound speed.
The helium glitch introduces an oscillatory component in the eigenfrequency pattern of the star, which is proportional to \citep{1988IAUS..123..175G, 1988IAUS..123..151V, 1990LNP...367..283G}

\begin{equation}
\label{eq:dnu_glitch_modulation}
\delta\nu \propto \sin\left( 4\pi\tauHeII\nunl + \Phi_{\mathrm{HeII}}\right),
\end{equation}
where $\phi_{\mathrm{HeII}}$ is the phase of the glitch modulation and $\tau_{\mathrm{HeII}}$ is the acoustic depth of the helium glitch relative to the surface of the star of radius $R_{*}$,

\begin{equation}
\label{eq:def_acoustic_depth}
\tau_{\mathrm{HeII}} = \int_{r_{\mathrm{HeII}}}^{R_{*}}{\frac{\diff r}{c_{\mathrm{s}}(r)}}.
\end{equation}
The modulation introduced in the local large separation  (Eq.~\ref{eq:Dnu_local}) can also be expressed in the form of Eq.~\ref{eq:dnu_glitch_modulation} with a phase shift compared to $\Phi_{\mathrm{HeII}}$. Consequently, the modulation period $\mathcal{G}$ and the acoustic depth $\tau_{\mathrm{HeII}}$ can be linked according to

\begin{equation}
\label{eq:link_G_tauHeII}
\tauHeII = \frac{1}{2\mathcal{G}\Dnu}.
\end{equation}
Furthermore, the total acoustic length of the stellar cavity is defined by
\begin{equation}
\label{eq:length_acoustic_cavity}
T_{0} = \frac{1}{2\Dnu},
\end{equation}
so that we can convert the acoustic depth $\tauHeII$ into the acoustic radius $\tHeII$ with the relation $\tHeII = T_{0} - \tauHeII$. This transformation allows us to reduce the biases that result from the unknown exact position of the stellar surface \citep{1995MNRAS.276..283C, 2004A&A...423.1051B}. Finally, the modulation period $\mathcal{G}$ can be inferred from the location of the helium second-ionisation zone with the expression

\begin{equation}
\label{eq:link_period_G_location_HeII}
\mathcal{G} = \frac{1}{1 - t_{\mathrm{HeII}}/T_{0}}.
\end{equation}

\section{Mixed-mode measurements in the eAGB phase}
\label{appendix:individual_fits_mixed_modes}

We selected individual stars identified as eAGB stars according to the classification method of \cite{2014A&A...572L...5M} and fitted their mixed dipole modes near $\numax$ to extract an estimate of the mode widths. The mixed dipole mode widths were then used to infer the pure-pressure dipole mode widths according to Eq.~\ref{eq:mixed_correction}.

\begin{figure*}[ht!]
        \begin{minipage}{1.\linewidth}  
                \rotatebox{0}{\includegraphics[width=0.5\linewidth]{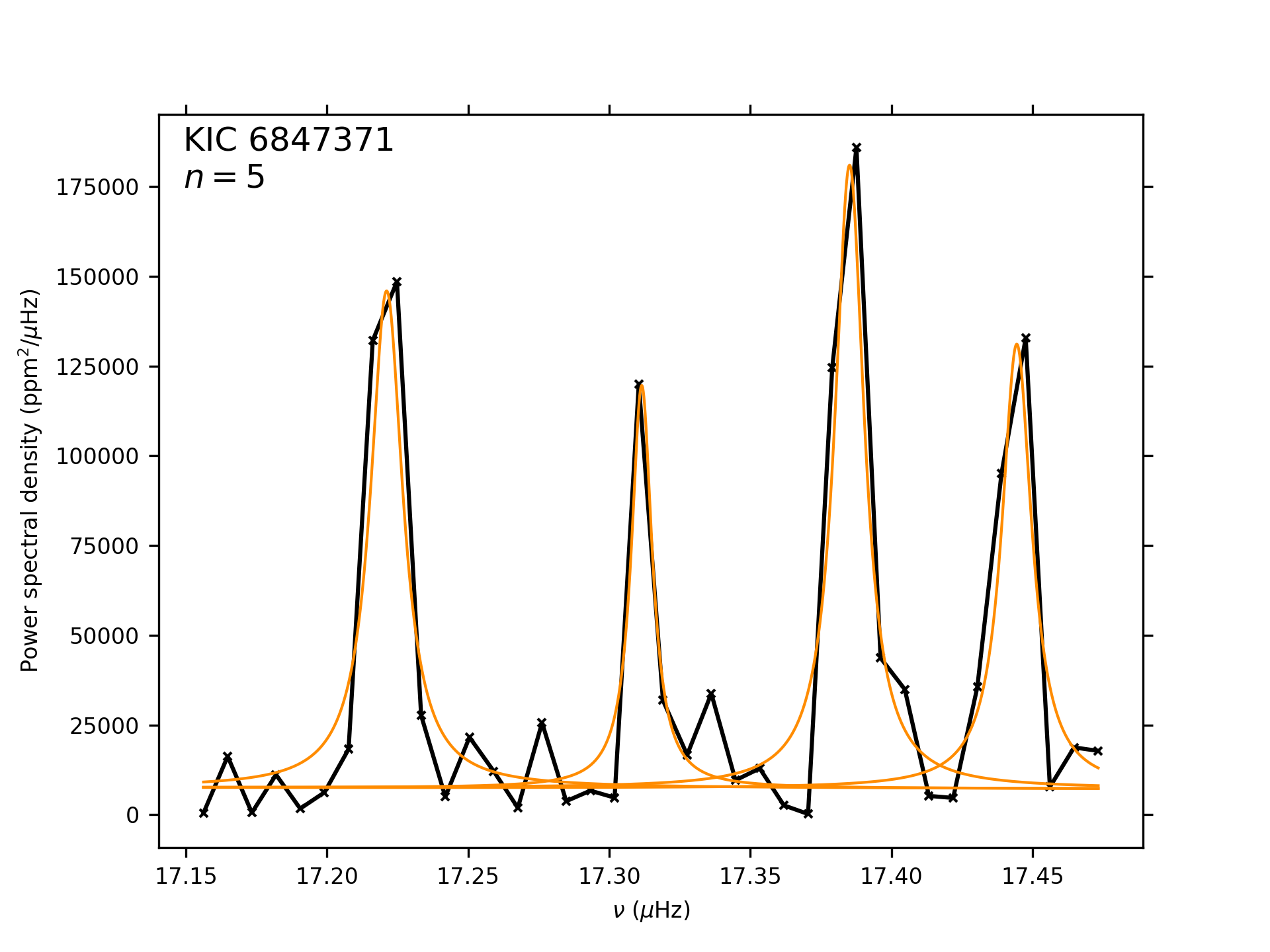}}
                \rotatebox{0}{\includegraphics[width=0.5\linewidth]{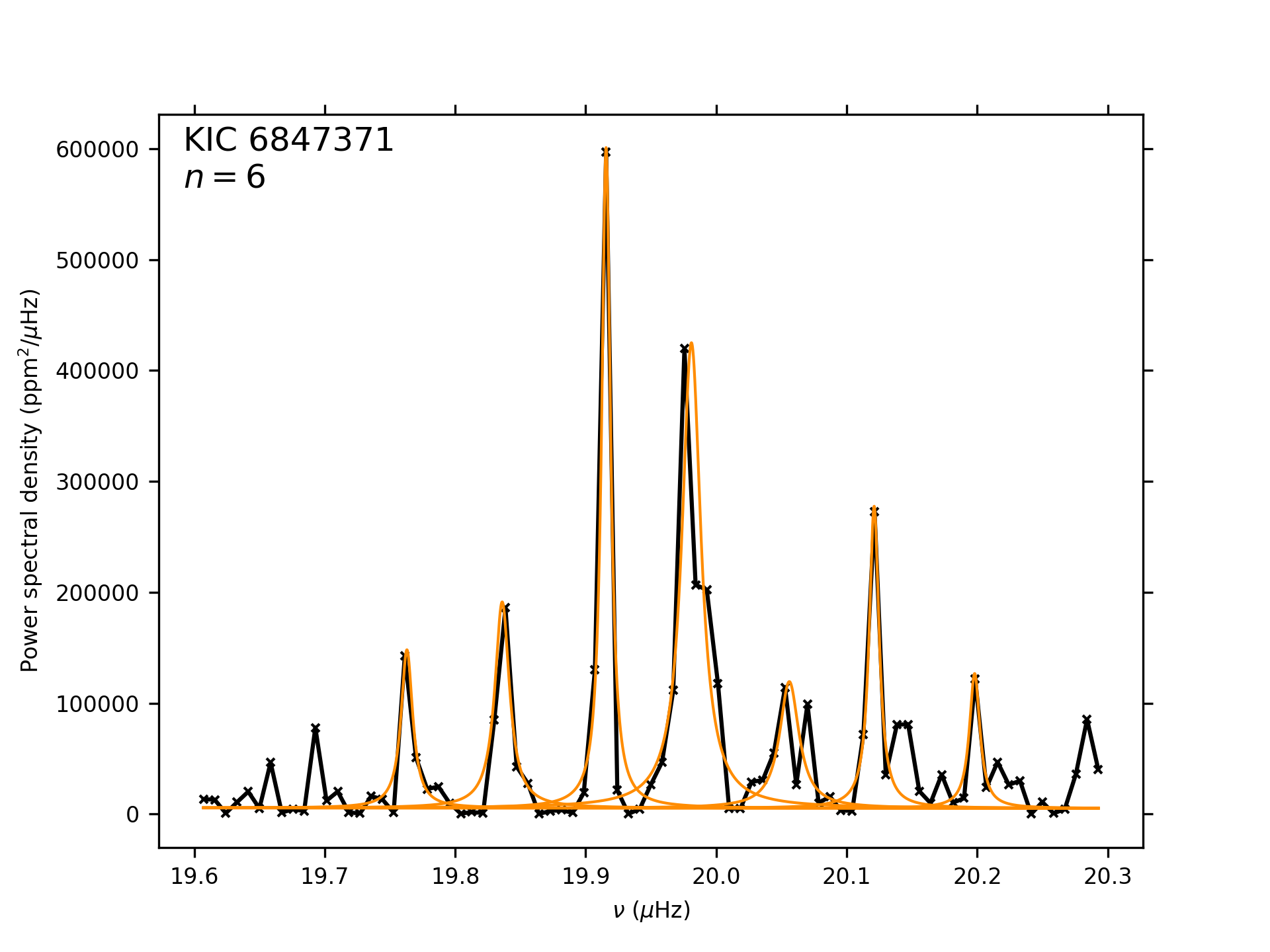}}
        \end{minipage}
        \begin{minipage}{1.\linewidth}  
                \rotatebox{0}{\includegraphics[width=0.5\linewidth]{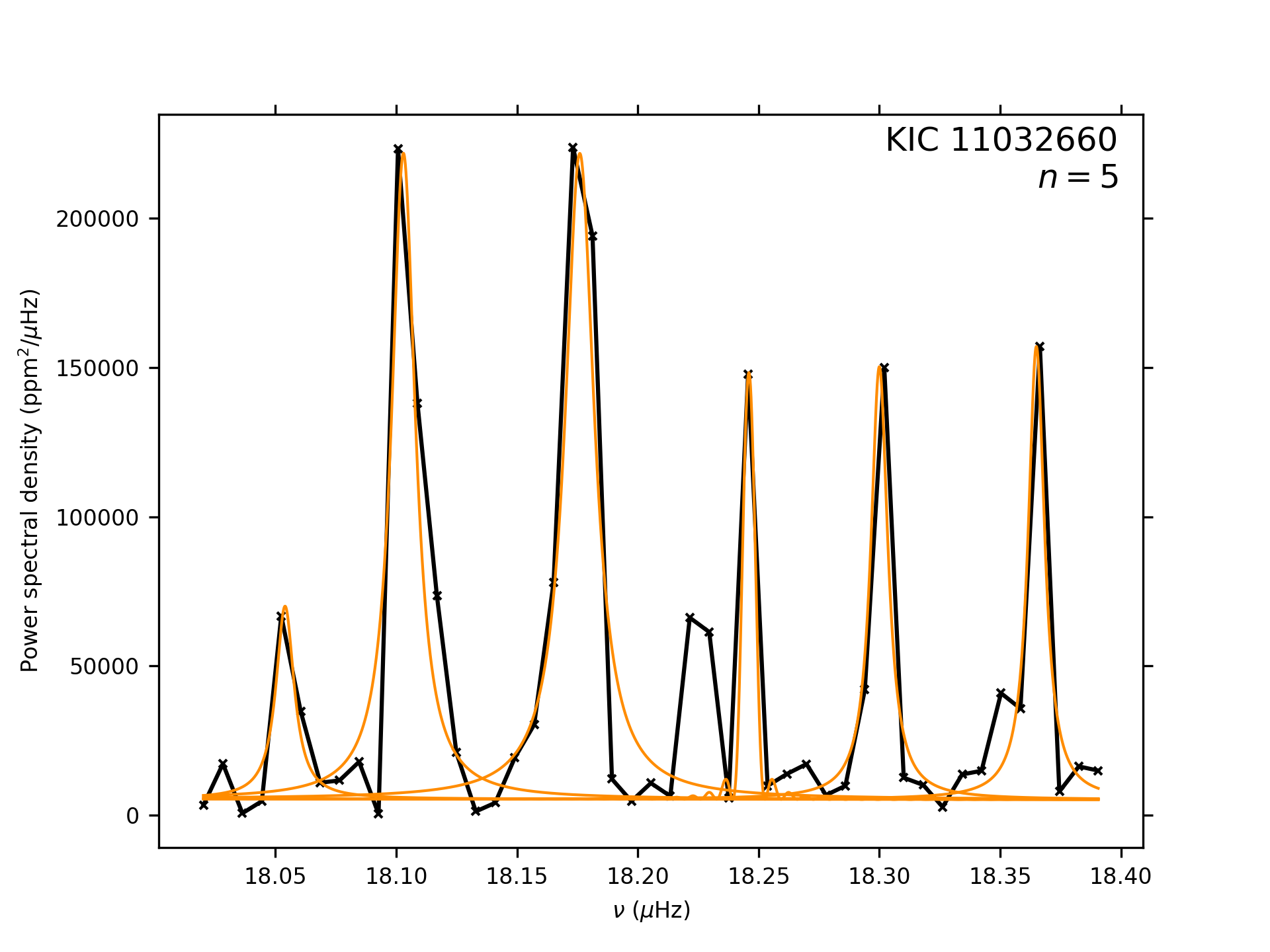}}
                \rotatebox{0}{\includegraphics[width=0.5\linewidth]{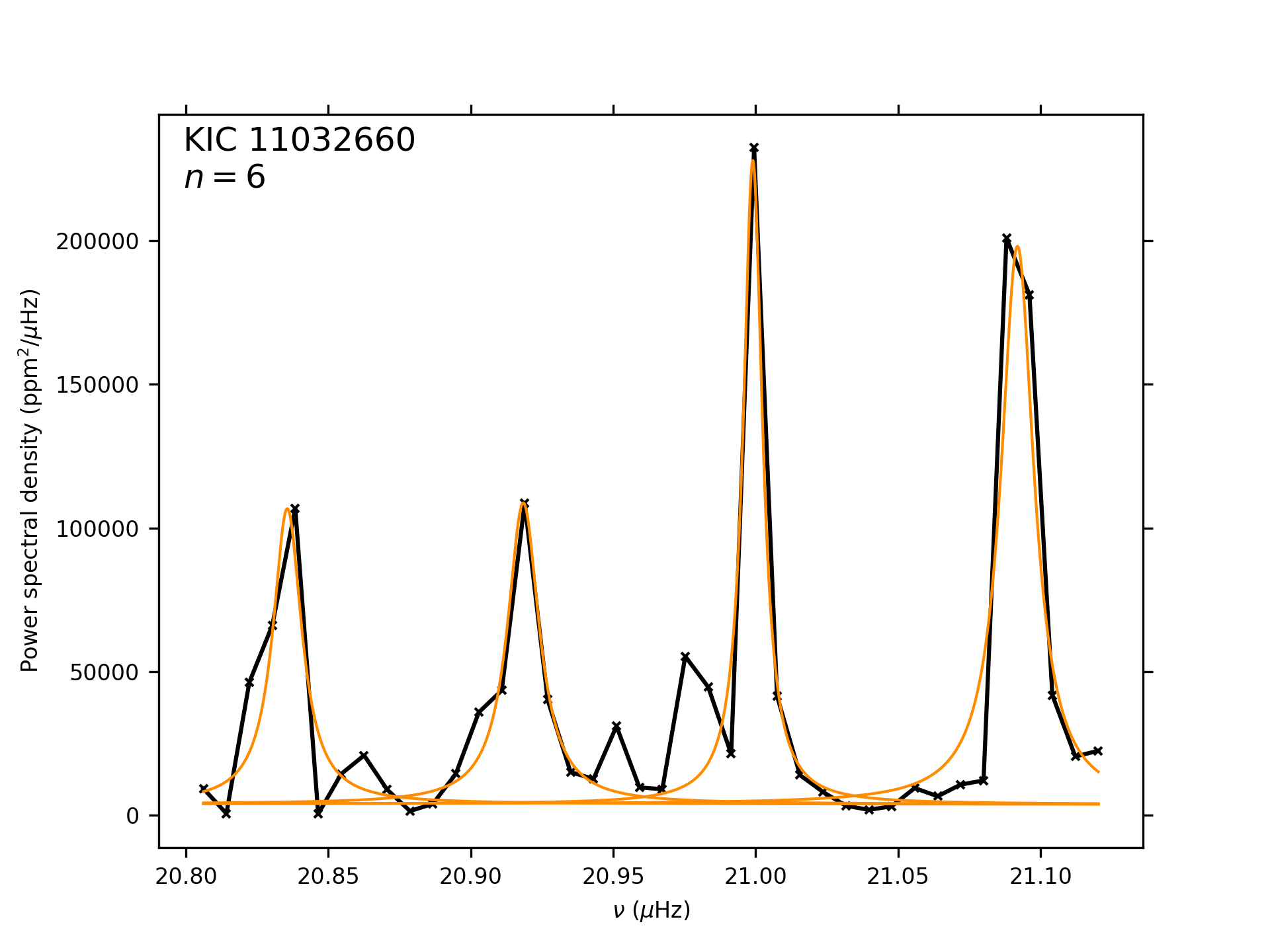}}
        \end{minipage}
        \begin{minipage}{1.\linewidth}  
        \end{minipage}
        \begin{minipage}{1.\linewidth}  
                \rotatebox{0}{\includegraphics[width=0.5\linewidth]{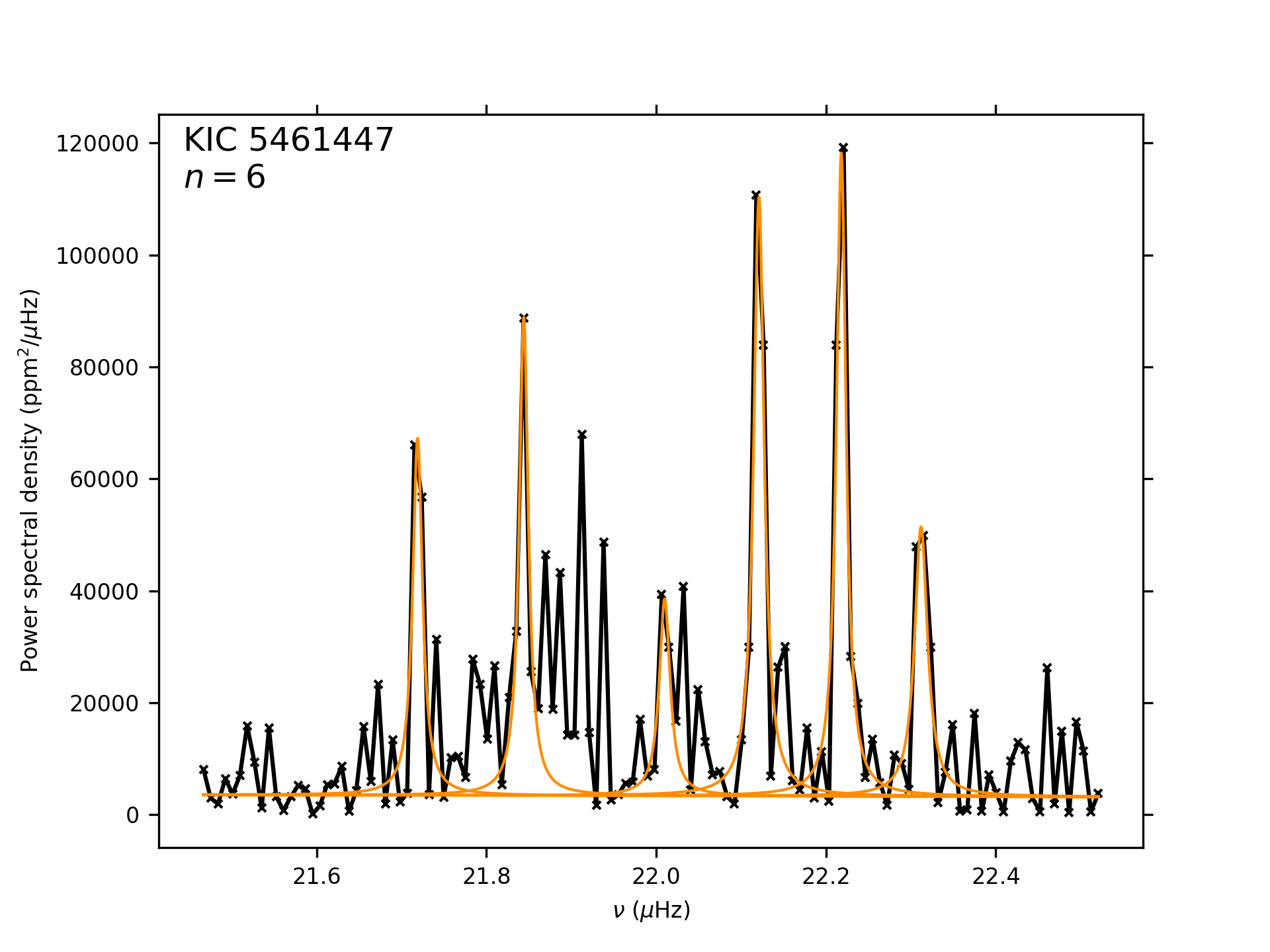}}
                \rotatebox{0}{\includegraphics[width=0.5\linewidth]{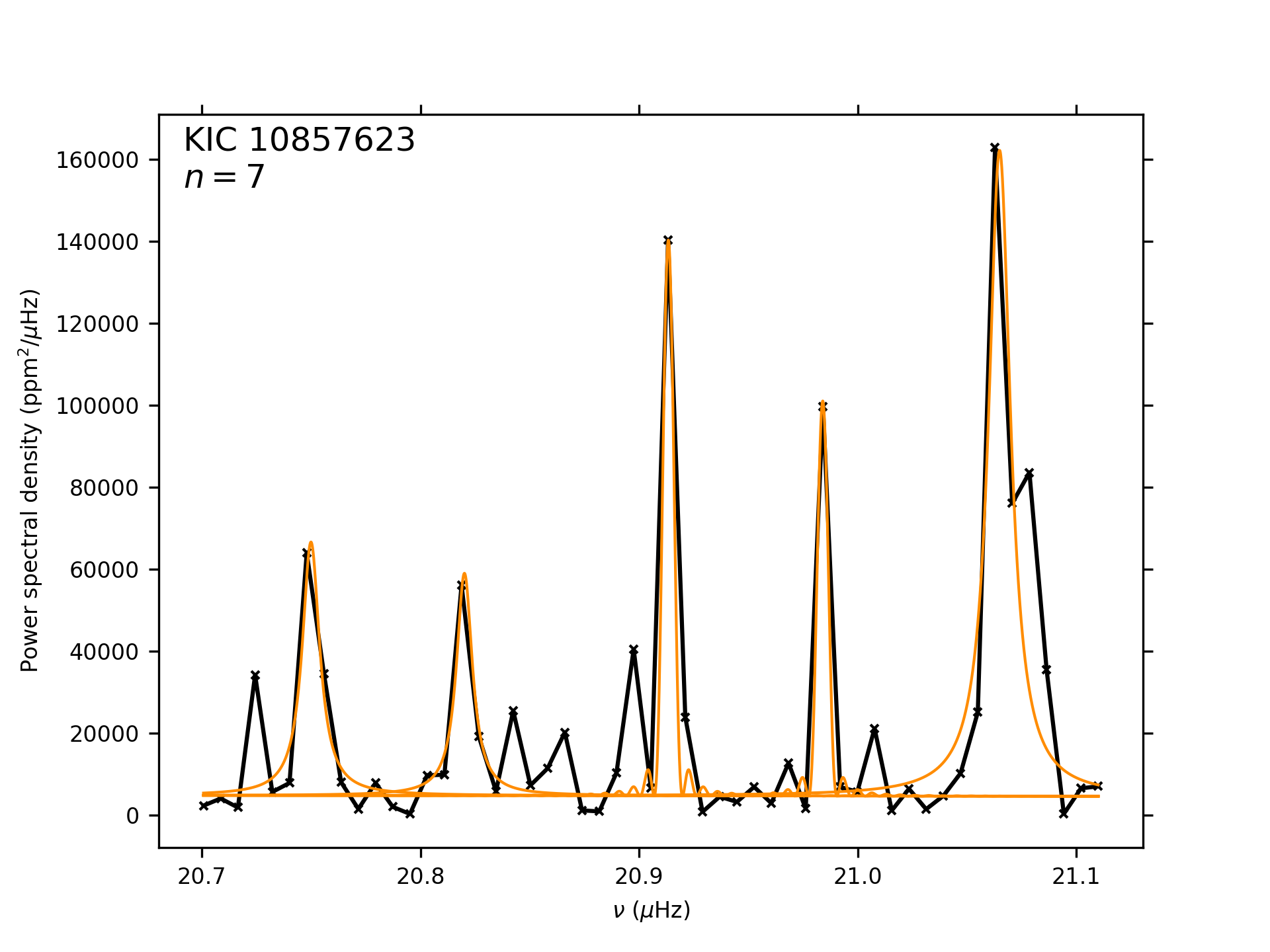}}
        \end{minipage}
        \caption{Mixed-mode pattern for KIC 6847371 ($V_{1}^{2} = 1.23 \pm 0.10$) at radial order $n = 5$ (top left) and at $n = 6$ (top right), for KIC 11032660 ($V_{1}^{2} = 1.14 \pm 0.10$) at $n = 5$ (middle left) and at $n = 6$ (middle right), for KIC 5461447 ($V_{1}^{2} = 1.15 \pm 0.15$) at $n = 6$ (bottom left), and for KIC 10857623 ($V_{1}^{2} = 1.01 \pm 0.12$) at $n = 7$ (bottom right). The stars are marked by stars in the lower right panel of Fig.~\ref{fig:Gamma_l_Dnu_T_eff} except for KIC 10857623 because we were unable to reliably extract its dipole mode widths following the method described in Sec~\ref{sec:mode_fitting_method}. Resolved modes are plotted by individual Lorentzians in blue, with the parameters given in Table~\ref{Table:mixed_modes_KIC_6847371}, while the unresolved modes at $\nu = 18.246~\mu$Hz (KIC 11032660), $\nu = 20.913~\mu$Hz, and $\nu = 20.984~\mu$Hz (KIC 10857623) are plotted by $\mathrm{sinc}^{2}$ functions.
        }
        \label{fig:mixed_modes_KIC_6847371}
\end{figure*}

\begin{figure}[htbp]
        \begin{minipage}{1.\linewidth}  
                \rotatebox{0}{\includegraphics[width=1.\linewidth]{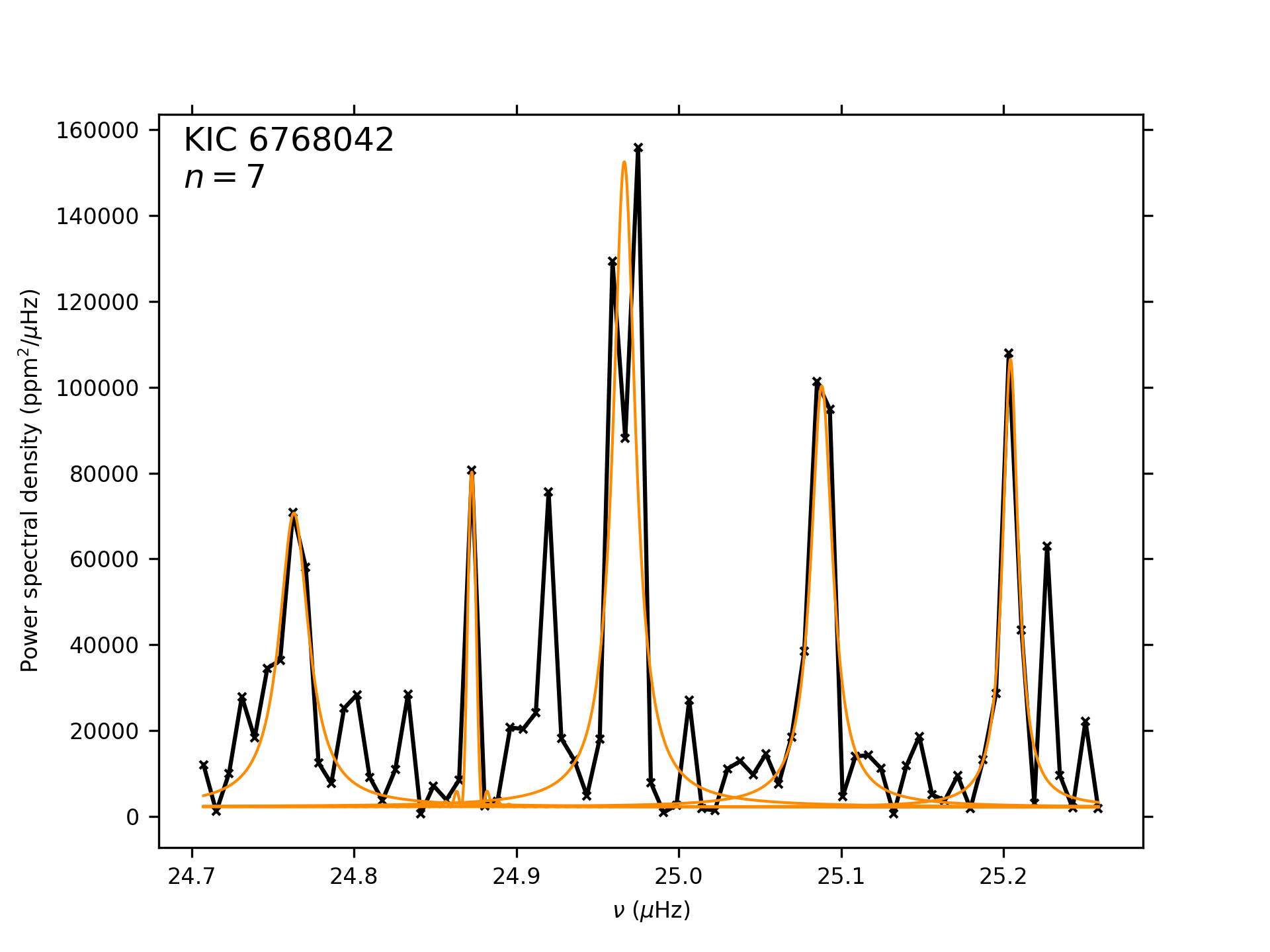}}
        \end{minipage}
        \begin{minipage}{1.\linewidth} 
                \rotatebox{0}{\includegraphics[width=1.\linewidth]{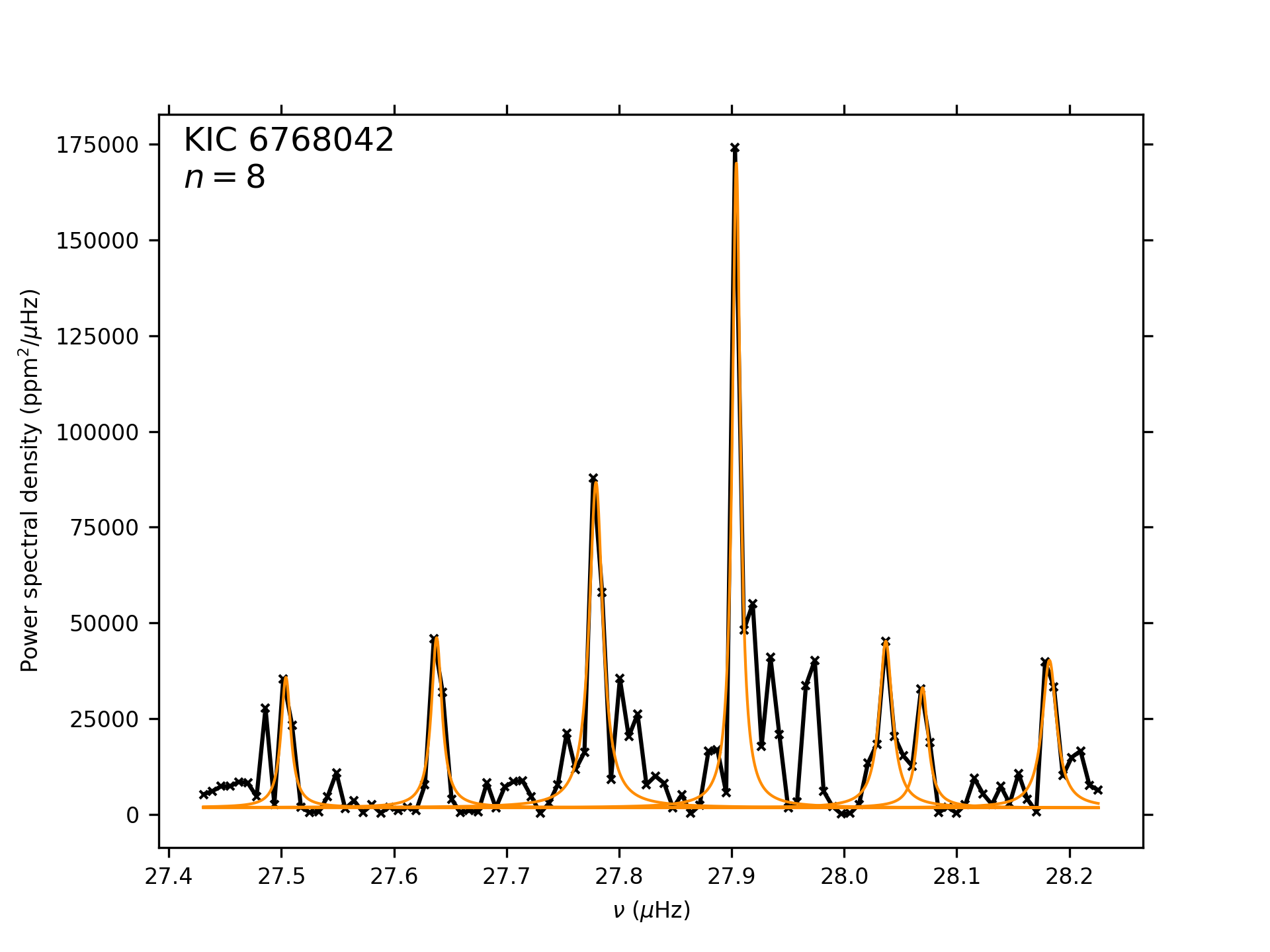}}
        \end{minipage}
        \caption{Same labels as in Fig.~\ref{fig:mixed_modes_KIC_6847371} for KIC 6768042 ($V_{1}^{2} = 1.38 \pm 0.13$) at radial order $n = 7$ (top) and at $n = 8$ (bottom). The mode located at $\nu = 24.873~\mu$Hz is unresolved and has been fitted by a $\mathrm{sinc}^{2}$ function.
        }
        \label{fig:mixed_modes_KIC_6768042}
\end{figure}

Results are shown in Figs.~\ref{fig:mixed_modes_KIC_6847371}, \ref{fig:mixed_modes_KIC_6768042}, and in the fifth column of Table~\ref{Table:mixed_modes_KIC_6847371}. Their average values are presented in Table~\ref{Table:from_mixed_modes_infer_pure_p}.

\begin{table}[htbp]

\caption{Pressure dipole mode widths compared to the radial mode widths}
\begin{tabular}{rrrrr}
\hline
\hline
KIC\ \ & $\Dnu$ & $\numax$ & $\langle\Gamma_{n,1}^{p}\rangle$ \ \ \ & $\Gammazero$ \ \ \ \ \ \\
& ($\mu$Hz) & ($\mu$Hz) & ($\mu$Hz) \ \ \ & ($\mu$Hz) \ \ \ \ \\
\hline
6847371 & $2.69$ & $19.84$ & $0.189 \pm 0.070$ & $0.128 \pm 0.032$ \\
11032660 & $2.83$ & $19.66$ & $0.142 \pm 0.051$ & $0.121 \pm 0.026$\\
5461447 & $2.97$ & $21.76$ & $0.163 \pm 0.051$ & $0.088 \pm 0.017$\\
10857623 & $2.49$ & $16.11$ & $0.083 \pm 0.029$ & $0.121 \pm 0.023$\\ 
6768042 & $2.96$ & $23.37$ & $0.138 \pm 0.069$ & $0.150 \pm 0.024$\\
  \hline
  \label{Table:from_mixed_modes_infer_pure_p}
\end{tabular}
\\
\textbf{Notes:} The average value of the dipole mode widths is computed as the arithmetic mean of the pressure mode widths $\Gamma_{n,1}^{p}$ presented in Table~\ref{Table:mixed_modes_KIC_6847371}.
\end{table}

\begin{table}[ht!]

\caption{Estimates of the ratio $\zeta$ between the mode inertia in the core and the total mode inertia, the mixed dipole mode widths $\Gamma_{n,1}$, heights $H_{n,1}$ and the pressure dipole mode widths $\Gamma_{n,1}^{p}$ for the stars KIC 6847371 at radial order $n = 5\text{and }6$ , KIC 11032660 at $n = 5 \text{ and }6$, KIC 5461447 at $n = 6$, KIC 10857623 at $n = 7$, and KIC 6768042 at $n = 7 \text{ and }8$.}
\begin{tabular}{rrrrrr}
\hline
\hline
KIC \ \  & $\nu$ \ \  &  $\zeta$ \ \ & $\Gamma_{n,1}$ & $\Gamma_{n,1}^{p}$ & $H_{n,1} \qquad$\\
& ($\mu$Hz)\ \ & & (nHz) & ($\mu$Hz) & (ppm$^{2}.\mu$Hz$^{-1}$) \\
\hline
6847371 & & & & & \\
& 17.221 & 0.912 & 14.0 & 0.173 & 148042$\qquad$\\
& 17.312 & 0.933 & 8.9 & 0.134 & 119977$\qquad$\\
& 17.385 & 0.951 & 12.6 & 0.256 & 185682$\qquad$\\
& 17.444 & 0.961 & 12.8 &  0.330 & 132426$\qquad$\\
& 19.763 & 0.917 & 10.8 & 0.130 & 142715$\qquad$\\
& 19.836 & 0.900 & 13.5 & 0.135 & 185918$\qquad$\\
& 19.916 & 0.895 & 8.1 & 0.077 & 595653$\qquad$\\
& 19.981 & 0.908 & 17.5 & 0.190 & 419721$\qquad$\\
& 20.056 & 0.929 & 18.0 & 0.254 & 114214$\qquad$\\
& 20.121 & 0.944 & 9.1 & 0.163 & 272377$\qquad$\\
& 20.198 & 0.958 & 9.8 & 0.233 & 121588$\qquad$\\
\hline
11032660 & & & & & \\
 & 18.054 & 0.901 & 9.1 & 0.092 & 66802$\qquad$ \\
 & 18.103 & 0.906 & 11.9 & 0.126 & 223603$\qquad$ \\
 & 18.176 & 0.922 & 14.7 & 0.189 & 223744$\qquad$ \\
 & 18.246 & 0.940 & 8.1* & 0.135 & 147870$\qquad$ \\
 & 18.300 & 0.956 & 8.5 & 0.191 & 149941$\qquad$ \\
 & 18.365 & 0.967 & 8.2 & 0.246 & 156975$\qquad$ \\
 & 20.836 & 0.892 & 12.3 & 0.113 & 106265$\qquad$ \\
 & 20.918 & 0.872 & 13.5 & 0.105 & 108367$\qquad$ \\
 & 20.999 & 0.881 & 8.2 & 0.069 & 231565$\qquad$ \\
 & 21.092 & 0.911 & 14.1 & 0.158 & 200770$\qquad$ \\
  \hline
5461447 & & & & & \\
  & 21.715 & 0.947 & 13.9 & 0.263 & 65961$\qquad$ \\
  & 21.844 & 0.902 & 12.9 & 0.131 & 88352$\qquad$ \\
  & 22.007 & 0.835 & 15.5 & 0.094 & 39418$\qquad$ \\
  & 22.121 & 0.879 & 14.9 & 0.123 & 110662$\qquad$ \\
  & 22.218 & 0.925 & 13.7 & 0.183 & 118859$\qquad$ \\
  & 22.315 & 0.954 & 18.4 & 0.398 & 49835$\qquad$ \\
  \hline
10857623 & & & & & \\
 & 20.745 & 0.924 & 9.7 & 0.127 & 63988$\qquad$ \\
 & 20.820 & 0.891 & 8.8 & 0.081 & 56137$\qquad$ \\
 & 20.913 & 0.848 &  7.8* & 0.051 & 140292$\qquad$ \\
 & 20.984 &  0.851 & 7.8* & 0.052 & 99708$\qquad$ \\
 & 21.065 & 0.882 & 12.1 & 0.103 & 163003$\qquad$ \\
  \hline
6768042 & & & & & \\
 & 24.763 & 0.839 & 22.1 & 0.138 & 70787$\qquad$ \\
 & 24.873 & 0.832 & 7.9* & 0.047 & 80691$\qquad$ \\
 & 24.967 & 0.878 & 16.6 & 0.136 & 155491$\qquad$ \\
 & 25.088 & 0.928 & 16.3 & 0.225 & 101482$\qquad$ \\
 & 25.204 & 0.953 & 11.5 & 0.243 & 107916$\qquad$ \\
 & 27.504 & 0.927 & 10.5 & 0.143 & 35161$\qquad$ \\
 & 27.639 & 0.863 & 11.3 & 0.082 & 45895$\qquad$ \\
 & 27.780 & 0.790 & 14.5 & 0.069 & 87860$\qquad$ \\
 & 27.904 & 0.820 & 8.7 & 0.048 & 174133$\qquad$ \\
 & 28.037 & 0.899 & 14.0 & 0.138 & 45005$\qquad$ \\
 & 28.070 & 0.909 & 11.7 & 0.129 & 32445$\qquad$ \\
 & 28.182 & 0.940 & 15.3 & 0.254 & 39866$\qquad$ \\
  \hline
  \label{Table:mixed_modes_KIC_6847371}
\end{tabular}
\\
\textbf{Notes:} The different estimates of $\Gamma_{6,1}^{p}$ are inferred from Eq.~\ref{eq:mixed_correction}. For these particular stars, the term $\zeta$ is not derived from scaling relations as described in Sect.~\ref{sec:Method}, but is extracted from the database of \citet{2018A&A...618A.109M}. (*) The measurement of the modes located at $\nu = 18.246~\mu$Hz (KIC 11032660), $\nu = 20.913~\mu$Hz and $\nu = 20.984~\mu$Hz (KIC 10857623), and $\nu = 24.873~\mu$Hz (KIC 6768042) are limited by the resolution.
\end{table}

\newpage
\section{Table available at the CDS with results of pressure mode-fitting}
\label{appendix:table_for_CDS}
The approach described in Sect.~\ref{sec:Method} allowed us to thoroughly characterise the pressure modes of 2103 evolved red giants observed by Kepler with $\Dnu \leq 4.0~\mu$Hz. A selection of the seismic parameters, obtained in Sect.~\ref{sec:results}, is shown in Table~\ref{appendix:table_for_CDS} for 25 stars. The complete set of global seismic parameters of the whole sample of stars is available at the CDS. The stellar mass and effective temperature were extracted from the APOKASC catalogue \citep{2014ApJS..215...19P}. For some stars, the stellar mass and the effective temperature are not listed in the APOKASC catalogue. It concerns roughly 5\% of our sample, with half of this fraction being associated to very low $\Dnu$-values (i.e. $\Dnu \leq 0.5\ \mu$Hz). For these stars, we nevertheless obtained rough estimates of the stellar mass and effective temperature using semi-empirical and empirical scaling relations implying both the frequency at the maximum oscillation power $\numax$ and large frequency separation $\Dnu$ \citep{1995A&A...293...87K, 2010A&A...509A..77K, 2010A&A...517A..22M}.

\begin{table*}
\caption{Seismic parameters.}
\begin{center}
\begin{tabular}{llllllllllllllllllllllllllllllllllllll}
\hline
\hline
\ \ \ KIC & \ \ \ $\Dnu$ & \ \ $M$ & \ $\Teff$ & \ \ $\varepsilon$ & \ $\delta\varepsilon$ & \ \ $d_{01}$ & $\delta d_{01}$ &\  \ $d_{02}$ & $\delta d_{02}$ & \ \ $d_{03}$ & $\delta d_{03}$ \\
& \ \ ($\mu$Hz) & ($M_{\odot}$) & \ (K) & & & & & & & & \\
\hline
01026309 & \ \ 1.944 & \ 2.58 & 4514 & 0.795     & 0.011     & -0.043    & 0.017     & 0.126    & 0.007     & 0.391    & 0.016 \\
01160789 & \ \ 3.524 & \ 0.86 & 4724 & 0.950     & 0.014     & -0.053    & 0.021     & 0.144    & 0.011     & 0.389    & 0.022 \\
01162746 & \ \ 3.804 & \ 0.85 & 4762 & 0.956     & 0.016     & -0.093    & 0.023     & 0.173    & 0.012     & 0.343    & 0.026 \\
01163359 & \ \ 2.644 & \ 1.67 & 4560 & 0.855     & 0.012     & -0.013    & 0.017     & 0.143    & 0.008     & 0.342    & 0.017 \\
01432587 & \ \ 1.082 & \ 0.85 & 4295 & 0.635     & 0.011     & -0.042    & 0.017     & 0.171    & 0.009     & 0.378    & 0.017 \\
01435573 & \ \ 3.587 & \ 0.90 & 4698 & 0.945     & 0.017     & -0.091    & 0.023     & 0.181    & 0.013     & 0.358    & 0.022 \\
01572780 & \ \ 2.693 & \ 0.97 & 4738 & 0.854     & 0.013     & -0.054    & 0.022     & 0.171    & 0.012     & 0.392    & 0.023 \\
01719297 & \ \ 1.215 & \ 1.29 & 4255 & 0.664     & 0.010     & -0.083    & 0.016     & 0.157    & 0.009     & 0.362    & 0.017 \\
01720425 & \ \ 3.667 & \ 1.09 & 4798 & 0.945     & 0.015     & -0.044    & 0.021     & 0.169    & 0.011     & 0.382    & 0.022 \\
01725552 & \ \ 1.221 & \ 1.53 & 4344 & 0.654     & 0.010     & -0.090    & 0.020     & 0.146    & 0.010     & 0.362    & 0.018 \\
01725732 & \ \ 0.707 & \ 0.87 & 4100 & 0.565     & 0.010     & -0.148    & 0.014     & 0.182    & 0.008     & 0.396    & 0.014 \\
01726211 & \ \ 3.720 & \ 1.32 & 4862 & 0.963     & 0.016     & -0.010    & 0.024     & 0.162    & 0.014     & 0.338    & 0.022 \\
01865595 & \ \ 1.815 & \ 1.34 & 4386 & 0.755     & 0.011     & -0.044   & 0.018     & 0.128    & 0.009     & 0.406    & 0.019 \\
01868101 & \ \ 3.785 & \ 1.28 & 4633 & 0.966     & 0.015     & -0.030    & 0.017     & 0.149    & 0.009     & 0.376    & 0.021 \\
01872517 & \ \ 3.299 & \ 1.14 & 4543 & 0.915     & 0.015     & -0.043    & 0.019     & 0.148    & 0.011     & 0.356    & 0.020 \\
01995358 & \ \ 3.238 & \ 1.17 & 4824 & 0.930     & 0.015     & -0.029    & 0.020     & 0.155    & 0.011     & 0.429    & 0.020 \\
02011582 & \ \ 3.863 & \ 2.15 & 4684 & 0.943     & 0.015     & -0.060    & 0.020     & 0.130    & 0.010     & 0.328    & 0.020 \\
02017541 & \ \ 1.457 & \ 1.33 & 4242 & 0.690     & 0.011     & -0.055    & 0.018     & 0.155    & 0.009     & 0.364    & 0.018 \\
02018392 & \ \ 3.789 & \ 1.52 & 4669 & 0.943     & 0.015     & -0.035    & 0.021     & 0.136    & 0.009     & 0.316    & 0.019 \\
02141932 & \ \ 3.013 & \ 1.37 & 4429 & 0.894     & 0.014     & -0.017    & 0.019     & 0.146    & 0.009     & 0.426    & 0.021 \\
02142095 & \ \ 3.694 & \ 1.17 & 4839 & 0.932     & 0.015     & -0.046    & 0.019     & 0.151    & 0.009     & 0.391    & 0.019 \\
02156178 & \ \ 3.824 & \ 0.95 & 4853 & 0.932     & 0.018     & -0.047    & 0.030     & 0.188    & 0.016     & 0.391    & 0.031 \\
02157059 & \ \ 3.002 & \ 1.27 & 4424 & 0.874     & 0.013     & -0.056    & 0.017     & 0.161    & 0.009     & 0.353    & 0.017 \\
02157901 & \ \ 3.795 & \ 1.05 & 4760 & 0.933     & 0.017     & -0.049    & 0.027     & 0.205    & 0.017     & 0.374    & 0.031 \\
02164874 & \ \ 1.779 & \ 1.45 & 4447 & 0.765     & 0.011     & -0.029    & 0.019     & 0.142    & 0.009     & 0.359    & 0.018 \\

\hline
\end{tabular}
\label{Table:data_CDS}
\end{center}
\textbf{Notes:} The columns correspond to, from left to right, the KIC number, the large separation $\Dnu$, the stellar mass $M$, the effective temperature $\Teff$, the acoustic offset $\varepsilon$, the uncertainty on $\varepsilon$, the reduced small separations $d_{0\ell}$ and the uncertainties on $d_{0\ell}$. The list of the full data set, including the glitch parameters, the mean mode widths, amplitudes, visibilities and the evolutionary stages, is available at the CDS.
\end{table*}
\end{appendix}

\newpage 
\bibliographystyle{./Outils_latex/aa} 
\bibliography{./article_RGB_AGB_stars} 


\end{document}